\documentclass{article}

\usepackage{arxiv}

\usepackage[utf8]{inputenc} 
\usepackage[T1]{fontenc}    
\usepackage[colorlinks=true, urlcolor=blue, linkcolor=red]{hyperref} 
\usepackage{xurl}           
\usepackage{booktabs}       
\usepackage{amsfonts}       
\usepackage{nicefrac}       
\usepackage{microtype}      
\usepackage{lipsum}		    
\usepackage{graphicx}
\usepackage{natbib}
    \setcitestyle{citesep={;}}  
\usepackage{doi}
\usepackage{tcolorbox}     
\usepackage{subcaption}    
\usepackage{fancyvrb}      
\usepackage{placeins}      
\usepackage[table]{xcolor} 
    \definecolor{softred}{RGB}{255, 179, 179}
    \definecolor{softgreen}{RGB}{179, 230, 179}
    \definecolor{softyellow}{RGB}{255, 242, 180}
\usepackage{longtable}     
\usepackage{array}         
\usepackage{subcaption}    
\usepackage{tikz}          
\usepackage{needspace}     

    \usepackage{titlesec}       
    \titleformat{\section}{\normalfont\LARGE\bfseries}{\thesection}{1em}{}
    \titleformat{\subsection}{\normalfont\Large\bfseries}{\thesubsection}{1em}{}
    \titleformat{\subsubsection}{\normalfont\large\bfseries}{\thesubsubsection}{1em}{}
    \titleformat{\paragraph}
    {\normalfont\fontsize{11}{13}\selectfont\bfseries}{\theparagraph}{1em}{}
    \titlespacing*{\paragraph}  
    {0pt}{1.5ex plus 1ex minus .2ex}{0.5ex plus .2ex}
\title{Defining AI Models and AI Systems:\\A Framework to Resolve the Boundary Problem}

\author{
  \textbf{Yuanyuan~Sun $^{1}$ \quad
  Timothy~Parker$^{7}$ \quad
  Lara~Gierschmann$^{7}$ \quad
  Sana~Shams$^{2}$ \quad 
  Teo~Canmetin$^{3}$} \\[0.5em]
  \textbf{Mathieu~Duteil$^{7}$ \quad
  Rokas~Gipiškis$^{4,5}$ \quad
  Ze~Shen~Chin$^{4,6}$}\thanks{Corresponding author: \texttt{zeshen@aistandardslab.org}} \\[2em]
  $^{1}$AI Governance Exchange \\
  $^{2}$University of British Columbia \\
  $^{3}$Oxford Internet Institute, University of Oxford \\
  $^{4}$AI Standards Lab \\
  $^{5}$Vilnius University \\
  $^{6}$Oxford Martin AI Governance Initiative \\
  $^{7}$Independent
}

\renewcommand{\undertitle}{} 
\renewcommand{\shorttitle}

\hypersetup{
pdftitle={A template for the arxiv style},
pdfsubject={q-bio.NC, q-bio.QM},
pdfauthor={David S.~Hippocampus, Elias D.~Striatum},
pdfkeywords={First keyword, Second keyword, More},
}

\begin{document}
\maketitle

\begin{abstract}

Emerging AI regulations assign distinct obligations to different actors along the AI value chain (e.g., the EU AI Act distinguishes providers and deployers for both AI models and AI systems), yet the foundational terms \textit{AI model} and \textit{AI system} lack clear, consistent definitions. Through a systematic review of 896 academic papers and a manual review of over 80 regulatory, standards, and technical or policy documents, we analyze existing definitions from multiple conceptual perspectives. We then trace definitional lineages and paradigm shifts over time, finding that most standards and regulatory definitions derive from the OECD's frameworks, which evolved in ways that compounded rather than resolved conceptual ambiguities. The ambiguity of the boundary between an AI model and an AI system creates practical difficulties in determining obligations for different actors, and raises questions on whether certain modifications performed are specific to the model as opposed to the non-model system components. We propose conceptual definitions grounded in the nature of models and systems and the relationship between them, then develop operational definitions for contemporary neural network-based machine-learning AI: models consist of trained parameters and architecture, while systems consist of the model plus additional components including an interface for processing inputs and outputs. Finally, we discuss implications for regulatory implementation and examine how our definitions contribute to resolving ambiguities in allocating responsibilities across the AI value chain, in both theoretical scenarios and case studies involving real-world incidents.

\end{abstract}



\section{Introduction}

The effective regulation of AI requires clear definitions for the terms contained within said regulation. This makes it easier to consistently enforce the regulation, while also providing clarity to AI developers regarding their responsibilities. Clarifying definitions allows for more meaningful academic and public discussion on this increasingly important subject. Two particularly fundamental definitions are \textit{AI model} and \textit{AI system}. Therefore, it is our intention to provide a thorough survey and categorisation of the existing definitions, along with a discussion of our findings and recommendations for the most important considerations when selecting which definitions to use.

The importance of clear definitions around AI has already been demonstrated multiple times in the real world. One example is the AlienChat dispute in China, where developers of the application were convicted for producing obscene materials \citep{Zhang2026Jailed}. In this case, the court decided that the fault lies with the developer of the system and not the underlying model, as the system-level components were found to have elicited this behavior. Another example is the 2023 case of the Italian Data Protection Authority banning ChatGPT from operating in Italy, citing the failure to follow General Data Protection Regulation (GDPR) in the collection and processing of the personal data used to train the GPT-3.5 model (the model powering ChatGPT at the time) \citep{Garante202303ChatGPT}. While service was reinstated after OpenAI implemented additional disclosure mechanisms \citep{Garante202304ChatGPT}, it highlighted a potential regulatory gap whereby restrictions were imposed at the system level (ChatGPT as a service), yet the relevant data may have already embedded within the model itself, beyond the scope of the imposed remedies. A third example is the case of Clearview AI, a US company that developed an AI facial recognition system trained on up to three billion images \citep{Jung2024Privacy}. Regulators argued that this system violated privacy rights, while Clearview responded that the facial recognition algorithm and the database of embeddings were separate technical components, making the algorithm a neutral backend model distinct from the data itself. As a result, Clearview was able to continue operating in most of the US, but not in the EU, Canada or Australia. While these jurisdictional outcomes were driven primarily by differences in privacy law frameworks, they also illustrate the kind of definitional ambiguity surrounding systems and models that this paper addresses. These case studies are discussed in more detail in Section \ref{subsec:CaseStudies} where we illustrate how our proposals may contribute towards resolving these cases.

Our review reveals that the definitional landscape for \textit{AI model} and \textit{AI system} is characterised by a concentrated chain of influence. The majority of regulatory and standards definitions trace back to a small number of foundational sources, most notably Russell and Norvig's textbook, \textit{Artificial Intelligence: A Modern Approach 3rd Edition} \citep{Russell2009AI}, and \textit{Scoping the OECD AI Principles} \citep{OECD2019Scoping}. While this convergence has facilitated a degree of international harmonisation, it has also propagated ambiguities.

Furthermore, it appears definitions of AI system are relatively well-developed and numerous, appearing across regulatory instruments ranging from the EU AI Act to ISO standards and NIST frameworks. In contrast, definitions of AI model remain sparse, fragmented, and often purely derivative, defined only in relation to systems or through specific subtypes such as general-purpose AI models without a clear base definition. This imbalance reflects the historical focus of governance discourse on system-level considerations, yet it creates practical difficulties as regulatory frameworks (such as the EU AI Act) seek to assign distinct obligations to model providers and system providers.

The rapid pace of AI development has further complicated definitional efforts. Terms that were developed to describe classical agent architectures or early machine learning systems strain under application to modern foundation models and large language models (LLMs). The very word \textit{model} carries semantic baggage from its origins in representing external phenomena. This framing fits poorly with current frontier LLMs, which are not obviously models of anything in the traditional sense (particularly post-fine-tuning). 

This paper addresses these challenges through a systematic examination of existing definitions and using what we have learned to propose our own improved definitions. We aim to serve multiple audiences: policymakers seeking to draft or interpret AI legislation, standards bodies working toward international harmonisation, technical and legal teams of AI developers seeking clarity on their compliance obligations, and researchers engaged in discourse on AI governance including those involved in the interpretation of case law. We recognise that effective definitions must be administrable by and comprehensible to non-specialists, while remaining coherent with how the technical community understands and uses these terms.

The remainder of this paper is organised as follows. In Section \ref{sec:related} (Related Work), we review prior scholarship on AI terminology and definitions, situating our contribution within the existing literature. Section \ref{sec:methods} (Methods) describes our dual approach combining a systematic literature review of academic sources with a manual review of regulatory and standards documents across major jurisdictions. In Section \ref{sec:findings} (Findings and Analysis), we gather the results of these reviews, including a survey of existing definitions from intergovernmental organisations, standards bodies, governmental actors, and NGOs, and provide an analysis of conceptual perspectives that organise definitions according to their underlying characterisations of models and systems. Subsequently, we describe shifting paradigms and examine how definitions have evolved over time, analyze the observed chain of influence across definitions, and identify key inconsistencies in foundational frameworks. Section \ref{sec:definitions} (Developing definitions for \textit{AI models} and \textit{AI systems}) presents our proposed conceptual and operational definitions, along with criteria for evaluating definitional quality and the challenges inherent in drawing boundaries between models and systems, as well as consider the implications of our proposals for regulatory implementation and examine case studies illustrating model-system boundary disputes. Section \ref{sec:limitations} (Limitations) acknowledges constraints on our analysis and remaining areas of ambiguity. Finally, Section \ref{sec:conclusions} (Conclusion) summarises our contributions and identifies directions for future work.


\section{Related Work}
\label{sec:related}

To the best of our knowledge, there is no previous work that systematically surveys the different definitions of AI models and systems. However, there are a number of works that are related to our paper through providing reviews of related terms, or discussing the importance of clear terminology around AI. Note that works that simply present definitions of AI models or AI systems without substantial analysis will be surveyed in Section \ref{sec:findings} (Findings and Analysis).

\citet{Maas2023Concepts} provides a comprehensive overview of many important terms and concepts related to the governance of advanced AI, alongside a taxonomy and preliminary analysis. He discusses three different purposes for seeking definitions of AI (technological, sociotechnical, and regulatory), the importance of such terminology in shaping AI policy and law, and then reviews a total of 101 definitions across 69 terms relevant to advanced AI. Finally, he reviews various concepts within AI governance. However, he does not provide definitions of \textit{AI model} or \textit{AI system}, though he does define more specific terms such as \textit{general-purpose AI system} and \textit{foundation model}. These terms were also missing in other taxonomies and overviews of definitions that we surveyed \citep{Bandi2025Agentic,Mukhamediev2022Review,Muller2021Industrial,Sposato2025Educational,Triguero2024GPAIS}.

\citet{Murdick2020Definitions} discuss the importance of clear definitions around AI, and propose key principles for writing such definitions, and present a definition of AI research that leverages judgments from AI experts using machine learning. They also discuss the implications of their definition for our understanding of the state of international AI competition.

\citet{Fernandez-Llorca2025Account} provide an interdisciplinary analysis of the terms and definitions used in official documentation around the EU AI Act (whereas our work surveys a broader set of literature), particularly the terms \textit{AI system}, \textit{general-purpose AI system}, \textit{foundation model} and \textit{generative AI}. They also address the distinction between \textit{AI system} and \textit{AI model} briefly, but do not discuss in detail the boundaries between the two. They also highlight the importance of making definitions accessible to both AI experts and lawyers in order for legislation to be effective.

In a similar vein, \citet{Trincado2024Legal} charts the history of the EU definition of \textit{AI system}, including the debate over whether the definition should be ``broad'' (any techniques considered to be AI plus other software) or ``narrow'' (only specific techniques) and how the definition evolved during the legislative process of the AI Act. He also surveys definitions from other sources, including the US, China and the OECD.

\citet{Tomic2025Octopus} argue that the definition of \textit{AI system} used in the EU AI Act is primarily a conceptual definition, and is thus insufficient for effective AI regulation in practice. As a complement to the current conceptual definition, they present an operational definition of AI systems, focusing on the immutable features that any AI system would be expected to possess: decision models, data, and an interface. They also discuss a number of areas where current EU AI regulation could be amended in light of their work.

Similarly, \citet{Burden2025Framework} provide a framework for identifying whether models qualify as general-purpose AI (GPAI) models based on their capabilities and generality. The framework is based on four core components of general-purpose models: attention and search, comprehension and compositional expression, conceptualisation, learning and abstraction; and quantitative and logical reasoning. They give examples of their framework with empirical cases from existing models, and make policy recommendations for thresholds and metrics when categorising GPAI models.

\citet{Krafft2019Defining} also highlight the dangers of ambiguous definitions around AI, and compare the definitions favoured by AI researchers and policymakers. They find that the definitions of AI researchers emphasise technical functionality, while those of policymakers use comparisons to human thinking and behavior.

Finally, both \citet{AlgorithmAudit2025AIAct} and the \citet{EC2025GPAI,EC2025Guidelines} have provided guidelines for the implementation of the EU AI Act focusing on the definition of an AI system, demonstrating demand for clarity on this topic.


\section{Methods}
\label{sec:methods}

First, the literature review was conducted in two ways: a manual review and a systematic literature review (SLR). Completing a manual review of the literature ensured that important papers in the legislative context were included, while the SLR allowed for the inclusion of academic findings and discussions on existing definitions, capturing both well-established definitions as well as variations across more domain-specific research. Next, we analysed the conceptual perspectives of these definitions, and traced how past sources have influenced later definitions. Finally, we outline the approach taken to propose our own definitions. 

\subsection{Manual Review of Global Regulatory and Standards Documents}
\label{subsec:manual}

Alongside the SLR, we conducted a manual review of how ``AI system'' and ``AI model'' are defined across government documents, regulations and policy documents. These include key documents related to the jurisdictions of the EU, United States, and China — published up to October 2025. Chinese-language documents were assessed and translated by a native Mandarin speaker with expertise in policy terminology, reducing ambiguities associated with machine translation. Because relevant materials are decentralised and fragmented across various internet archives, websites, repositories, and legal publishers, we opted for manual search rather than automation when reviewing texts. Keywords such as ``definition'', ``glossary'', ``system'', ``model'' were used to identify relevant sections across documents. These sections were then screened against our inclusion and exclusion criteria, similar to that being used for the systematic literature review as described in Section \ref{subsec:SLR} (Systematic Literature Review). For example, the use of the word ``model'' outside of contexts related to AI were excluded. We grouped institutions into four categories: (1) national/state-level government bodies, (2) intergovernmental organisations, (3) standards bodies, and (4) NGOs. Verbatim definitions were extracted and tabulated with source, date, and organisational category.

\subsection{Systematic Literature Review}
\label{subsec:SLR}

To identify how \textit{AI model} and \textit{AI system} are defined in academic literature, we conducted a systematic literature review following \citet{Carrera-Rivera2022Review}. We searched Scopus, Web of Science, and IEEE Xplore for papers from 2012 onwards, marking the beginning of the modern deep learning era with AlexNet \citep{Krizhevsky2012ImageNet}. To ensure papers actually discussed definitions rather than merely using these terms, we employed proximity operators (NEAR/4 or W/4) requiring definitional terms (``definition'', ``ontology'', ``taxonomy'', etc.) to appear within four words of ``AI model'', ``AI system'', or their synonyms. After excluding papers focused on domain-specific applications (e.g., medical diagnosis, educational contexts) to maintain focus on general conceptual definitions, 896 papers remained after deduplication. Papers were then assessed for definition clarity and completeness, with particular attention to whether definitions were original or cited from other sources. Full details including search strings, PICOC criteria, and complete inclusion/exclusion criteria are provided in Appendix \ref{subsec:SLROverview} Overview of the Systematic Literature Review. 

\subsection{Analysis of the Definitions}

Using the definitions compiled from the systematic literature review and the manual review, we analysed the different conceptual perspectives of these definitions. We identified three key dimensions along which definitions vary: (i) parent categories (subordinate concept class), (ii) features (production, function, and for systems, mechanism), and (iii) relationships between models and systems. We also traced citation patterns to identify influential sources and analysed how past definitions have influenced recent regulatory frameworks, particularly the EU AI Act.

\subsection{Development of Proposed Definitions}

We developed evaluation criteria for proposed definitions by drawing on principles from ISO 704:2022 (technical concept definition) \citep{ISO704:2022}, legal drafting scholarship, and AI governance literature. These sources informed criteria related to structural clarity, technical coherence, and legal operability. Using these criteria, we developed both conceptual definitions (clarifying fundamental nature and relationships) and operational definitions (enabling practical regulatory application). For operational definitions, we specifically addressed contemporary neural network-based machine learning AI, where we systematically analysed aspects of a transformer-based LLM (e.g. weights, prompts, activations, RAG, fine-tuning adapters, filters, etc.) to establish if these aspects are part of a model or a system.


\section{Findings and Analysis}
\label{sec:findings}

This section presents the results of both manual and systematic literature reviews examining how \textit{AI models} and \textit{AI systems} are defined across academic and regulatory sources. These findings provide the empirical foundation for understanding the definitional landscape and the relationships between different formulations of \textit{AI model} and \textit{AI system} definitions. Then, we discuss paradigm shifts that lead to an evolution of the meanings of these terms. Finally, we explore definitional lineages in major academic references and regulatory frameworks, and investigate how changes in definitions in the OECD frameworks have affected later definitions.

\subsection{Conceptual Perspectives on Existing Definitions}
\label{subsec:conceptual}

As described in Section \ref{subsec:manual} (Manual Review of Global Regulatory and Standards Documents), we manually gathered over 80 definitions from four categories of institutions: (i) intergovernmental organisations, (ii) standards bodies, (iii) national or state-level governmental organisations, and (iv) NGOs. Across documents produced by governments, regulatory agencies, inter-governmental and non-governmental organisations, we observed that AI definitions often remain ambiguous and inconsistent, particularly at the boundary between an AI model and an AI system. We provide a compilation of these definitions in Appendix \ref{subsec:CompilationDefinitionManualReview} (Compilation of Definitions from the Manual Review).

Additionally, based on the methods described in Section \ref{subsec:SLR} (Systematic Literature Review), we conducted an SLR which screened 896 records from the literature resulting in 37 studies included. The process and findings of this SLR is further described in Appendix \ref{subsec:SLROverview} (Overview of the Systematic Literature Review), and a compilation of these definitions are included in Appendix \ref{subsec:CompilationDefinitionManualReview} (Compilation of Definitions from the Systematic Literature Review).

Collectively, the definitions gathered through both the manual and systematic literature review can be interpreted from several perspectives. While many definitions appear similar, they can nonetheless be considered from different angles that highlight different conceptual emphases. Overall, when including both the papers found through the SLR and the manual review, more definitions were found for \textit{AI system} and related terms (83 instances), than for \textit{AI model} and related terms (43 instances), without excluding double-counting for instances that appear in both the manual review and the SLR. Included in the following conceptual perspectives were the terms \textit{AI model}, \textit{machine learning model}, \textit{AI system}, and \textit{machine learning system}, as well as sources providing clarifications on autonomous systems and intelligent systems.

\begin{figure}[h]
    \centering
    \includegraphics[width=1\linewidth]{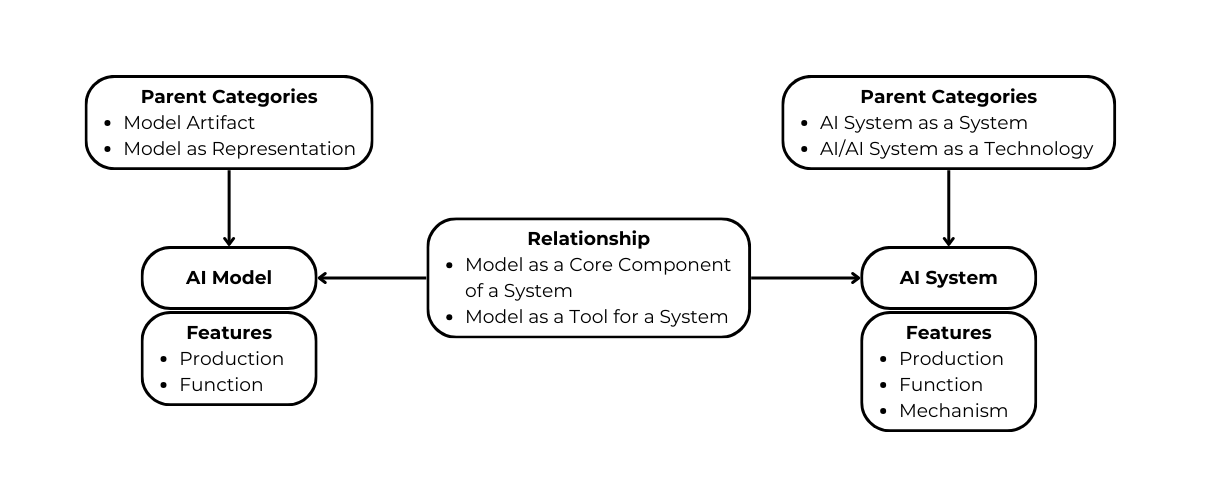}
    \caption{Overview of conceptual perspectives for definitions of \textit{AI model} and \textit{AI system}.}
    \label{fig:OverviewConceptualDefinitions}
\end{figure}

We organise these conceptual perspectives along three dimensions (Figure \ref{fig:OverviewConceptualDefinitions}). The first captures what kind of ``thing'' an AI model or system is understood to be; for example, a representation, a component, or a system. This is about identifying the subordinate concept of a term, which ISO 704:2022 \citep{ISO704:2022} notes as an important component of a definition. We refer to this dimension as ``parent categories'', as it reflects the broader conceptual class to which the definition assigns the term.

The second dimension addresses the features of the model or system referenced in the definitions. For both models and systems, this addresses their production and function, analysing how models and systems are made, and what they do respectively. Additionally, for AI systems, this dimension also addresses its mechanism in order to define how the system operates.

Third, we also consider the relational dimension, looking at the relationship between AI models and AI systems and how they are distinguished. Having a closer look at this difference provides further clarity on both terms. We refer to this dimension as ``relationship between AI model and AI system''. 

As an example of these dimensions, one source might define an AI model as the result of machine learning algorithms (parent category), used to make predictions (features), and referred to as the reasoning component of an AI system (relationship). Any of these perspectives can overlap within a single definition. For example, several definitions define an AI model as both the core reasoning or knowledge component of an AI system (relationship), and also as a representation of some external structure (parent category).

The following sections elaborates on these different perspectives with several examples. A more comprehensive list of references behind these perspectives can be found in Appendix \ref{subsec:ReferencesConceptualDefinitions} (References on Conceptual Perspectives of Definitions).

\subsubsection{AI Model}

We first consider how the term \textit{AI model} is conceptualised across definitions. The definitions can be roughly divided into two parent categories: model artifact and model as representation. Definitions under model artifact describe models as the result of building or training, specifying the required elements a model possesses after building (e.g. weights, parameters, or the specific architecture). Definitions describing models as representations emphasise not the way a model has been built, but its purpose of providing a representation of some external structure (e.g. representing laws of the external world or the data it has been trained on).

\vspace{0.3cm}
\textbf{Parent Categories:}

\begin{enumerate}
    \item \textbf{Model artifact (result of building/training)}: Focusing on weights, parameters, or architecture.
        \begin{enumerate}
            \item ``A model can be based on data and/or expert knowledge, by humans and/or by automated tools like machine learning algorithms.'' \citep[p. 7]{OECD2019Scoping}
        \end{enumerate}
    \item \textbf{Model as representation}: A model represents some external structure (e.g. the laws of the external world or the distribution of token sequences in a dataset).
        \begin{enumerate}
            \item ``AI models are mathematical representations that capture, in a set of parameters, the patterns underlying their training personal data'' \citep[p. 9]{EDPS2025Risk}
            \item ``physical, mathematical or otherwise logical representation of a system, entity, phenomenon, process or data'' \citep[ISO/IEC 22989:2022 3.1.23]{ISOIEC22989:2022-3.1.23}
            \item ``A Model is an actionable representation of all or part of the external environment of an AI system that describes the environment’s structure and/or dynamics'' \citep[p. 7]{OECD2019Scoping}
        \end{enumerate}
\end{enumerate}

Next, the features of an AI model are split into two perspectives: production and function. In accordance with the model life cycle, we first consider how models are built, and next what they are used for after being built. Accordingly, the production perspective contains all definitions referring to how a model is built or trained, including specific techniques or approaches. For example, a model can be described as being built using machine learning. The function perspective addresses definitions describing what a model is used for after building. This can be its intended use, specifications on its purpose, and describing the capabilities and characteristics it possesses to fulfil its function.

\vspace{0.3cm}
\textbf{Features:}

\begin{enumerate}
    \item \textbf{Production (how it is built/trained)}: Considering the techniques or approach used to develop or create the model (e.g. use machine learning with data or hard-coded expert knowledge).
        \begin{enumerate}
            \item ``uses computational, statistical, or machine-learning techniques to produce outputs from a given set of inputs'' \citep[Sec. 3(c)]{EOPresident2023Safe}
        \end{enumerate}
    \item \textbf{Function (what it is used for after building)}: 
        \begin{enumerate}
            \item Emphasising what the model is intended to do (e.g. inference, prediction, pattern recognition).
                \begin{enumerate}
                    \item ``used to make inferences from inputs to produce outputs'' \citep[p. 8]{OECD2024Explanatory}
                    \item ``can make predictions/decisions over data'' \citep[p. 43]{EstevezAlmenzar2022Glossary}
                \end{enumerate}
            \item Wide purpose or narrow purpose: Describing its characteristics and capabilities (adaptable across a wide range of tasks or specific to one).
                \begin{enumerate}
                        \item ``every AI Model possesses the potential to adapt to other tasks, some (for example, neural networks) even being universal approximator'' \citep[p. 1]{Li2022SAIBench}
                \end{enumerate}
        \end{enumerate}
\end{enumerate}

These perspectives are non-exclusive and orthogonal, meaning that definitions are not separated into different categories, but can fall into several different perspectives. Furthermore, perspectives on the features of a model only suggest different focal points of a definition. Therefore, different definitions grouped into these features may have other differences, as long as they share the same focal point of either production or function. For example, one definition might say a model is built using machine learning, while another might say it is built using hard-coded expert knowledge; but they both are defined with similar functions. It is also important to note that while two definitions may occupy different points along the first dimension (different parent categories of \textit{AI model}), they can agree on the same definition along the second dimension (the same features are referenced in the definition). For example, one definition might describe a model as the result of machine learning, while the other describes it as a representation of the external world, resulting in different parental categories. However, they might both describe the function of an AI model as making inferences.

\subsubsection{AI System}

In contrast to AI models, AI systems are defined more frequently and comprehensively. As with models, definitions can be viewed from two different dimensions: the parent categories and the features. Definitions of \textit{AI system} can be divided into two different parent categories: AI  systems as a type of system or as a type of technology. Many definitions describe AI systems as a type of system, specifying for example an AI system as a kind of machine-based system or an engineered system. Definitions falling into the technology category are focused on a broader category, either describing AI systems as technologies, or defining AI as a system, suggesting that the terms \textit{AI} and \textit{AI system} are often used interchangeably.

\vspace{0.3cm}
\textbf{Parent Categories:}

\begin{enumerate}
    \item \textbf{AI system as a specific system}: AI systems as a specific type of system.
        \begin{enumerate}
            \item ``An AI system is a machine-based system'' \citep[p. 4]{OECD2024Explanatory}
            \item ``an engineered system'' \citep[ISO/IEC 22989:2022 3.1.4]{ISOIEC22989:2022-3.1.4}
            \item ``Artificial intelligence (AI) systems are software (and possibly also hardware) systems'' \citep[p. 36]{HLEGAI2019Ethics}
        \end{enumerate}
    \item \textbf{AI/AI system as a technology}: AI systems as a technology and AI as AI-based systems in general.
        \begin{enumerate}
            \item ``AI systems are information-processing technologies'' \citep[p. 10]{UNESCO2022Ethics}
            \item ``Artificial Intelligence (AI) refers to systems'' \citep[p. 7]{HLEGAI2019Definition}
        \end{enumerate}
\end{enumerate}

The features of an AI system are grouped into similar perspectives as for an AI model. The first angle, considering the production, addresses what a system is made of and which techniques were used to build it. For example, a system can be described as requiring different components, such as an AI model, data pipelines, and an architecture. The second perspective describes the function of an AI system, entailing definitions that describe what an AI system does (e.g. giving predictions) and what kind of action it is conducting (e.g. acting autonomously or affecting the physical world). In addition to the perspectives of production and function, definitions of AI system have also been found to describe its mechanism. This perspective answers how the system does what it is created to do. For example, if a system is built to provide recommendations (function), this perspective may describe it as generating recommendations by inferring an output from the input it receives, or by perceiving, planning and reflecting.

\vspace{0.3cm}
\textbf{Features:}

\begin{enumerate}
    \item \textbf{Production (how it is built)}: What a system is made of and the techniques used to build it.
        \begin{enumerate}
            \item ``can either use symbolic rules or learn a numeric model'' \citep[p. 36]{HLEGAI2019Ethics}
            \item ``more `grown' or `trained' than they are programmed [...] produced by rewarding an AI for `good behavior' and punishing it for `bad behavior''' \citep{CHT2024Society}
            \item ``Bringing one or more AI models to a real-world application is not immediate, as it implies the effort of integrating them in a functional system (i.e., an AI system), including the necessary infrastructure, user interfaces, data pipelines, and other components required for the application to operate effectively in a production environment'' \citep[p. 3]{Hupont2024Cards}
        \end{enumerate}
    \item \textbf{Function (what it does)}: What is the system intended to do and what kind of actions it performs (intelligent action, impactful action). 
        \begin{enumerate}
            \item ``generate outputs such as predictions, content, recommendations, or decisions'' \citep[Article 3(1)]{EUParliament2024AIAct}
            \item ``vary in their levels of autonomy'' \citep[p. 4]{OECD2024Explanatory}
            \item ``influence physical or virtual environments`` \citep[Article 3(1)]{EUParliament2024AIAct}
        \end{enumerate}
    \item \textbf{Mechanism (how it does it)}: How does the system fulfil its function.
        \begin{enumerate}
            \item ``It does so by utilising machine and/or human-based inputs/data'' \citep[p. 7]{OECD2019Scoping}
            \item ``[AI systems act] by perceiving their environment through data acquisition, interpreting the collected structured or unstructured data, reasoning on the knowledge, or processing the information, derived from this data and deciding the best action(s) to take to achieve the given goal.'' \citep[p. 36]{HLEGAI2019Ethics}
        \end{enumerate}
\end{enumerate}

\subsubsection{Relationship between AI Models and AI Systems}

Another dimension to be considered is the relationship between AI models and AI systems. This relationship has been split into two perspectives. ``Model as a core component of a system'' encompasses definitions that describe models as the core of an AI system, often performing a system’s main function. ``Model as a tool for a system'' takes a broader view on the relationship of an AI model and an AI system, where an AI model is defined according to its function as a tool for an AI system.

\vspace{0.3cm}
\textbf{Relationship between AI models and AI systems:}

\begin{enumerate}
    \item \textbf{Model as a core component of a system}: A model defined as the core of an AI system, performing functions such as reasoning and knowledge recall (e.g. frontier language models in modern chatbots).
        \begin{enumerate}
            \item ``A model characterizes an input-output transformation intended to perform a core computational task of the AI system'' \citep[p. 12]{Dunietz2024EUUSTaxonomy}
        \end{enumerate}
    \item \textbf{Model as a tool for a system}: Focusing on the function of a model as a tool within an AI system.
        \begin{enumerate}
            \item ``Al systems contain a model which they use to produce predictions [...]'' \citep[ISO/IEC 22989:2022 7.1]{ISOIEC22989:2022}
            \item ``AI systems are information-processing technologies that integrate models and algorithms that produce a capacity to learn and to perform cognitive tasks'' \citep[p. 10]{UNESCO2022Ethics}
        \end{enumerate}
\end{enumerate}

\subsection{Paradigm Shifts in the Meaning of AI Models and AI Systems}
\label{subsec:ParadigmShiftsAIModelandAISystem}

Having laid out the different conceptual perspectives of \textit{AI models} and \textit{AI systems}, we now describe the observed trends and evolutions of definitions of these terms over time. This section is not intended as a comprehensive history. Its purpose is to show that shifts in the meaning of \textit{AI model} and \textit{AI system} track changes in technical paradigms rather than deliberate definitional refinement, which become inadequate when applied to modern general-purpose models.

\subsubsection{AI Models}

\paragraph{Conceptual Senses of AI models}
\label{para:ConceptualSenseAIModel}

The \citet{OEDNDModel} entry distinguishes four senses of \textit{model} as a noun: a representation of structure, an object of imitation, a type or design, and other uses (obsolete historical senses). This corresponds to the sense of ``model as representation'' that we identified in our literature review. It is also compatible with ``model artifact (result of building/training)'' as its parent category and ``model as a tool for a system'' as its relationship with the system. However, this definition of \textit{model} is not clearly compatible with ``model as a core component of a system'', since reasoning and knowledge recall are not capabilities typically associated with representations.

The notions of ``model as representation'' and ``model as a particular tool'' are closely tied to the concept of a ``world model'' which is a key component of some (typically older) archetypes of AI systems or agents. The idea is that since most AI agents will only have partial observability of the world around them, they need some way to deduce or predict changes in the world that they cannot directly observe, such as food decaying in a closed fridge, or the outcomes of actions that the agent has yet to perform. Thus, an AI agent will require some component that ``models'' the exterior world and allows them to make better-informed decisions. For example, \textit{Artificial Intelligence: A Modern Approach} \citep{Russell2009AI} says that ``knowledge about ‘how the world works’ [...] is called a model of the world'' and ``an agent that uses such a model is called a model-based agent.'' Crucially, under this paradigm the ``model'' is not expected to decide what action the agent should take, merely to predict what the likely effects of different actions would be.

In the field of modern AI, LLMs are typically referred to as \textit{models} despite not seeming to be a model of anything, and are relied upon to perform tasks and make decisions by themselves, rather than being used to inform a decision-making process. Therefore the use of the term \textit{model} to describe these entities appears to be due to the ancestry of current frontier models, since nearly all such models are (primarily) neural networks using transformer architecture \citep{Zhao2025Survey}. Neural networks were originally proposed as a model of networks of neurons in the human brain \citep{McCulloch1943Calculus}, and are known to be capable of approximating (or ``modelling'') a wide range of mathematical functions \citep{Cybenko1989Sigmoidal,Hornik1989Approximators}. Furthermore, transformers were originally proposed in the field of computational language \textbf{modelling} \citep{Vaswani2023Attention}. For these reasons, it is commonplace to refer to any neural network with transformer architecture as a \textit{model}, even if it is not clear what, if anything, it is modelling (as is the case in the neural networks that form the core of modern frontier LLMs).

As transformer-based neural networks have become an increasingly dominant force in AI, particularly in the most capable and general systems, and as more classical agent architectures have become correspondingly less popular, we argue that the usage of the term \textit{AI system} has shifted. This can be observed in the OECD definitions. The 2022 definition states ``AI models are actionable representations of all or part of the external context or environment of an AI system (encompassing, for example, processes, objects, ideas, people and/or interactions taking place in context)'' \citep{OECD2022Classification}, which corresponds to the first (representation) sense of \textit{AI model}. The 2024 definition states ``An AI model is a core component of an AI system used to make inferences from inputs to produce outputs […] while the parameters of an AI model change during the build phase, they usually remain fixed after deployment once the build phase has concluded'' \citep{OECD2024Explanatory}, corresponding to a different second sense of a model, which views the model primarily as a component instead of a representation.

While frontier models are not typically used for modelling purposes, they are still built from an architecture (transformers) that was intended to model something (natural language). What, if anything therefore, can frontier models be said to be models of? The pre-training of GPT-3 attempts to maximise its accuracy at modelling its training dataset \citep{Brown2020FewShot}, meaning that pre-trained GPT-3 can meaningfully be said to be a model of whatever corpus of text it was trained on. The fine-tuning will, strictly speaking, make it less representative of its training data (for example, it will produce less offensive language than contained in its dataset). However, it will move the model closer to being a useful AI assistant. Therefore, if a fine-tuned GPT-3 can be meaningfully said to be a model of anything, it is an (imprecise) model of the perfect AI assistant.

\paragraph{Historical Shift towards Generality}

Beyond a shift in the different senses of an AI model, there was also a notable shift from AI models that have specific purposes to being general-purpose during the machine-learning era.

\vspace{0.3cm}
\textbf{2016-2020: Statistical Methods, Specific Tasks}

When specifically searching for definitions of \textit{AI model} in the systematic review, some early definitions characterised AI models predominantly through the lens of machine learning and statistical models, emphasizing their task-specific nature. One 2016 definition described machine learning models as ``non-deterministic statistical approximations'' bound to function correctly only ``a certain portion of the time,'' emphasizing their probabilistic nature and limited scope \citep{Ramanathan2016Symbolic}. Following the conventions of this era, \citet[ISO/IEC 22989:2022]{ISOIEC22989:2022} used the term ``AI model'' with reference to two separate definitions for ``AI'' and ``model'', where model is defined in part as a ``mathematical or otherwise logical representation of a system''. This pattern in the retrieved literature suggests that the \textit{AI model} terminology was closely associated with traditional machine learning paradigms during this period.

\vspace{0.3cm}
\textbf{2021-2022: Foundation Models}

By 2021, definitions had started to include structured components such as algorithms, training datasets, and performance metrics while remaining fundamentally task-oriented. The concept of a foundation model was introduced in 2021, marking a shift in moving away from machine learning for specific applications to focus on versatility. These definitions focus on models being ``pretrained on massive amounts of data'' and capable of adaptation ``to perform a wide variety of tasks through fine-tuning'' \citep{Bommasani2022Opportunities}. This period marked the transition from AI as specialised tools to AI as general-purpose capabilities. The characteristics of foundation models would later be recognised under the term ``general-purpose'' after its use in the EU AI Act in 2023. 

\vspace{0.3cm}
\textbf{2023-2024: Formalisation of General-purpose}

By 2023, the shift in capability influenced the definition of \textit{general-purpose AI models} within formal regulatory recognition. Article 3(63) of the EU AI Act defined these as models ``capable of competently performing a wide range of distinct tasks regardless of the way the model is placed on the market'' \citep{EUParliament2024AIAct}. This trend reflects a fundamental reconceptualisation of AI from being application-specific to general-purpose, which played a significant role in expanding the scope of how systems were to be defined.

\subsubsection{AI Systems}

From the systematic literature review and manual extraction, several observations emerge regarding the evolution of the definition of AI systems. Unlike the evolution observed in the definitions of AI models, the trends in the definitions of AI systems emerged in parallel, hence a lack of distinct timelines.

\paragraph{The Emergence of Autonomy and Adaptiveness as Core Attributes}

Initial definitions of autonomy in AI systems and/or autonomous systems focused on independence from human supervision. The 2016 characterisation emphasised systems ``capable of performing certain tasks in an unstructured environment without human supervision,'' with self-awareness defined as the ability to ``analyze the situation and do the decision-making on its own'' \citep{Helle2016Testing}. The OECD's 2019 definition introduced the idea that AI systems are ``designed to operate with varying levels of autonomy,'' acknowledging autonomy as a spectrum rather than a binary state \citep{OECD2019Scoping}.

The period from 2019 to 2024 saw increasing sophistication in how autonomy was conceptualised within the definitions of systems. The 2024 EU AI Act definition represented this evolution, noting that the autonomy offers the possibility of continual adaptation, acknowledging systems ``designed to operate with varying levels of autonomy […] that may exhibit adaptiveness after deployment'' \citep[Article 3(1)]{EUParliament2024AIAct}. 

The concept of adaptiveness introduced an additional temporal dimension: systems could change their behavior after deployment through ``self-learning capabilities, allowing the system to change while in use'' \citep[Recital 12]{EUParliament2024AIAct}. This post-deployment evolution distinguished modern AI systems from earlier conceptions of AI as a program for symbolic logic, and traditional software which operates according to fixed programming.

\paragraph{Recognition of Unintended Capabilities}

Early definitions implicitly assumed that AI systems would perform tasks for which they were explicitly designed and trained. Capabilities were understood as direct results of intentional design choices, training data selection, and optimisation objectives.

Later definitions explicitly recognised that modern AI systems exhibit capabilities beyond those explicitly programmed or trained for. The 2022 definitions of general-purpose AI systems could have ``multiple intended and unintended purposes'' \citep{FLI2022GPAI}. By 2023, definitions acknowledged systems that could ``accomplish a range of distinct tasks, including some for which it was not intentionally and specifically trained'' \citep{Gutierrez2022Definition}.

Overall, the recognition of emergent capabilities signals an acknowledgement that traditional approaches based on comprehensive specification and validation become vulnerable to rapid technological change.

\paragraph{Legal Frameworks and Regulatory Motivations}

Pre-2021 definitions originated primarily from academic and technical communities, focusing on capabilities, methods, and performance characteristics. These definitions were adequate for descriptive purposes in technical research, but were not written with regulatory needs in mind. 

The \textit{OECD AI Principles} \citep{OECD2019Scoping}, adopted in May 2019, played a foundational role in shaping subsequent regulatory standards. Along with their supplementary documentation that define their scope, they established a common reference point for the EU, ISO, and NIST. 

The EU AI Act \citep{EUParliament2024AIAct}, developed between 2021 and 2024, uses a risk-based classification framework (unacceptable, high, limited, and minimal risk) rather than attempting comprehensive definitional precision. 

The 2023 American Executive Order on ``Safe, Secure, and Trustworthy Development and Use of Artificial Intelligence'' (later rescinded in 2024) provided an alternative regulatory approach, defining AI systems broadly as ``any data system, software, hardware, application, tools, or utility that operates in whole or in part using AI,'' demonstrating generalised and all-encompassing approach \citep{EOPresident2023Safe}. 

Overall, the emergence of legally operative definitions represents the maturation of the field and its growing significance. These definitions establish frameworks for governance and compliance. However, as the analysis above demonstrates, approaches to definitional precision vary considerably, reflecting competing priorities between regulatory clarity and future-proofing. 

\subsubsection{Overall Trends}

In general, the trends observed in the definition of AI systems may also apply to definitions of AI models. However, trends identified for AI models were derived primarily from changes in explicit definitional language across the sources retrieved from our search. In contrast, several trends associated with AI systems were observed in their definition and accompanying explanatory context and regulatory framing. Definitions of AI models were typically accompanied by far less contextual elaboration, reflecting the larger volume of system-level definitions retrieved in the review. Moreover, governance discourse mostly focused on the concept of systems over models. As such, these system-level trends should not be interpreted as uniformly applying to AI models, though both systems and models may share broader conceptual paradigms. 

\subsection{Definitional Lineages and Influence}

In this section we consider more broadly how the various definitions from our surveys have influenced each other. Our analysis reveals that many secondary definitions (summarisations, syntheses, and adaptations), and most definitions proposed by major organisations, can be traced back to a particularly concentrated set of foundational sources, including OECD frameworks. We also find that these OECD frameworks have evolved in ways that integrate previously distinct concepts, blurring the boundary between models as representations and models as components of AI systems. We examine these issues in depth, tracing the chain of influence through various sources, analysing how ambiguities have compounded through successive iterations.

\subsubsection{Overall Chain of Influence}

\begin{figure}[h]
    \centering
    \includegraphics[width=1\linewidth]{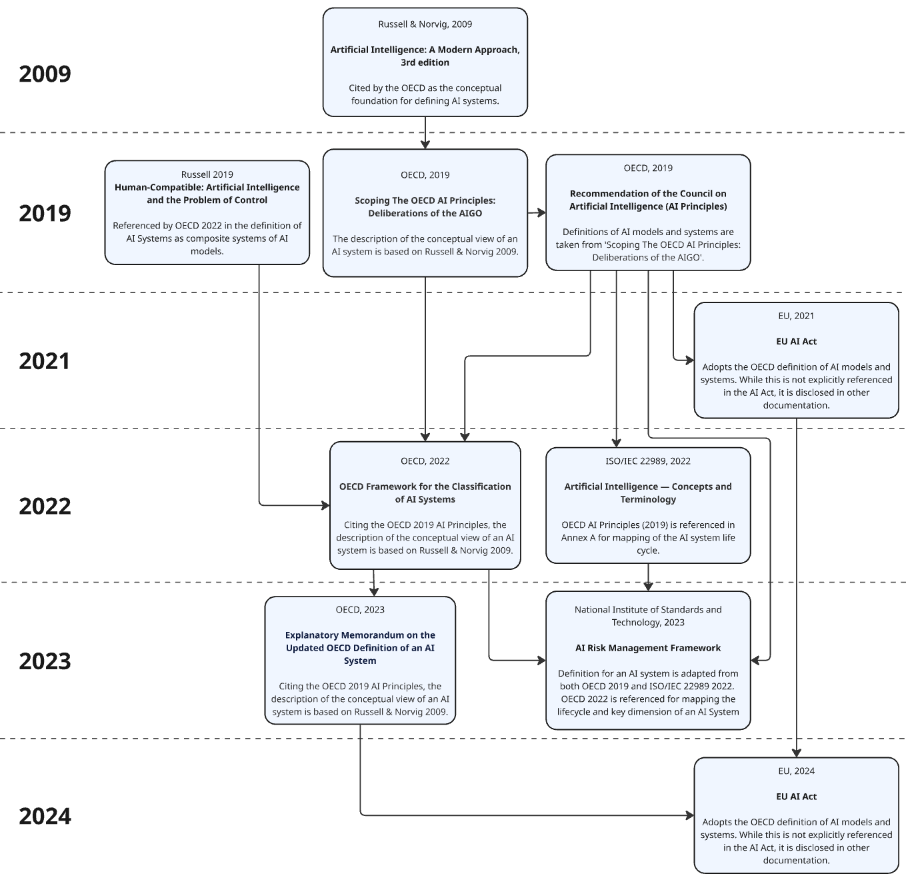}
    \caption{Definitional lineage of terms related to \textit{AI models} and \textit{AI systems}}
    \label{fig:DefinitionLineage}
\end{figure}

As described, much of the definitions used in major regulatory frameworks have their lineages traced back to the OECD definitions, which in turn are heavily based on the work of Stuart Russell and Peter Norvig, as illustrated by (Figure \ref{fig:DefinitionLineage}).

The lineage begins with Russell and Norvig’s \textit{Artificial Intelligence: A Modern Approach} \citep[3rd edition]{Russell2009AI}, which articulates the agent-based architecture of perception, representation, reasoning, and action. This framework is adopted directly into OECD’s \textit{Scoping the OECD AI Principles: Deliberations of the Expert Group on Artificial Intelligence at the OECD (AIGO)} \citep{OECD2019Scoping}, which identifies Russell and Norvig as the conceptual basis for understanding AI systems as composite systems of interacting models. AIGO deliberations informed the \textit{OECD Recommendation on Artificial Intelligence} \citep[AI Principles]{OECD2019Recommendation}. \citet{OECD2019Scoping} established key definitional elements including ``machine-based system,'' ``varying levels of autonomy,'' and the capacity to ``influence the environment,'' which would propagate through subsequent regulatory instruments. Notably, while the 2019 OECD documents \citep{OECD2019Scoping, OECD2019Recommendation} incorporated the concept of \textit{model} as a component of AI systems, they did not provide a formal standalone definition of \textit{AI model} at this stage.

In 2022, the OECD published the \textit{Framework for the Classification of AI Systems} \citep{OECD2022Classification}, which built upon \citet{OECD2019Scoping} by embedding the same conceptual architecture into a more granular classification taxonomy. This document formally introduced and defined the term \textit{AI model} as a distinct concept, establishing a clearer conceptual boundary between models and systems.

Parallel to this, ISO/IEC 22989:2022 \textit{Artificial Intelligence—Concepts and Terminology} \citep{ISOIEC22989:2022} adopted and detailed AI terminology for international technical standardisation. While ISO drew on its own pre-existing general definition of \textit{model} from other standards, the document explicitly references \citet{OECD2019Scoping} in Annex A when mapping the AI system lifecycle.

By 2023, the U.S. National Institute of Standards and Technology (NIST) published its \textit{AI Risk Management Framework (AI RMF 1.0)} \citep{NIST2023RMF}. NIST explicitly notes that its definition of an AI system is ``adapted from: OECD Recommendation on AI: 2019; ISO/IEC 22989:2022,'' and it uses the \citet{OECD2022Classification} to structure its lifecycle and dimensional analysis.

Also in 2024, the OECD published the \textit{Explanatory Memorandum on the Updated OECD Definition of an AI System}\citep{OECD2024Explanatory}, revising the 2019 definition. The memorandum states that the updated definition was shaped to ``support broad alignment of the definitions of AI systems in several ongoing processes in the European Union'' as well as other emerging regulatory frameworks.

Subsequently, the \citet{EC2021Proposal} adapted and expanded upon the \citet{OECD2019Scoping}. Although the OECD is not explicitly cited in the legislative text itself, the influence is acknowledged in recitals and supplementary documentation. The final EU AI Act \citep[Article 3(1)]{EUParliament2024AIAct} demonstrates convergence with the updated 2024 OECD definition, incorporating OECD’s new additions: ``machine-based system,'' ``explicit or implicit objectives,'' ``varying levels of autonomy,'' ``infers from input,'' and ``influence physical or virtual environments.''

This represents a bidirectional feedback loop: \citet{OECD2024Explanatory} notes that its revised definition was developed with awareness of the evolving EU AI Act, while the EU AI Act's final text converged toward the OECD formulation.

\subsubsection{Analysis of OECD Frameworks}
\label{subsubsec:AnalysisOECDFrameworks}

The OECD's AI taxonomy has become one of the most influential frameworks for AI governance globally. Its definitions have been adopted or adapted by numerous jurisdictions, In particular, definitions originating in the OECD framework have influenced ISO standards, the NIST AI Risk Management Framework, and most significantly the EU AI Act.

This section assesses whether successive OECD revisions preserve a stable conceptual boundary between \textit{AI model} and \textit{AI system}. We observe that across the 2019, 2022, and 2024 iterations, the framework progressively expands the meaning of \textit{AI model} without sufficient conceptual clarification. As a result, the distinction between model and system becomes increasingly flexible, which may create challenges when used for regulatory allocation of obligations.

\paragraph{\texorpdfstring{\citet{OECD2019Scoping}}{OECD (2019a)}: The Original Conceptual Framework}

\begin{figure}[h]
    \centering
    \begin{subfigure}{1\linewidth}
        \centering
        \includegraphics[width=1\linewidth]{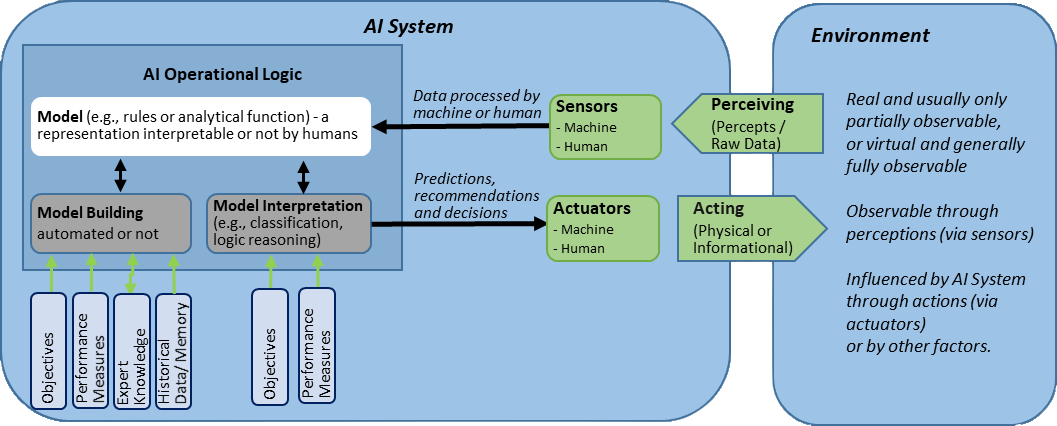}
    \end{subfigure}
    \vspace{4pt}
    \begin{subfigure}{1\linewidth}
        \centering
        \includegraphics[width=1\linewidth]{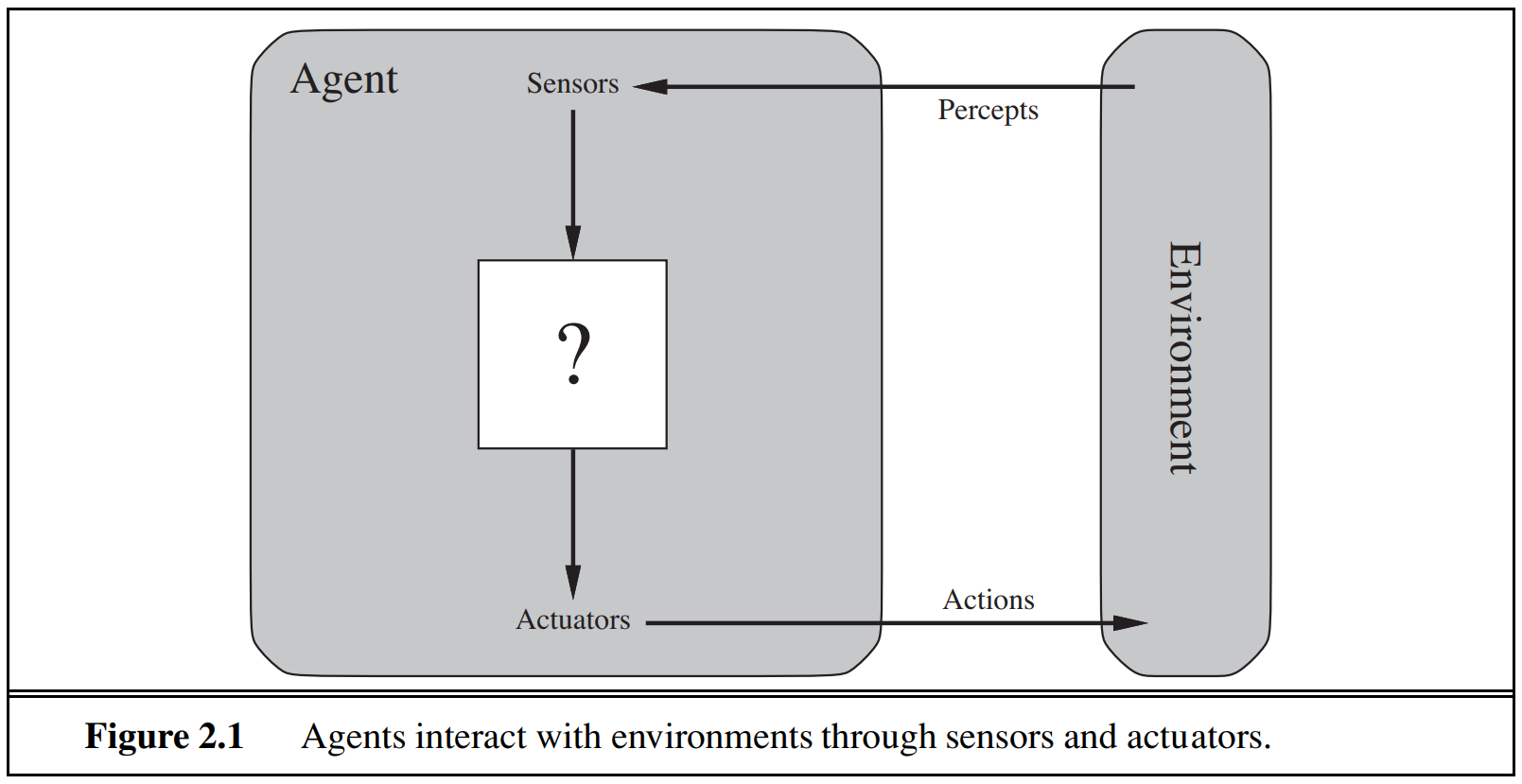}
    \end{subfigure}
    \caption{\citet{OECD2019Scoping} (top) versus \citet{Russell2009AI} (bottom)}
    \label{fig:OECD2019vsRussell&Norvig}
\end{figure}

\citet{OECD2019Scoping} noted its adaptation of \citet{Russell2009AI} for its definitions, a connection made clear by comparing the two conceptual diagrams, as shown in Figure \ref{fig:OECD2019vsRussell&Norvig}.

The OECD's 2019 report \textit{Recommendation of the Council on Artificial Intelligence} \citep{OECD2019Recommendation} established a comprehensive taxonomy with three core concepts. These definitions appear to draw heavily from Russell and Norvig's agent-based framework, while introducing several additional terms and conceptual reinterpretations \citep{Russell2009AI}. The main relevant definitions are AI system (or intelligent agent), AI operational logic, and model, as summarised in Table \ref{table:OECDDefinitions2019Rec}.

\Needspace{12\baselineskip}
\renewcommand{\arraystretch}{1.2}
\begin{longtable}{|p{3cm}|p{11cm}|}
\caption{Key definitions related to AI models and AI systems from \citet{OECD2019Scoping}}
\label{table:OECDDefinitions2019Rec}\\
\hline
    \textbf{AI system / Intelligent Agent} &
    
    ``An AI system is a machine-based system that is capable of influencing the Environment by making recommendations, predictions or decisions for a given set of Objectives.

    It does so by utilising machine and/or human-based inputs/data to: 
    
    i) perceive real and/or virtual environments; 
    
    ii) abstract such perceptions into models manually or automatically; and 
    
    iii) use Model Interpretations to formulate options for outcomes.'' \citep[p. 7]{OECD2019Scoping}\\
\hline
    \textbf{AI Operational Logic} (p. 6) & 
    ``The key power of an AI system resides in its Operational Logic, which, for a given set of objectives and based on input data from Sensors, provides output for the Actuators --- as recommendations, predictions or decisions --- that are capable of influencing the state of the Environment.'' \citep[p. 6]{OECD2019Scoping}\\
\hline
    \textbf{Model} (p. 6-7)& 
    ``A Model is an actionable representation of all or part of the external environment of an AI system that describes the environment’s structure and/or dynamics. 
    
    The model represents the core of an AI system. 
    
    A model can be based on data and/or expert knowledge, by humans and/or by automated tools like machine learning algorithms. 
    
    Model Interpretation is the process of deriving an outcome from a model.'' \citep[p. 7]{OECD2019Scoping}

    ``Model (a data object constructed by the Model Building process)'' \citep[p. 6]{OECD2019Scoping}\\
\hline
\end{longtable}

\citet{OECD2019Scoping} defines an AI system as comprising two parts: its function and its mechanisms (including both building and using). The latter two terms are defined by their relationship to the AI system itself: the AI operational logic maps the ``input from sensor'' to the ``output for actuator,'' while the model represents the ``external environment'' of the AI system.

This taxonomy presents three conceptual challenges that may affect its application in regulatory contexts:

\vspace{0.3cm}
\textbf{1. Ambiguous Inclusion or Exclusion of Human Involvement}

The report states: ``An AI system consists of three main elements: Sensors, Operational Logic and Actuators'' \citep[p. 8]{OECD2019Scoping}. Under this strict interpretation, a software program that generates a credit score would not, by itself, constitute an AI system, as it lacks a physical actuator (e.g., the bank staff who ultimately approve or deny the loan). However, the \citet{OECD2019Scoping}’s conceptual diagram introduces flexibility by noting that sensors and actuators can be human. Consequently, a credit scoring system would be considered a hybrid of human and machine, which is an interpretation that may differ from common public understanding.

This ambiguity marks the beginning of a broader, recurring issue in subsequent years, where human involvement is often reconciled by classifying such systems as AI with a ``low level of autonomy.''

\vspace{0.3cm}
\textbf{2. The Conflation of Real, Digital, and Abstract Objects}

This framework illustrates a conflation of object types. It describes an AI system as being made of Sensors, Actuators, and Operational Logic. While sensors and actuators are tangible, physical components, the operational logic is presented as a pure, abstract object described as a ``logical mapping'' with no reference to the physical computational substrate (e.g., hardware like GPUs) required to implement it.

Take an autonomous vehicle as an example. Under this description, the self-driving car is said to be ``made of'' the radar (sensor), the engine (actuator), and the logical mapping between them. The computing hardware that physically instantiates this logic is not explicitly referenced in the definition, which leaves the relationship between abstract function and real-world implementation less clearly specified.

\citet[p. 35]{Russell2009AI} maintain strict separation: ``The agent function is an abstract mathematical description; the agent program is a concrete implementation, running within some physical system.'' 

They formalise this as:

\begin{displaymath}
    Agent = Architecture + Program
\end{displaymath}
\begin{displaymath}
    Architecture = Sensor + Actuator + Computing\ Device
\end{displaymath}

Therefore:

\begin{displaymath}
    Agent = (Sensor + Actuator + Computing\ Device) + Program
\end{displaymath}

The OECD framework does not explicitly articulate this distinction, instead presenting the system as encompassing both tangible components and abstract functional elements.

\vspace{0.3cm}
\textbf{3. The Ambiguity of ``Actionable Representation''}

The term ``actionable representation'' is not formally defined within \citet{OECD2019Scoping}’s taxonomy. Nevertheless, there are academic sources which define it. For example, \citet[p. 1]{Ghosh2018Learning} defines it as the following:

\begin{enumerate}
    \item Is ``aware of the dynamics of the environment.''
    \item Captures ``only the elements of the observation that are necessary for decision making rather than all factors of variation, eliminating the need for explicit reconstruction.''
    \item Is typically ``obtained from a goal-conditioned policy—a policy that knows how to reach arbitrary goal states from a given state.''
\end{enumerate}

This narrower technical interpretation requires a ``goal-conditioned policy'' which may not align with all AI systems encompassed by the OECD framework, potentially creating ambiguity between terminology and intended scope. The OECD appears to use ``actionable'' colloquially to mean ``can inform actions,'' but this divergence from technical usage creates ambiguity about what exactly a ``model'' is supposed to be.

\vspace{0.3cm}
\textbf{Terminological Relationships}

\citet{OECD2019Scoping}’s taxonomy introduces the non-standard concept of AI operational logic, which is rarely found in other literature. Its definition, describing the mapping of outputs for actuators from inputs received from sensors, is conceptually almost identical to Russell and Norvig’s agent function, which is defined as a description that ``maps any given percept sequence to an action'' \citep[p.35]{Russell2009AI}.

The relationship and differences between the core concepts in the two frameworks are summarised in Table \ref{table:RelationshipDifferencesRussellNovigAndOECD}:

\renewcommand{\arraystretch}{1.8}
\begin{longtable}{|>{\centering\arraybackslash}p{3cm}|>{\centering\arraybackslash}p{3cm}|>{\centering\arraybackslash}p{4cm}|>{\centering\arraybackslash}p{4cm}|}
\caption{Relationship and differences between \citet{Russell2009AI} term and \citet{OECD2019Scoping}}
\label{table:RelationshipDifferencesRussellNovigAndOECD}\\
    \hline
        \textbf{Russell \& Norvig’s Term} & \textbf{\citet{OECD2019Scoping} Equivalent} & \textbf{Key Difference in Taxonomy} & \textbf{Primary Object Type} \\
    \hline
        \textbf{Agent / Rational Agent / Intelligent Agent} & \textbf{AI System / Intelligent Agent} & The definitions are almost identical. & OECD: A mixture of real, digital, and abstract objects. \\ 
    \hline
        \textbf{Agent Function} & \textbf{AI Operational Logic} & Near-identical definitions (mapping inputs to outputs) & Abstract \\
    \hline
        \textbf{Agent Program} & \textbf{(Most analogous to) Model} & Agent program is concrete software; OECD model is ``data object'' & Digital \\
    \hline
        \textbf{World Model} (of a model-based agent) & A specific type of Model & \citet{Russell2009AI}: specific to subset of agents (model-based); OECD: implied for all systems & Digital or Abstract \\
    \hline
\end{longtable}


\paragraph{\texorpdfstring{\citet{OECD2022Classification}}{OECD (2022)}: Compounding the Conceptual Confusion}

\begin{figure}[h]
    \centering
    \begin{subfigure}{0.48\linewidth}
        \includegraphics[width=\linewidth]{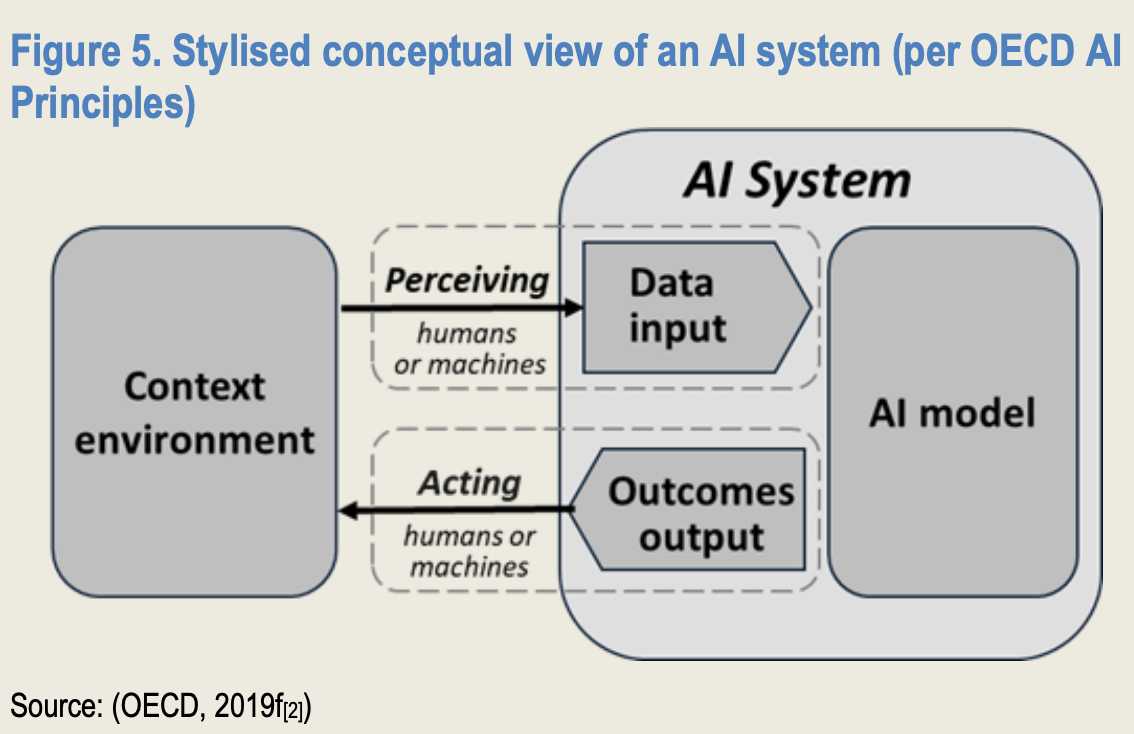}
    \end{subfigure}
    \hfill
    \begin{subfigure}{0.48\linewidth}
        \includegraphics[width=\linewidth]{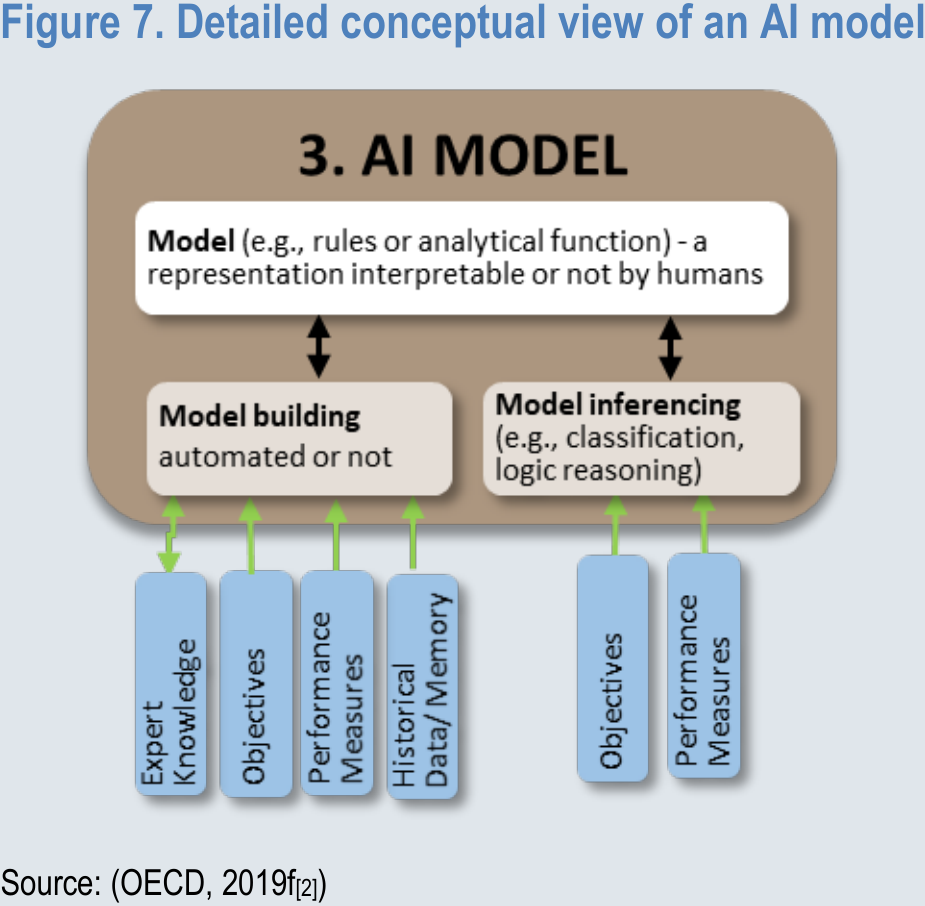}
    \end{subfigure}
    \caption{\cite{OECD2022Classification}'s conceptualisation of the term \textit{AI model}}
    \label{fig:OECD2022AIModel}
\end{figure}

As an update to the previous version, \citet{OECD2022Classification} did not fully resolve the conceptual questions identified in 2019; instead, it consolidated previously distinct elements into a unified framework. The framework merged the already problematic AI operational logic and model into a single, ill-defined term: ``AI model''. This move effectively formalises the earlier conflation, redefining the model as the component that dictates the system's operational logic while simultaneously treating it as a representation of the environment. The new conceptual diagrams are shown in Figure \ref{fig:OECD2022AIModel}.

The original rationale provided was a desire for simplification \citep[p. 23]{OECD2022Classification}:

\begin{quote}
    ``It should be noted that in the present classification framework, the ‘Perceiving’ (data collection) and ‘data/input’ elements – that were separate elements in the original OECD work presented in Figure 5 – have been combined in an effort to simplify the framework (see dotted lines in Figure 4). The ‘Outcomes’ and ‘Acting’ elements have also been combined.''
\end{quote}

This simplification creates significant ambiguity. Combining ``Perceiving'' with ``Data/Input'' blurs the boundary between the AI model and the AI system. Inputs now flow directly from the environment into the AI model, bypassing the conceptual layer of the system and its sensors that would typically collect and pre-process data. The AI Model is not only expanded to encompass what was previously called the AI’s operational logic, but the operational logic itself is broadened relative to the system, where its input becomes the system's input. This, in turn, narrows the definition of the AI System itself, potentially excluding essential components like data pre-processing modules.

For this taxonomy to remain coherent, there is a logical pressure to treat the \textbf{AI model as an abstraction of the entire AI system} --- a tendency that becomes explicit in the 2024 documents, which we discuss further in Section \ref{para:OECD2024Expansion} (\citet{OECD2024Explanatory}: Expansion and Functional Broadening). If both the AI model and the AI system are conceived as concrete entities, the conceptual boundary between them becomes less clearly defined. This interpretation may not fully capture the structure of certain real-world deployments. For example, a recommendation model abstracts user preferences for items, not the entire recommendation system, which requires additional components to collect live user data and integrate business rules (e.g., paid promotions) that ultimately determine how the model’s outputs are applied.

\vspace{0.3cm}
\textbf{Internal Inconsistencies: Diagram versus Definition}

The conceptual diagram places the label ``AI model'' precisely where ``AI Operational Logic'' was previously located. Yet, the textual definitions of an ``AI model'' remain closer to the \citet{OECD2019Scoping} notion of a ``model'' (a representation of the environment) rather than to operational logic. A key inconsistency is that the diagram depicts system inputs and outputs flowing directly to and from the AI model, while the definition states a model may represent ``all or part'' of the external environment. This implies the model's inputs/outputs need not align with the system's.

\renewcommand{\arraystretch}{1.8}
\begin{longtable}{|p{3cm}|p{3cm}|p{4cm}|p{4cm}|}
\caption{Differences and similarities between terminologies in \citet{OECD2019Scoping} and \citet{OECD2022Classification}}
\label{table:DifferencesOECD2019andOECD2022}\\
    \hline
        \textbf{\centering Aspect\arraybackslash} &
        \textbf{\centering \citet{OECD2019Scoping}\arraybackslash} &
        \textbf{\centering \citet{OECD2022Classification}\arraybackslash} &
        \textbf{\centering Consequences\arraybackslash} \\
    \hline
        \textbf{Core Definition} & 
        An actionable representation of all or part of the \textbf{external environment} of an AI system. & 
        An actionable representation of all or part of the \textbf{external context or environment} of an AI system. & 
        Slight expansion from ``environment'' to ``context.'' \\ 
    \hline
        \textbf{Relationship with Environment} & 
        The model [...] \textbf{describes} the environment’s structure and/or dynamics. & 
        The model \textbf{represents, describes, and interacts with} real or virtual environments. & Introduces the incorrect idea that a model can ``interact.'' \\
    \hline
        \textbf{Relation to System} & 
        Clearly a component \textit{within} the system, representing its environment. & 
        Blurred; diagrammatically equated with the system's core logic, definitionally remains a representation. & 
        Fundamental tension between the visual framework and its textual definitions. \\
    \hline
\end{longtable}

The \citet{OECD2022Classification} claim that a model can ``interact with'' environments is conceptually flawed and contradicts common usage of the term ``actionable representation,'' where ``actionable'' should mean the representation's outputs can inform actions taken by \textit{another component} (e.g. an actuator), not that the model itself acts. This misinterpretation is evident in the report's own examples. It states that actions are taken ``based on the outcomes of an AI system, e.g. by autonomous vehicles,'' confirming the system provides an output that is actualised by the vehicle. Similarly, an AI gamer's policy model generates action probabilities; the system architecture executes the actions. The capacity to act resides in the system, not in the model.

\paragraph{\texorpdfstring{\citet{OECD2024Explanatory}}{OECD (2024)}: Expansion and Functional Broadening}
\label{para:OECD2024Expansion}

\begin{figure}[h]
    \centering
    \includegraphics[width=0.8\linewidth]{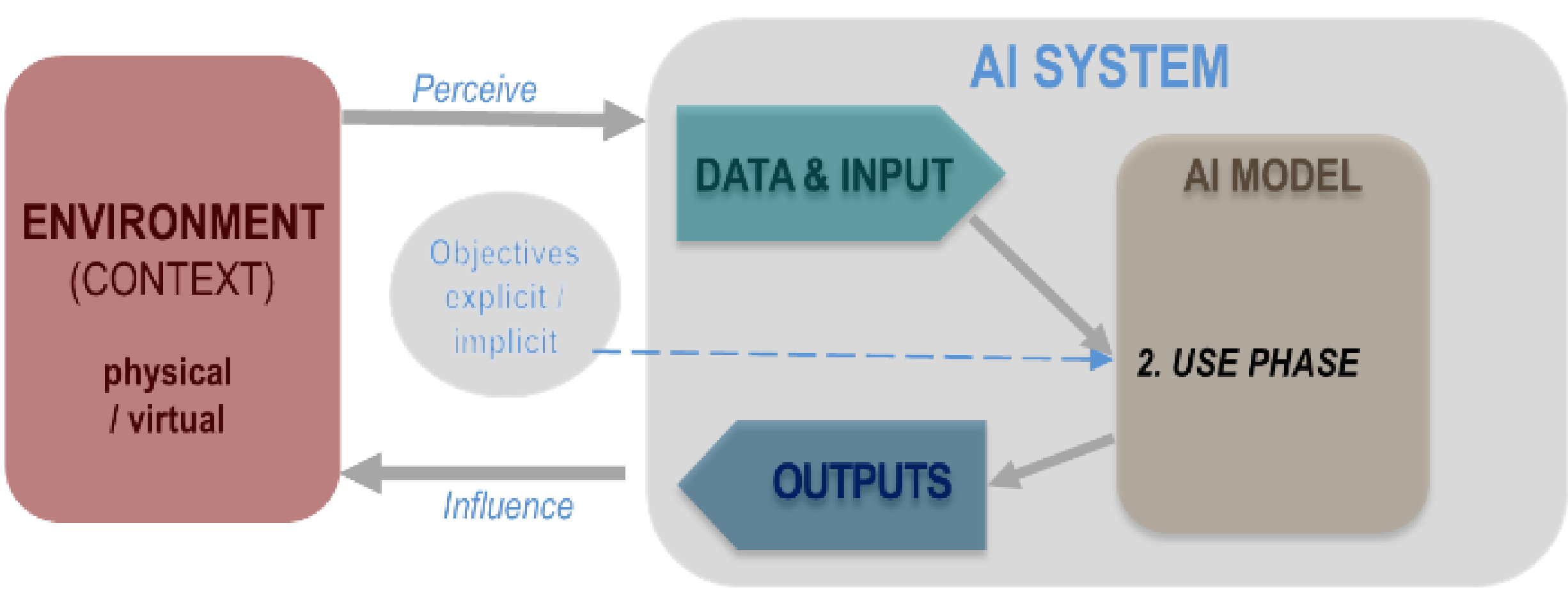}
    \caption{\citet{OECD2024Explanatory}'s conceptual diagram}
    \label{fig:OECD2024Diagram}
\end{figure}

Subsequently, \citet{OECD2024Explanatory} illustrates the extent to which the taxonomy has evolved in scope and application. In its latest conceptual diagram as seen in Figure \ref{fig:OECD2024Diagram}, the term ``influence'' replaces ``acting,'' a change that further abstracts the system's relationship with its environment without resolving core ambiguities.

Critically, \citet{OECD2024Explanatory} appears to have recognised a foundational flaw, where many modern AI systems, such as those using model-free reinforcement learning, operate without an explicit ``model'' as previously defined. Instead of revising the earlier conceptual structure, the organisation expanded the definition of \textit{model} to retroactively include these systems, thereby altering the term's meaning and its relationship to the AI system.

This has resulted in a set of scattered, often conflicting definitions for \textit{AI model} within the same document:

\begin{enumerate}
    \item \textbf{A system component}: ``[…] an AI model is a core component of an AI system used to make inferences from inputs to produce outputs.'' \citep[p. 8]{OECD2024Explanatory}
    \item \textbf{A representation (per ISO)}: ``A model is defined as a `physical, mathematical or otherwise logical representation of a system, entity, phenomenon, process or data' […]'' \citep[p. 8]{OECD2024Explanatory}
    \item \textbf{An input-output mapping}: ``AI models include, among others, various kinds of input-output functions (such as decision trees and neural networks).'' \citep[p. 8]{OECD2024Explanatory}
    \item \textbf{A world model}: ``An AI model can represent the transition dynamics of the environment, allowing an AI system to select actions by examining
their possible consequences using the model.'' \citep[p. 8]{OECD2024Explanatory}
    \item \textbf{A policy or Q-function}: ``In such cases [‘model-free reinforcement learning’], the policy or Q-function would be an AI model in the more general sense of an input-output function.'' \citep[p. 11]{OECD2024Explanatory}
\end{enumerate}

These definitions create a high degree of contextual and internal conflict. For instance, in a model-based agent, the AI system uses the world model to determine actions. In a model-free agent, the AI model (the policy/Q-function) is now said to determine the action itself. This contradiction highlights an underlying shift: the model is inconsistently framed as a component within the system, a functional abstraction of the system, and the system's driving logic.

This new, expansive definition finally aligns the term \textit{AI model} with what was originally termed \textit{AI operational logic} or \citet{Russell2009AI}’s \textit{agent function}. The taxonomy now reflects a substantially broadened conception of \textit{AI model} across contexts, as summarised in Table \ref{table:ConceptionsAIModelDifferentFrameworks}, which tracks the term's shifting applicability across frameworks.

\renewcommand{\arraystretch}{1.8}
\begin{longtable}{|p{4cm}|>{\centering\arraybackslash}p{2cm}|>{\centering\arraybackslash}p{2cm}|>{\centering\arraybackslash}p{2cm}|>{\centering\arraybackslash}p{2cm}|>{\centering\arraybackslash}p{2cm}|}
\caption{Conceptions of \textit{AI model} across different frameworks}
\label{table:ConceptionsAIModelDifferentFrameworks}\\
    \hline
        \textbf{Concept} & 
        \textbf{\citet{Russell2009AI}} & 
        \textbf{\citet{OECD2019Scoping}} & 
        \textbf{\citet{OECD2022Classification} (Text)} &
        \textbf{\citet{OECD2022Classification} (Image)} &
        \textbf{\citet{OECD2024Explanatory}} \\
    \hline
        World model of a model-based agent & 
        \cellcolor{softgreen}Yes & 
        \cellcolor{softgreen}Yes & 
        \cellcolor{softgreen}Yes &
        \cellcolor{softyellow}Sometimes &
        \cellcolor{softyellow}Sometimes \\ 
    \hline
        Representation of (part of) the external environment & 
        \cellcolor{softgreen}Yes & 
        \cellcolor{softgreen}Yes & 
        \cellcolor{softgreen}Yes &
        \cellcolor{softyellow}Sometimes &
        \cellcolor{softyellow}Sometimes \\ 
    \hline
        Core component of an AI system & 
        \cellcolor{softyellow}Sometimes (model-based only) & 
        \cellcolor{softgreen}Yes & 
        \cellcolor{softgreen}Yes &
        \cellcolor{softgreen}Yes &
        \cellcolor{softgreen}Yes \\ 
    \hline
        Abstract mapping between system input and output & 
        \cellcolor{softred}No (Agent Function) & 
        \cellcolor{softred}No (AI Operational Logic) & 
        \cellcolor{softyellow}Sometimes &
        \cellcolor{softgreen}Yes &
        \cellcolor{softgreen}Yes \\ 
    \hline
        Policy/Q-function of a model-free agent & 
        \cellcolor{softred}No & 
        \cellcolor{softred}No & 
        \cellcolor{softred}No &
        \cellcolor{softgreen}Yes &
        \cellcolor{softgreen}Yes \\ 
    \hline
\end{longtable}

\textbf{Regulatory Implications}

The OECD itself acknowledges the indeterminacy surrounding core concepts, noting that ``different interpretations of the word ‘model’ exist.'' While such open-textured definitions are appropriate within a non-binding, standards-setting context, they acquire regulatory significance when subsequently referenced in binding legal instruments.

In the same document, the OECD explicitly disclaims any engagement with liability allocation, stating that its definition of an AI system ``intentionally does not address the issue of liability and responsibility for AI systems,'' which it leaves to human actors and jurisdiction-specific regulatory choices.

This creates a structural challenge when OECD-derived taxonomies are repurposed within the EU AI Act, which must operationalise distinctions between \textit{AI models} and \textit{AI systems} in order to assign legal obligations. The result is a potential misalignment when a deliberately non-normative technical taxonomy is referenced within a normative legal framework.

Overall, the progression from 2019 to 2024 demonstrates how well-intentioned simplification can increase conceptual ambiguity. OECD's acknowledgment of ``different interpretations'' may sit in tension with the precise boundaries that legal instruments required to allocate liability, determine compliance, and enable enforcement. 

This analysis motivates our proposed framework in Section \ref{sec:definitions} (Developing Definitions for AI Models and AI Systems), which provides technically accurate and legally robust definitions capable of withstanding regulatory scrutiny while reflecting the architectural realities of modern AI systems.


\section{Developing Definitions for AI Models and AI Systems}
\label{sec:definitions}

We now turn to the constructive task of developing improved definitions for \textit{AI model} and \textit{AI system}. Section \ref{sec:findings} (Findings and Analysis), particularly Section \ref{subsubsec:AnalysisOECDFrameworks} (Analysis of OECD Frameworks) identified several deficiencies in existing definitions: the conflation of distinct conceptual categories (representation versus operational component), internal inconsistencies between textual definitions and accompanying diagrams, the progressive expansion of terms to accommodate technological developments without corresponding conceptual clarification, and the absence of clear criteria for distinguishing where a model ends and a system begins.

In the case of both \textit{AI model} and \textit{AI system} we argue that these problems cannot be solved with a single, all-encompassing definition, and instead propose two separate but complementary definitions. First, we provide conceptual definitions that capture what AI models and systems fundamentally are, providing clarity about the nature of these entities and their relationship to one another. Second, we provide operational definitions that enable consistent, practical application in the real world, allowing non-specialists to determine whether a given artifact falls within the scope of a particular definition. This two-level approach draws inspiration from \citet{Tomic2025Octopus}, who distinguish between conceptual and operational definitions in the context of AI regulation. However, their analysis is primarily concerned with operationalizing the concept of an AI system itself by identifying its invariant features, whereas our focus is to clarify the boundary between AI models and AI systems.

The following sections develop both types of definition, beginning with an examination of the criteria that good definitions should satisfy, proceeding to our proposed conceptual definitions, followed by our operational definitions specifically tailored to the dominant paradigm of neural network-based machine learning systems, and concluding by discussing regulatory implications and case studies.

\subsection{Criteria for Technically and Legally Robust Definitions}

Definitions exist for a variety of purposes. They can be meant for coordinating usage and ensuring that interlocutors are referring to the same phenomenon \citep{Delancey2017Logic,Copi2014Logic}. They can be used to stipulate a term’s meaning for the purposes of a specific inquiry or to refine an imprecise expression into a form suitable for analytical or regulatory reasoning, as discussed in analytic philosophy and legislative drafting scholarship \citep{Zalta2023Definitions}. Definitions can also be used to create or stabilise a concept by naming it, turning a diffuse or loosely recognised phenomenon into an object of inquiry \citep{Burgess2020Conceptual}. They can clarify a concept’s content by identifying the features that constitute it and distinguishing it from related concepts \citep{Margolis2023Concepts}. Finally, in law and regulation, definitions serve a normative and operational function: they simplify language, secure consistent interpretation across a text, and determine the boundaries of obligations, rights, and responsibilities \citep{Jopek-Bosiacka2011Law,Bailey2025Clarity}. Because AI-related terms sit at the intersection of these different functions (descriptive, conceptual, and normative) this section focuses on the latter purposes and on how they can best contribute to clearer regulation.

ISO 704:2022 \citep{ISO704:2022} outlines principles for defining technical concepts in a precise, systematically structured way. It requires definitions to identify indispensable features that distinguish a concept from related ones, while excluding variable non-essential attributes. Therefore, according to ISO 704:2002, extensional definitions that attempt to specify a concept by listing its instances are inadequate for technical and conceptual clarity. The norm requires that the distinguishing characters of similar or adjacent terms are explicitly listed, clarifying the boundaries within the broader conceptual structural ontology of the domain (what ISO 704:2002 calls a ``concept system''). Specifically, definitions should follow a specific pattern where the broader category to which the concept belongs is first identified, then differentiating characteristics from neighboring concepts is further specified.

Definitions can be written for various purposes, such as clarifying research scope or explaining technical concepts. A technical definition primarily captures meaning and facilitates communication within a domain, and might not fit the criteria we can expect from a legal definition, which must establish clear, enforceable boundaries that support consistent application in practice. Below, we describe properties of a definition suitable for this regulatory function, particularly when distinguishing between AI models and AI systems.

\vspace{0.3cm}
\textbf{Structural Clarity and Definitional Boundaries}

A legal definition should express necessary and sufficient conditions in plain, unambiguous language, avoiding circularity and unstated background assumptions \citep{Dickerson1977Drafting,House2015Style}. It must enable a non-specialist decision-maker to determine inclusion or exclusion from scope using the text alone. At the same time, a definition should establish the outer boundary of application while leaving detailed classifications or risk tiers to secondary instruments. This layered and calibrated structure allows precision to evolve without reopening primary legislation and prevents the definition from becoming overly rigid or granular \citep{Ebers2024Risk,Renda2023Name}.

\vspace{0.3cm}
\textbf{Technical Coherence and Adaptability}

Legal definitions in fast-changing domains must both reflect and withstand technological evolution. They should be technology-neutral, describing functions rather than methods. For example, they should focus on what a system does rather than how it operates \citep{Reed2007Neutrality,Greenberg2016Neutrality,Birnhack2012Reverse}. Because underlying technologies develop rapidly, definitions should also include mechanisms for future-proofing: delegated standards, guidance documents, or periodic reviews. These mechanisms should enable reinterpretation without statutory amendment \citep{Ebers2024Risk}, and remain coherent with the technical object it describes, even as that object changes over time.

\vspace{0.3cm}
\textbf{Legal Operability and Policy Alignment}

The point of a good legal definition is not only to describe phenomena accurately, but also to work as a legal instrument. It should be administrable and enforceable, enabling authorities and courts to apply it consistently without deep technical expertise. Each definitional element should correspond to observable or documentable features (for instance, the presence of a learning process or adaptive behavior), allowing compliance to be verified objectively. \citet{Schuett2021Scope} observes that many AI definitions fail such tests because they rely on vague or colloquial descriptors. It should also be interoperable with existing technical and international vocabularies to minimise conflict between legal and engineering interpretations. The EU AI Act’s definition of an AI system, for instance, was intentionally aligned with the OECD wording \citep{Fernandez-Llorca2025Account,Nahra2024Parliament}. Finally, the definition should exhibit policy fit and proportionality, including only the systems the law intends to govern and excluding those that would make enforcement disproportionate \citep{Ebers2024Risk,Renda2023Name}. In practice, this means drawing the boundary of AI systems where risk-based obligations attach, not where the term is used informally.

Additionally, where appropriate, providing examples and explicit carve-outs clarifies intent and prevents disputes at the margins. Legislative drafting manuals recommend this practice to ensure shared understanding among implementers \citep{Dickerson1977Drafting,House2015Style}.

These criteria show that the purpose of a legal definition for a technical concept such as \textit{AI model} or \textit{AI system} go beyond describing, as they must also establish enforceable boundaries. A good definition has to draw stable, policy-aligned lines that regulators can administer consistently and that remain intelligible and applicable as technology develops.

\subsection{Conceptual Definitions}

As described in the Section \ref{sec:findings} (Findings and Analysis), we can see that the conceptual frameworks behind AI models and systems have changed over time. In this section, we attempt to flesh out useful conceptual definitions of AI models and AI systems. Note that these proposed definitions are not operational definitions which we will discuss further in Section \ref{subsec:OperationalDefinitionContemporaryAI} (Operational Definitions for Contemporary AI). Our conceptual definitions are not recommended for technical use, but help to explain the motivation behind our operational definitions, and outline the principles according to which the operational definitions could be further extended if needed. We propose the following conceptual definitions, largely adopted from \citet{OECD2024Explanatory}:

\begin{quote}
    \textbf{AI model}: A computational construct that makes inferences from inputs to produce outputs and forms (or is intended to form) the reasoning or knowledge core of an AI system.

    \vspace{0.2cm}
    \textbf{AI system}: A machine-based system that makes inferences from inputs from physical or virtual environments to produce outputs that influence physical or virtual environments.
\end{quote}

Our definition of \textit{AI model} represents what we see as the conceptual shift in the usage of the term from ``model as representation'' to ``model as core reasoning/knowledge component'' as discussed in more detail in Section \ref{para:ConceptualSenseAIModel} (Conceptual Senses of \textit{AI model}). In particular, we feel that the current usage of the term in industry (particularly as it relates to frontier models, the kind of models where effective legislation is by far the most important) has shifted almost entirely to the  ``model as core reasoning/knowledge component'' perspective. This is primarily because most of the post-training  ``fine-tuning'' that is done to these entities makes them less accurate representations of the datasets on which they were originally trained (for example, they produce far less offensive text than randomly sampling from the dataset). The fact that this does not appear to be considered as making the entity ``less of a model'' strongly indicates that the sense of ``model'' being used in frontier AI development now has very little to do with the notion of ``representation''.

In principle we allow that a strict subset of an AI model may be an AI model, and a strict subset of an AI system may itself be an AI system. Whether real-world techniques such as mixture-of-experts (MoE) architectures should be considered as nested models or not will be discussed in the Section \ref{subsec:OperationalDefinitionContemporaryAI} (Operational Definitions for Contemporary AI). In addition, we allow for the possibility that an AI system may contain multiple AI models. In principle, we consider it possible that an AI system may not clearly contain a distinct component that can be called the AI model, however for all important contemporary systems, such a component clearly exists.

We now explain the reasoning behind each element of these definitions and their implications for the model-system boundary. In practice, given the current paradigm of large neural networks based on transformer architecture, most AI models will start out as representations of some dataset. However, in order for our definition to function as intended, it is important that an entity continues to be an AI model even if it is not used for the purpose of representation, and even if it is modified in ways that make it a less accurate representation.

According to our definition, AI models are core components of AI systems. However, the key difference is that AI systems receive input from and influence external (real or virtual) environments, which AI models cannot do. Therefore, according to our definition, an AI model can never be a system by itself, since it lacks the necessary properties. This puts us in line with the EU’s view on models and systems \citep[Recital 97]{EUParliament2024AIAct}.

Our definition of AI system is a simplified version of the definition found in \citet{OECD2024Explanatory}. We used OECD as a basis because their definition of AI system already effectively captured the core concept behind the term, and that it is sensible to re-use existing definitions where possible, to aid compatibility with existing work. We simplified the definition slightly (mostly by removing examples) to make it easier to read. We are not concerned by the loss of technical precision or applicability, since this is the role of the operational definition, which will be presented later

We recognise that the boundary between AI systems and systems that are not considered AI can be fuzzy. Since the Dartmouth Workshop in 1956, which is said to be the inception of AI \citep{Russell2009AI}, there have been several paradigms and trends of AI development, some of which we have described in Section \ref{subsec:ParadigmShiftsAIModelandAISystem} (Paradigm Shifts in the Meaning of \textit{AI model} and \textit{AI system}). Notably, an earlier paradigm known as Good Old-Fashioned AI (GOFAI) mainly uses symbol manipulation explicitly defined by rules, and AIs developed in this manner are referred to as symbolic AI \citet[ISO/IEC 22989:2022 3.1.33]{ISOIEC22989:2022-3.1.33}. After the 21st century, especially after the creation of AlexNet in 2012, the dominant paradigm has been machine learning, where model parameters are optimised through computational techniques \citep[ISO/IEC 22989:2022 3.3.5]{ISOIEC22989:2022-3.3.5}. We elaborate on these paradigms below. 

These two main paradigms of AI do not have a clean boundary between them, as they have evolved slowly over decades. In addition, modern hybrid systems (e.g., neurosymbolic AI) increasingly combine learned statistical models with symbolic reasoning mechanisms, further blurring the boundaries between these approaches. Our intention is that our conceptual definition of \textit{AI system} covers both approaches, making it as widely applicable and future-proof as possible. Our operational definition (given in the next section) focuses more heavily on the current paradigm of AI.

Given the historical context of the term AI, which is to create intelligence artificially (i.e. non-human intelligence), and with the different paradigms that the field of AI went through, we find it challenging to come up with a clean conceptual definition of AI systems and AI models. 

Nevertheless, for AI systems, we believe it is important that, in line with cybernetic principles which preceded the field of AI, AI systems should carry the concept of producing outputs from inputs, where it would interact with its environment. 

\subsection{Operational Definitions for Contemporary AI}
\label{subsec:OperationalDefinitionContemporaryAI}

Having a conceptual definition helps us understand what AI systems and AI models are, but it does not necessarily draw clean boundaries in a way that makes it clear what counts as AI systems or AI models. In this section, we attempt to develop an operational definition such that the boundary between an AI model and an AI system is clear.

According to the \citet{APA2018OperationalDefinition}, an operational definition is ``a description of something in terms of the operations (procedures, actions, or processes) by which it could be observed and measured''. For example, Alan Turing’s proposed ``imitation game'' \citep{Turing1950Computing}, now widely known as the Turing test, is considered to be an operational definition of machine intelligence \citep{French2012BeyondTuring}. In other words, while a conceptual definition helps us understand the ideas behind a concept, an operational definition allows us to reliably identify instances of that concept in the real world.

As discussed previously, there are many types of AIs, including symbolic AI, machine learning, and neurosymbolic systems. The question of where the boundary between the model and the system lies is thus applicable across the different types of AI. However, because the dominant paradigm in current high-impact and general-purpose AI deployments is neural network-based machine learning AI such as LLMs, the operational definitions developed here are explicitly scoped to this class of systems, since at present these are the systems most in need of effective regulation. We therefore focus on contemporary general-purpose AI systems such as transformer-based LLMs. These operational definitions are not intended to exhaustively characterise model-system boundaries for other types of AIs. This means that our operational definition is much narrower and less future-proof than our conceptual definition, while being much easier and clearer to apply in practice. This is why we provide both a conceptual and operational definition, as the conceptual definition can be used to guide future updates to the operational definition.

We propose our operational definition in this section, summarised as follows:

\begin{quote}
    \textbf{AI model} (for contemporary AI): Trained parameters and their architecture, understood as the computational specification necessary to use them for inference.

    \vspace{0.2cm}
    \textbf{AI system} (for contemporary AI): One or more AI models, an interface for receiving inputs from and delivering outputs to an environment, and the configuration connecting these and any additional components.
\end{quote}

In the rest of this section, we explain the rationale behind establishing these operational definitions.

Based on both the systematic literature review and the manual review as well as the observed definitional developments in Section \ref{sec:findings} (Findings and Analysis), we have found the OECD reports to have been the most informative and influential in developing such definitions. Hence, here we attempt to use definitions from the OECD reports as a starting point, and further operationalise it to clarify the boundary between AI systems and AI models.

\citet{OECD2024Explanatory} defines AI system as follows:

\begin{quote}
    ``An AI system is a machine-based system that, for explicit or implicit objectives, infers, from the input it receives, how to generate outputs such as predictions, content, recommendations, or decisions that can influence physical or virtual environments. Different AI systems vary in their levels of autonomy and adaptiveness after deployment.'' (p. 4)
\end{quote}

Although the report does not explicitly provide a definition of an AI model, they describe it in several paragraphs, including the following:

\begin{quote}
    ``An AI model is a core component of an AI system used to make inferences from inputs to produce outputs. It is important to note that while the parameters of an AI model change during the build phase, they usually remain fixed after deployment once the build phase has concluded'' (p. 8)
\end{quote}

Several points follow from the OECD description. First, an AI system is described as inferring ``from the input it receives,'' implying that inputs originate outside the system and are not constitutive parts of it. Second, an AI model is characterised as ``a core component of an AI system'' which means it is considered a subset of the system rather than a representation of it. This means all parts of the model are necessarily parts of the system, but not vice-versa. Third, the model is the component responsible for performing inference. While the OECD's phrasing is somewhat ambiguous (it is unclear if the AI model or the AI system is the one making inference), we take this to mean that the model is the component responsible for performing inference, and that the system possesses this capability by virtue of containing the model.

\begin{figure}[h]
    \centering
    \includegraphics[width=0.6\linewidth]{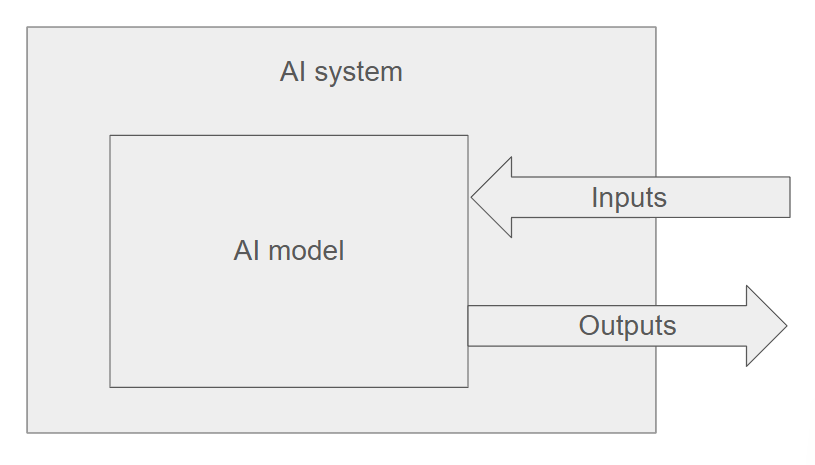}
    \caption{AI system containing an AI model}
    \label{fig:InputsOutputsAISystemAIModel}
\end{figure}

Nevertheless, we can draw relationships between an AI system, AI model, and the inputs and outputs, as shown in Figure \ref{fig:InputsOutputsAISystemAIModel}. This would be similar to the diagrams that OECD uses.

\begin{figure}[h]
    \centering
    \includegraphics[width=0.6\linewidth]{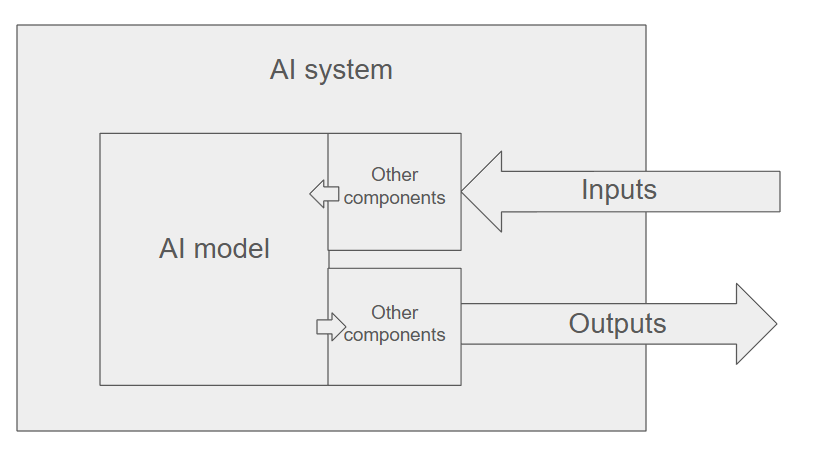}
    \caption{AI system containing an AI model with other non-model components}
    \label{fig:InputsOutputsAISystemAIModelOtherComponents}
\end{figure}

In most cases, however, the AI system would contain other components that are not part of the AI model, which we elaborate further below. This can be illustrated in Figure \ref{fig:InputsOutputsAISystemAIModelOtherComponents}.

The description of an AI model being ``a core component of an AI system'' alone does not disambiguate what parts of a system are a part of the model. We must then rely on the additional elaboration, where the OECD notes that ``while the parameters of an AI model change during the build phase, they usually remain fixed after deployment once the build phase has concluded''. This suggests two additional properties that can be operationalised as criteria for AI models: its parameters change during the build phase, and that these parameters usually remain fixed after deployment.

OECD does not specify an operational test for ``core component.'' In applying the definition, we interpret ``core'' as being necessary for the AI system to perform inference, and ``component'' as that it is part of the system’s configuration rather than a transient computational state.

Also referenced above are the ``parameters'' and the ``build phase.'' While the ``build phase'' is not defined in the same \citet{OECD2024Explanatory} framework, it is described in \citet{OECD2022Classification} which references \cite{OECD2019Scoping}, which states ``the model-building process is often called ``training'' or ``optimisation'' \citep[p. 46]{OECD2022Classification}.

In the context of machine learning, the process of training updates the model parameters to achieve the training objectives. Parameters can also be defined as the mathematical objects that can be changed by the gradient descent process during training \citep{LeCun2015DeepLearning}. 

Interpreting the OECD description operationally yields three candidate criteria for identifying what counts as part of an AI model:

\begin{enumerate}
    \item Core component of AI system to make inferences
    \item Parameters change during the build phase, i.e. training phase
    \item Usually remain fixed after deployment
\end{enumerate}

The concepts of ``build phase,'' ``parameters,'' and ``after deployment'' are compatible with machine-learning-based AI. We therefore use the three criteria above as operational criteria for determining whether a given element belongs to the AI model or not. We apply these rules below to elements of a transformer-based LLM system, including weights, input prompts, activations, system prompts, retrieval-augmented generation (RAG), temperature settings, input or output filters, and fine-tuning.

\begin{enumerate}
    \item \textbf{Weights (and biases, where available)}: In the transformer architecture, these include embeddings and unembeddings weights, attention weights, feed-forward network weights, and layer normalisation weights. They are certainly a core component of AI systems, where these parameters change during the build phase, and usually remain fixed after deployment (unless the model undergoes continuous learning post-deployment). As such, weights are certainly part of the AI model, and they are often referred to as ``model weights''.
    \item \textbf{Input prompt}: Inputs are processed by the system but are not part of the system’s configuration. They are runtime data rather than system components and are not made of parameters. Thus, they are not part of the AI model.
    \item \textbf{Activations}: These are intermediate computational states produced during forward passes as a function of inputs and parameters. Although they are necessary for performing inference, they are not themselves trained parameters and do not persist as part of the system after computation. While they are usually referred to as ``model activations'', they are not part of the system’s configuration, as they are results of intermediate calculations performed during model inference. 
    \item \textbf{System prompt}: As inputs to the AI model (and correspondingly, the AI system), system prompts can be important components of an AI system as it greatly influences how it makes inferences. A system prompt stored as part of deployment configuration can be a core component of the system, but it is not made of parameters and thus not part of the AI model.
    \item \textbf{Retrieval-augmented generation (RAG)}: As a technique that takes in additional sources beyond the input or system prompts to perform model inference, it does not consist of parameters and therefore is not part of the AI model. The associated knowledge base or retrieval database may constitute a component of the AI system if it is integrated into deployment, but it contains data rather than trained parameters. Hence, RAG databases are treated as system components but not as parts of the AI model.
    \item \textbf{Temperature}: As a setting that determines which tokens is being selected as output, this is not a component but a model setting. They do not contain parameters, and can be changed after deployment.
    \item \textbf{Input or output filters}: These can be said to be components that modify inputs before it is being processed by the model, or modify model results before it is being output by the system. These filters can be rule-based or produced using a training process. In the latter case, they would then contain parameters that are updated during training. 
    \item \textbf{Fine-tuning}: As a training process that modifies model parameters, it is not itself a component. However, certain parameter-efficient fine-tuning (PEFT) methods such as low-rank adaptation (LoRA), produce trained parameters in an adapter that is then ``attached'' to a frozen base model. These adapters can be argued as a core component of an AI system, which contains parameters that are trained and usually do not change after deployment. 
\end{enumerate}

These are summarised in Table \ref{table:AssessmentElementPartOECD2024}.

\renewcommand{\arraystretch}{1.8}
\begin{longtable}{|p{3cm}|>{\centering\arraybackslash}p{3cm}|>{\centering\arraybackslash}p{3cm}|>{\centering\arraybackslash}p{3cm}|>{\centering\arraybackslash}p{2cm}|}
\caption{Assessment of whether an element is part of an AI model according to the three candidate criteria from \citet{OECD2024Explanatory}}
\label{table:AssessmentElementPartOECD2024}\\
    \hline
         & \textbf{Core component of AI system to make inferences} & \textbf{Parameters change during the build phase} & \textbf{Usually remain fixed after deployment} & \textbf{Part of AI model}? \\
    \hline
        \textbf{Model weights} & Yes & Yes & Yes & Yes \\ 
    \hline
        \textbf{Input prompt} & Yes & Not parameters & Not parameters & No \\
    \hline
        \textbf{Activations} & Not a component - they are dynamic intermediate computation results & Not parameters & Not parameters & No \\
    \hline
        \textbf{System prompt} & Not a component - it’s an input of the model & Not parameters & Not parameters & No \\
    \hline
        \textbf{RAG} & Not a component - it’s a technique; but the associated knowledge base used by RAG can be a core component of the system & Not parameters & Not parameters & No \\
    \hline
        \textbf{Temperature setting} & Not a component - it’s a setting & Not parameters & Not parameters & No \\
    \hline
        \textbf{Input / output filters} & Component, but ambiguous if core & It depends on whether it has parameters & Yes if they are parameters & Maybe, depending on whether it has parameters \\
    \hline
        \textbf{Parameter-efficient fine-tuning (PEFT), e.g. LoRA} & Yes & Yes, even though it may not be built together with the rest of the model & Yes & Yes \\
    \hline
\end{longtable}

We can now see a fairly principled way of determining if something is part of an AI model or not. Nevertheless, we believe these conditions can be simplified. With regards to the point on ``parameters change during the build phase'', where the ``build phase'' refers to model training, we can simplify the criteria of being considered part of the AI model as being ``trained parameters''. In short, the definition of an AI model would contain two criteria: being a component of the AI system, and being trained parameters.

\renewcommand{\arraystretch}{1.8}
\begin{longtable}{|p{3cm}|>{\centering\arraybackslash}p{3cm}|>{\centering\arraybackslash}p{3cm}|>{\centering\arraybackslash}p{3cm}|}
\caption{ Assessment of whether an element is part of an AI model according to our candidate criteria}
\label{table:AssessmentElementPartCriteria}\\
    \hline
         & \textbf{Component of AI system} & \textbf{Trained parameters} & \textbf{Part of AI model}? \\
    \hline
        \textbf{Model weights} & Yes & Yes & Yes \\ 
    \hline
        \textbf{Input prompt} & No & No & No \\
    \hline
        \textbf{Activations} & Not really & No & No \\
    \hline
        \textbf{System prompt} & No & No & No \\
    \hline
        \textbf{RAG} & No for the technique itself; yes for the associated knowledge base, if implemented as part of the system & No & No \\
    \hline
        \textbf{Temperature setting} & No & No & No \\
    \hline
        \textbf{Input / output filters} & Yes & Yes in some cases & Yes in some cases \\
    \hline
        \textbf{Dumb classifier} & Yes & No & No \\
    \hline
        \textbf{Parameter-efficient fine-tuning (PEFT), e.g. LoRA} & Yes, if it is a component e.g. adapter & Yes, if it is a component e.g. adapter & Yes, if it is a component e.g. adapter \\
    \hline
\end{longtable}

A notable omission from the list above is model architecture. Architecture is not itself a set of trained parameters, and in its broad sense (e.g., ``transformer'' or ``convolutional neural network'') it denotes only a general design family rather than a specific model. However, in a narrower technical sense, architecture refers to the full computational graph specification, including the precise topology, connections, and operations that determine how trained parameters transform inputs into outputs. It is this narrower sense that is relevant here.

Under our conceptual definition, an AI model is something capable of performing inference. Trained parameters alone cannot perform inference without a computational specification that determines how they are applied. Accordingly, model architecture in this technical sense must be treated as part of the AI model, as it provides the necessary computational structure through which trained parameters are used to generate outputs.

This definition provides a principled way of determining whether something is part of an AI model. It also helps clarify when multiple trained parts belong to the same model. When several trained parts work together to produce a single inference and are not usable on their own to perform that inference, they are treated as parts of one AI model rather than as separate models.

Under this approach, PEFT adapters such as LoRA are considered to be part of the AI model, since they adjust the model’s parameters and are not used on their own. Likewise, instances of mixture-of-experts (MoE) architectures (including weights of all the experts) can be treated as a single AI model, because the experts and routing mechanisms work together to produce one inference and are not designed to operate separately.

A further implication of treating trained parameters as constitutive of the model is that the underlying data used to produce those parameters (i.e., the training data) should be considered as falling within the scope of the model, since it directly shapes the model's trained parameters. By contrast, data that is merely connected to or retrieved by the model at runtime as part of the wider system (e.g., via RAG) should be considered as falling within the scope of the wider system rather than the model itself.

Stepping back to the system level, the distinction between models and systems becomes clearer. An AI model consists of trained parameters and their architecture and is capable of performing inference. An AI system, by contrast, includes not only one or more such models but also the surrounding components that receive inputs and produce outputs in relation to an environment. This includes interfaces, configuration choices, and any additional components that are part of the system’s overall functioning.

This distinction becomes especially clear in the case of AI agents, which are typically designed to interact with an environment, maintain state or memory across steps, and select actions in pursuit of goals. Such systems generally include one or more AI models alongside additional components such as memory stores, planning modules, or tool-use mechanisms. Under the definitions proposed here, these assemblies clearly qualify as AI systems, as they combine models with interfaces and other components that enable interaction with an environment.

\subsection{Comparison Against Existing Definitions}

As per the previous sections, our proposed definitions and clarifications on the terminologies are relevant to some of the conceptual perspectives and not others. We summarise our proposed definitions in Table \ref{table:ProposedConceptualSummarisedDefinitions}.

\begin{longtable}{|p{2cm}|p{6cm}|p{7cm}|}
\caption{Our proposed conceptual and summarised operational definitions of \textit{AI model} and \textit{AI system}}
\label{table:ProposedConceptualSummarisedDefinitions}\\
    \hline
        & \textbf{Conceptual definition} & \textbf{Summarised operational definition for contemporary AI} \\
    \hline
        \textbf{AI model} & 
        A computational construct that makes inferences from inputs to produce outputs and forms (or is intended to form) the reasoning or knowledge core of an AI system. &
        Trained parameters and their architecture, understood as the computational specification necessary to use them for inference. \\
    \hline
        \textbf{AI system} & 
        A machine-based system that makes inferences from inputs from physical or virtual environments to produce outputs that influence physical or virtual environments. &
        One or more AI models, an interface for receiving inputs from and delivering outputs to an environment, and the configuration connecting these and any additional components. \\
    \hline
\end{longtable}

In defining an AI model, we use the conceptual perspective of treating the model as an artifact resulting from training (i.e. trained parameters) instead of treating the model as a representation of some external structure. As we are mainly concerned with contemporary general-purpose AI which are predominantly ML models, we operationally define a model as the result of the training process, which may be a definition that may not be appropriate for AI models that are not built through a training process. 

Additionally, we do not define an AI model in terms of its task-level function (e.g., prediction, recommendation, classification), since these functions are often properties of the system as a whole. While specifying particular functions may be suitable for defining specific AI systems, such as how Article 3(1) of the EU AI Act defines an AI system as being able to generate outputs ``such as predictions, content, recommendations, or decisions'', this specificity may not be helpful for defining AI models. This is because task-level functions of a system may arise from the interaction between the model and other system components and may not arise solely from the model itself. Hence, we define a model only by the general function of being able to perform inference.

For the relationship between the AI model and the AI system, we use one as a subset of the other, where the model is the subset of the system, and the system is the one that deals with the environment by taking in inputs and spitting out outputs. This is similar to but in contrast with \citet{Sharkey2024Causal} who defines an AI model as a component of the AI system, but states that AI models can also be AI systems that have no other system parameters. We posit a model, considered in isolation, does not constitute a perception-action loop with an environment, as environmental interaction arises only at the system level. This is in line with Recital 97 of the EU AI Act which says ``although AI models are essential components of AI systems, they do not constitute AI systems on their own. AI models require the addition of further components, such as for example a user interface, to become AI systems'' \citep{EUParliament2024AIAct}.

\subsection{Regulatory Implications and Recommendations}

In this section, we discuss regulatory implications and our recommendations both generally as well as specifically in the EU-context.

\subsubsection{General Implications and Recommendations}

Regulatory frameworks implicitly assume a supply-chain structure. The distinction between AI models and AI systems is valuable insofar as it corresponds to separable stages of design and creation of components, integration into products, and deployment into consumer markets that are often handled by different actors with different capacities to foresee and mitigate risk. Its legal relevance therefore lies not in the terminology itself, but in how it helps map obligations onto real divisions of responsibility in the AI value chain. 

While some jurisdictions explicitly rely on a distinction between AI models and AI systems, notably the EU, other jurisdictions may frame similar separations albeit using different categories such as components, services, or applications. This issue becomes more salient with the rise of AI agents and agentic AI, which are often implemented by layering tools, memory, and other components onto existing or augmented models or systems \citep{Sapkota2025AIAgents}, often by different actors in a modular manner. Across these variations in terminology and architecture, the key regulatory question is whether an artefact constitutes a separable or additional unit associated with a distinct actor and a distinct locus of control.

Ambiguity in the boundary between AI models and AI systems or similar terminologies thus creates predictable regulatory failure modes, including misallocated obligations, duplicative compliance burdens, and enforcement gaps. As such, definitions should have sufficient clarity to result in regulation that clearly spells out what is regulated, which actor is responsible, and how obligations are triggered when specific modifications are performed.

Our proposed operational definition of an AI model is deliberately narrow, identifying only the trained parameters and their architecture. In practice, however, commercial providers often place on the market a product that bundles the bare model with pre-specified components such as safety filters, content moderation layers, and system prompts, and refer to the whole package as a ``model.'' This terminological variation is reflected even in provider documentation which have evolved over time. OpenAI published a ``Model Card'' for GPT-3 \citep{OpenAI2020GPT-3} but shifted to ``System Card'' from GPT-4 onward \citep{OpenAI2023GPT-4}; Anthropic similarly used ``Model Card'' through Claude 3.5 \citep{Anthropic2024Claude3.5} before adopting ``System Card'' from Claude 3.7 \citep{Anthropic2025Claude3.7}. The terminologies for these documentations are updated as their products appear to expand from bare models to include other system components. On the other hand, Google DeepMind's Gemini 3 Pro ``Model Card'' explicitly describes ``product-level mitigations such as safety filtering,'' acknowledging that what is being documented extends beyond the model while still labelling the document as a Model Card \citep{Google2025Gemini3Pro}.

More broadly, different parties along the value chain can and should draw their own contractual lines to allocate responsibilities. This is acceptable as long as some identifiable entity is designated as bearing each relevant obligation, and the precise boundary between a model and a system is only instrumental to ensuring no responsibilities or obligations fall between the cracks. Definitions should not be treated as ends in themselves; they serve a regulatory purpose, and if technological practice shifts enough that a given boundary no longer maps onto meaningful differences in control or responsibility, the definition should be revised or abandoned.

Finally, we offer three principles for how regulatory definitions should relate to technical terminology. First, regulatory definitions should not deviate from technical consensus. Legal categories ultimately map onto technical artefacts in the real world, and a definition that is internally coherent from a drafting perspective may fail its purpose if it does not correspond to how real-world systems are actually built. We approached our proposed definitions with this consideration as a priority. Second, where a technically grounded definition is too narrow for a regulation's objectives, the appropriate response is to expand the regulatory scope explicitly through carve-ins, special provisions, or adapted definitions, rather than to distort the underlying technical term. For example, if regulation targets systems based on autonomy or societal impact, some recommendation systems may fall within scope even without a clearly identifiable AI model. This should be handled through explicit provisions, not by stretching what \textit{model} means. Third, where no existing technical term cleanly captures the intended regulatory object, it may be better to coin a new or more specific term purpose-built for the regulatory context rather than forcing an ill-fitting label into service.

\subsubsection{Implications specific to the EU}

The EU AI Act \citep{EUParliament2024AIAct} imposes obligations for providers of both high-risk AI systems (Article 16) and general-purpose AI models (Article 53). Providers are not limited to entities who first developed an AI system or an AI model. For high-risk AI systems, Article 25(1)(b) states that entities who make substantial modifications to a high-risk AI system shall be considered to be a provider of a high-risk system and become subject to provider obligations. For general-purpose AI models, a similar implication follows from two provisions. Recital 97 states that general-purpose AI models may be further modified or fine-tuned into new models, and Article 3(3) defines a provider to include entities who develop a general-purpose AI model and place it on the market. Together, these provisions support the view that placing a modified model on the market may confer provider status.

However, ambiguity remains as to when a modifying entity qualifies as a provider. This stems from two related uncertainties: what counts as an AI model or AI system, and what degree of change counts as a substantial modification or as creating a new GPAI model. As such, the European Commission issued guidelines for providers of GPAI models to clarify the scope of these obligations \citep[Article 96]{EC2025Guidelines}, alongside criteria for identifying AI models as GPAI models (Section 2.1) and for determining when an entity is considered a GPAI model provider (Section 3). However, because they do not clearly define what constitutes an AI model, uncertainty about modification and provider status persists, creating ambiguities on whether certain modified models are unregulated \citep{Holtman2025GPAI}.

Several authors have attempted to address the ambiguity surrounding when a modification yields a new model. For example, building on work by \citet{Burden2025Framework}, \citet{Pacchiardi2025Framework} proposes a framework for treating modified GPAI models as new ones by assessing behavioral changes, either by direct measurement or by using proxy metrics. On the other hand, \citet{Hacker2025Regulation} proposes categorizing modifications into insubstantial and substantial modifications, where insubstantial modifications include RAG, fine-tuning with minor adjustments, system prompts; while substantial modifications include fine-tuning with substantial adjustments, reinforcement learning with human or AI feedback (RLHF/RLAIF), and jailbreaking and related interventions. Unfortunately, categorizing modifications by substantiveness merely shifts the ambiguity to the question of when certain modifications are considered substantial. In both proposals, whether a change is a modification of a model is evaluated through proxies concerning intervention or behavior, without first specifying what constitutes the model whose modification is at issue. Our definitions address this prior question of what the model is, even though we do not purport to resolve when a modification is substantial enough to create a new model that comes with the relevant provider obligations.

Our proposed operational definitions aim to help address these ambiguities and support more consistent allocation of responsibilities across the AI value chain. Current definitional gaps create uncertainty about what constitutes an AI model versus general-purpose software, where system boundaries lie when multiple components are integrated, and which entities bear provider obligations in complex deployment scenarios. For instance, when an entity combines a retrieval system with a generative model to create a retrieval-augmented generation application, our operational definition clarifies that the retrieval component alone (which merely retrieves data) does not constitute an AI model, whereas the generative component (which consists of trained parameters) does, thereby clarifying which aspects of modification ought to trigger relevant provider obligations. Similarly, the definitions help allocate responsibilities when models are accessed remotely via APIs and integrated into downstream applications. In such cases, even though a model in itself may have passed through rigorous safety assessments, the system into which it is integrated could behave in ways that are inconsistent with the model’s intended performance once additional components are added to the system. Hence, the obligations of the provider of this system can thus be considered separately from those of the underlying model provider.

A related source of ambiguity concerns safety-enhancing components bundled with a model prior to market placement. Article 55(1)(b) of the EU AI Act requires providers of GPAI models with systemic risk to ``assess and mitigate possible systemic risks'', and Measure 5.1 of the GPAI Code of Practice explicitly lists system-level interventions among appropriate safety mitigations, including input/output filtering, staged API access, and tools for downstream actors \citep{EC2025GPAI}. This makes clear that it is acceptable and even expected for model providers to deploy system-level components to mitigate model-level risks. Under our operational definition, such additions (e.g., output filters, guardrails) are system-level components rather than parts of the model itself. However, the provider may package and market the entire bundle as a single ``model.'' This does not, in our view, change the underlying technical composition, any more than packaging changes the nature of the packaged product. For regulatory purposes, this bundling should not affect the provider's status: if the product being offered clearly contains a model, the provider remains subject to model-provider obligations regardless of what additional components are included. Our recommendation is in-line with that of \citet{Chin2025WhitePaper}, who propose as a general principle that the provider of a new model should unilaterally and pre-emptively define in their documentation exactly what they consider the boundaries of their model to be.

Notably, our conceptual and operational definition of \textit{AI model} and \textit{AI system} is built upon the definition of OECD 2024, which also served as the basis of the definition in the EU AI Act \citep{Nessler2024Response}. As a result, these definitions can easily be incorporated into existing regulations simply by providing additional clarification rather than requiring amendments. In the EU context, the most practical mechanism for incorporating these refined definitions would be through updated Commission guidelines issued under Article 96 of the AI Act, such as the guidelines for providers of general-purpose AI models \citep{EC2025Guidelines} issued in July 2025. We believe additional clarity will be helpful for various actors in the value chain, such as providers and deployers for AI systems and GPAI models. 

\subsection{Case Studies}
\label{subsec:CaseStudies}

The following case studies illustrate practical consequences of ambiguities in distinguishing between models and systems, and demonstrate how our proposed operational definitions may contribute to resolving these ambiguities.

\subsubsection{AlienChat Dispute}

In September 2025, the Xuhui District People’s Court in Shanghai convicted the two developers of the AlienChat app of ``producing obscene materials for profit'' under Chinese criminal law \citep{Zhang2026Jailed}. AlienChat integrated a third-party large language model to generate conversational responses and monetised access through subscriptions. The court found that the chatbot routinely generated sexually explicit content at scale.

At the core of the dispute was not whether the underlying model was capable of producing explicit language, but why such content was regularly produced at scale in deployment. The defendants argued that the ability to generate sexual content was an inherent property of the large language model they had integrated, and that the application merely exposed this pre-existing capability. The prosecution and the court, by contrast, focused on system-level design and deployment choices, including modifications to system prompts, the configuration of safety constraints, and the overall interaction design of the application. These were treated as evidence that the app developers had intentionally shaped the chatbot’s behaviour to elicit and sustain explicit interactions, rather than merely passing through the default neutral model outputs.

Our proposed definitions would arrive at the same conclusion, where the addition of non-model components of the system had changed the behavior of the system, even though modifications were not performed on the model. Nevertheless, in this case, the primary contribution of clearer definitions is analytic clarity and evidentiary burden. They help distinguish between harms arising from deployment choices and those arising from latent model capabilities or propensities, which can structure legal and factual analysis with regards to the allocation of liability.

\subsubsection{ChatGPT Geofencing in Italy}

In 2023, the Italian Data Protection Authority (\textit{Garante per la Protezione dei Dati Personali, Garante}) issued a temporary limitation order on ChatGPT, which operated on OpenAI’s GPT-3.5 model, citing concerns about compliance with multiple provisions of the GDPR. Garante found that OpenAI lacked an appropriate legal basis for the collection and processing of personal data used to train the system’s underlying algorithms, raising potential infringement issues under the GDPR \citep{Braun2023Italy}. One issue highlighted in public debate around the case was the difficulty of applying data subject rights, where the relationship between model training data and ongoing service-level processing became conceptually contested. 

OpenAI was subsequently fined a total of €15 million in relation to GDPR violations connected to ChatGPT’s data processing \citep{Laher2025OpenAI}. Although the service was reinstated after OpenAI implemented additional transparency and user-rights mechanisms, the company did not publicly clarify whether personal data used in prior training had been removed or technically isolated. This uncertainty highlights a structural tension in data protection law: regulatory restrictions were applied to the deployed service, while some of the underlying concerns related to historical training practices that shape model behavior.

The order targeted the system level (ChatGPT service in Italy), but the contested data processing was reflected in the model parameters through training. While in this particular case, the provider of both the AI model and the AI system was the same entity i.e. OpenAI, a lack of clarity between the model and system may prove legally challenging had they been provided by different entities. Under our proposed definitions, disputes about personal data used to train the model would concern the model itself, whereas disputes about personal data used as part of the system’s operation or accessed by the system at runtime (e.g., via RAG databases) would concern the system. This distinction does not by itself resolve how regulators should respond when personal data are already incorporated into model parameters and cannot be selectively removed. However, it clarifies whether a regulatory measure is directed at data used in model training or system-level data use, which in turn affects which actors are in a position to comply.

\subsubsection{Clearview AI Litigation}

Clearview AI, a facial recognition company based in the United States, developed a search engine built on an AI system trained using data scraped from publicly available websites, including social media platforms. By 2024, the company’s database contained over 50 billion facial images, which were converted into biometric identifiers and stored as feature vectors, which are numerical embeddings that allow the system to match uploaded images against the database \citep{Jung2024Privacy}. Regulators across Europe determined that this end-to-end pipeline, from mass data collection to the embedding of facial features, violated privacy rights under GDPR. Similar legal scrutiny followed in Canada, Australia, and the United States. Under the terms of the \textit{ACLU v. Clearview AI} settlement, the restrictions applied to access to the faceprint database, while the facial recognition algorithm itself was not subject to the injunction \citep{ACLU2022Clearview}. Although banned from operating in the EU, Australia, and Canada, and barred under the ACLU settlement from providing database access to most private entities nationwide and to any entity in Illinois for five years, Clearview retained the ability to sell its facial recognition algorithm without the accompanying embeddings database to public and private entities elsewhere in the US \citep{ACLU2022Clearview}. In the \textit{ACLU v. Clearview AI} settlement, the company admitted no liability \citep{Clearview2022Settlement}. 

This divergence in regulatory outcomes illustrates that jurisdictions do not consistently regulate both the AI model and the broader system in which it operates. The US settlement left the algorithm untouched, restricting only access to the faceprint database, effectively treating the model and database as separable for remedial purposes. However, authorities in the EU, Canada, and Australia concluded that the embeddings themselves constitute biometric data that remain linked to identifiable individuals, and therefore subject to data protection law.

Our proposed operational definitions do not determine whether embeddings qualify as biometric data, which is a separate legal question. They instead clarify what is being referred to when actors and regulators speak of the \textit{model} versus the \textit{system}, which can matter in disputes involving modular architectures or multiple responsible parties. In this case, regulators differed in whether they treated the database at the system level or the trained model as the object of regulation. Making these layers explicit helps describe such differences in regulatory practice more precisely.


\section{Limitations}
\label{sec:limitations}

In this section, we discuss the limitations of the manual and systematic literature review process, as well as limitations of our proposed conceptual and operational definitions. 

\subsection{Methodological Limitations}

Several limitations emerged throughout the literature review process during the different stages of the review that affect the completeness of the collected dataset and the interpretive certainty of the findings. This section discusses the challenges and limitations of both the manual review and the systematic literature review.

\subsubsection{Manual Review: Limitations in Screening and Inclusion}

\textbf{Incorrect References}

In some cases, different versions of the same regulatory or standards documents were inadvertently referenced, leading to inconsistencies in citations and uncertainty about which definition was operative at the time of publication. This issue is exacerbated by the frequent revision cycles of regulatory drafts and newly updated standards. Relatedly, definitions were occasionally misattributed, due to publications reusing earlier definitions without direct credit. These incorrect references may have impacted the analysis and comparison of different definitions.

\vspace{0.2cm}
\textbf{Ambiguity in Definitional Boundaries}

In many cases, there was a lack of explicit definitional sections in the documents. Several sources discuss AI models and systems descriptively but do not formally define them, and descriptions were often dispersed throughout the text. This would require interpretive judgments about whether a passage constituted a definition or was merely contextual commentary. This may have introduced subjectivity into inclusion decisions and may have led to the exclusion of borderline cases.

\vspace{0.2cm}
\textbf{Identifying Original Contributions}

During the manual extraction of definitions, there was some difficulty in tracing the chain of influence across definitions. Many definitions that appear as standalone entries in different sources are in fact replications or paraphrases of earlier definitions. This limited the ability to assess definitional originality and may have artificially inflated counts of ``unique'' definitions.

\subsubsection{Systematic Literature Review: Limitations in Search and Corpus Construction}

\textbf{Language Constraints and Jurisdictional Representation}

The scope of the corpus was constrained by the inclusion and exclusion criteria, which limited the search to materials written in English. While this ensured terminological consistency given the study’s focus on the specific phrases \textit{AI model} and \textit{AI system}, it necessarily excluded regulations, standards, and academic literature from jurisdictions where English is not the primary language of publication. Consequently, the final corpus exhibits a predominantly Eurocentric representation, reflecting jurisdictions such as the EU and OECD member states, whose materials are both accessible in English and where these terms are most explicitly formalised. It is important to acknowledge, however, that non-English jurisdictions may use alternative terminology or conceptual framings that could contribute meaningfully to future analyses of how these definitions evolve.

\vspace{0.2cm}
\textbf{Source Fragmentation}

It was observed that relevant regulatory and standards documents were distributed across agency websites, standards repositories, and institutional portals, many of which lacked unified indexing or persistent identifiers. This decentralisation made it difficult to retrieve materials through database-driven searches and often required supplementary manual search and retrieval, leading to reduced reproducibility. As a consequence, results may be skewed towards more accessible or well-indexed documents.

\vspace{0.2cm}
\textbf{Access Barriers}

Certain materials, such as technical standards from ISO were paywalled, preventing full-text access and verification. For these materials, analysis retrieval was usually conducted through limited previews, or references in other sources including excerpts or summaries. This may have resulted in the overrepresentation of open-access publications, whereas closed-access materials yielded fewer definitional artifacts despite being relevant. 

\vspace{0.2cm}
\textbf{Imbalance between AI model vs AI system}

Definitions of \textit{AI systems} appeared far more frequently than definitions of \textit{AI models}. As a result, the corpus is inherently skewed toward system-level descriptions. This limited the comparative analysis across the two terms and their definitional boundary, and nuanced changes in the conceptualisation of AI models may have been overlooked.

\subsection{Definitional Limitations and Future Work}

This work aimed to clarify an important distinction between two concepts that have been used almost interchangeably for the past few years, despite being given different roles in the legislation. The main challenge was to align with common usage in the field, when it is this usage that has been ambiguous. Words do not have an inherent definition, as their meaning comes from usage, and the boundaries we have placed could be deemed arbitrary. Our focus was that these new definitions could disambiguate the concepts of AI models and AI systems, so that regulatory frameworks referencing these concepts could be applied with greater clarity.

However, the definitions do not resolve all ambiguities. Years of inconsistent and overlapping usage of the terms AI models and AI systems, be it in technical literature, standards, and industry practice mean that it is not always clear where the boundaries of these concepts should lie. While improved definitions can reduce ambiguity in many cases, they cannot provide definitive answers to all edge cases. Some questions may not have a single ``correct'' solution at present, as the field would first have to converge on shared understanding of certain fundamental distinctions. Rather than claiming to provide definitive answers to all edge cases, our work aims to propose operational definitions that reduce ambiguity where possible, while highlighting remaining areas where continued interpretation and debate among regulators, implementers, and the technical community will be necessary.

Furthermore, despite an attempt to be technology neutral and future-proof, it is impossible to predict whether radical shifts in AI paradigms could give rise to new forms that would not fit neatly within the boundaries drawn by our definitions. Our proposed conceptual and operational definitions are grounded in current understanding of how AI models and systems are structured and deployed. While expect this framework to remain functional for at least 5-10 years considering current AI trajectories and past AI paradigm cycles \citep{Dhar2023Paradigm}, future developments may be transformative enough to represent fundamental breaks from current paradigms, and could require reconceptualisation of these fundamental categories. In consequence, mechanisms for periodic review and refinement of regulatory definitions into the governance frameworks that rely on them will be necessary.

Several categories of ambiguity remain unresolved. First, questions about compositional architectures persist: In agent-based systems, should the tools available to the agent be considered part of the system? Likewise, should the data used by a RAG system be considered as part of the system, or external to the system? Second, determining when modifications create a new model rather than an updated version of an existing model remains challenging. While the EU AI Act provides specific thresholds, such as the guideline that modifications using one-third of the original training compute may constitute a new model \citep{EC2025Guidelines}, it remains unclear whether compute-based metrics adequately capture the relevant distinctions. The relationship between computational resources expended and functional changes is not straightforward: shallow fine-tuning with minimal compute can produce significant behavioral shifts \citep{Volkov2024Badllama}, such as circumventing safety guardrails, while extensive parameter updates might produce relatively smaller functional changes. Moreover, as the scale of AI training continues to grow, fixed compute thresholds may become less meaningful indicators of whether a model has been sufficiently transformed to warrant treatment as a new regulatory object. Third, technical questions about model identity remain open: does changing activation functions without modifying weights constitute a new model? More fundamentally, when regulatory obligations hinge on these distinctions, what criteria should determine whether a modified system triggers new compliance requirements?


\section{Conclusions}
\label{sec:conclusions}

This paper provides a comprehensive survey and analysis of how \textit{AI model} and \textit{AI system} are defined across regulatory frameworks, standards bodies, and academic literature. Through both systematic and manual literature reviews, we examined over 80 definitions from sources including the OECD, EU AI Act, ISO standards, NIST, and numerous academic publications, tracing the evolution of these terms from 2012 to the present.

Our findings reveal several key insights. First, definitions of \textit{AI system} are far more numerous and developed than those of \textit{AI model}, reflecting the historical focus of governance discourse on system-level considerations. Second, we identified a concentrated chain of influence flowing from Russell and Norvig's foundational textbook into OECD frameworks, and further into ISO, NIST and EU definitions. Third, we documented how the OECD's own definitions evolved from 2019 to 2024, with the term \textit{AI model} expanding from a representation-focused definition to one that considers a model’s role as the reasoning and knowledge core of an AI system.

In our conceptual analysis of the existing literature, we found that definitions of \textit{AI models} and \textit{AI systems} can be organised along two main axes: parent categories (what kind of object they are taken to be) and features (how they are produced and what functions they serve). For \textit{AI models}, we identified two principal parent categories: model as artifact, which emphasises the result of building or training processes such as weights and parameters, and model as representation, which focuses on the model's purpose of representing some external structure. The representation is rooted in the history of AI (world models in classical agent architectures, neural networks as models of biological neurons, transformers originating in computational language modelling, etc). However, this perspective has become increasingly strained when applied to contemporary systems. Large language models, for instance, are routinely called \textit{models} despite not obviously being models of anything (particularly post fine-training). We argue that their designation as models stems from architectural lineage rather than from any representational function they perform. Recognising and accounting for this semantic drift is important to ensure that our definitions resemble the usage of the terms by frontier AI developers, helping those developers to clearly understand their legal responsibilities.

Our proposed definitions of the terms \textit{AI models} and \textit{AI systems} improve on the current state of the art while remaining compatible with existing regulatory frameworks. Our operational definition takes the form of an extensional definition grounded in technical practice, ensuring that it can be straightforwardly applied in real-world settings. Specifically, for neural network-based machine learning AI, we propose that an AI model consists of trained parameters and their architecture, understood as the computational specification necessary to use them for inference. This resolves questions about whether components such as system prompts, temperature settings, or retrieval databases constitute part of the model. Furthermore, our conceptual definition of AI models, while less suitable for real-world application, helps to explain the conceptual motivation behind our work, and should serve as a useful guide in case future AI paradigms require an adjustment in our operational definition. For AI systems and other definitions such as GPAI models, we find that the existing definitions are generally adequate, so we only recommend light changes to existing standards, if necessary. 

Our work has several implications for policymakers and standards bodies. By design, our definitions and recommendations can be incorporated into existing legal or technical frameworks without requiring substantial amendment. By clarifying the boundary between models and systems, these definitions support more consistent allocation of responsibilities across the AI value chain. This is a matter of increasing practical importance as models are deployed through APIs and integrated into downstream applications by parties other than the original developers.

While we believe that our work represents a significant contribution to the field, it is not without limitations, which we have outlined and discussed in the paper. For example, our corpus exhibits a predominantly English-language and Eurocentric representation, potentially missing alternative conceptual framings from other jurisdictions. Additionally, while we have attempted to future-proof our definitions, transformative developments in AI could necessitate reconceptualisation of these fundamental categories. Future work should extend this analysis to non-English regulatory frameworks and examine how definitions are operationalised in enforcement actions and compliance assessments. As AI systems become more complex and modular, empirical studies of how boundary disputes are resolved in practice would provide valuable guidance for refining definitional frameworks. The periodic review and refinement of regulatory definitions should be built into governance frameworks as a matter of course.

The effective regulation of AI requires shared understanding of fundamental terms. By documenting the landscape of existing definitions, analysing their conceptual underpinnings, and proposing clarifications grounded in both technical practice and regulatory needs, we hope to contribute towards this goal.


\section{Acknowledgements}
\label{sec:acknowledgements}
This work was carried out through the Supervised Program for Alignment Research (SPAR), whose program structure and infrastructure support made the research possible. We thank Koen Holtman, Antonio Puertas Gallardo, Yi-Yang Chua, Zifeng Wang, and Adrian Regenfuß for helpful comments on earlier drafts. All errors remain our own.


\section{References}
\renewcommand{\refname}{}
\vspace{-3em}  
\bibliographystyle{unsrtnat}
\bibliography{references}


\appendix
\label{sec:appendix}

\section{Overview of the Systematic Literature Review}
\label{subsec:SLROverview}

A systematic literature review was conducted to obtain definitions of the terms \textit{AI model} and \textit{AI system}. The proposed structure by \citet{Carrera-Rivera2022Review} regarding the planning and conducting of a systematic literature review was followed. Specifically, we adopted the planning-phase steps outlined in their framework, adapted to the present topic, including defining the research objective using the Population, Intervention, Comparison, Outcome, and Context (PICOC) structure \citep{Carrera-Rivera2022Review}, selecting databases, establishing inclusion and exclusion criteria, conducting quality assessment, and performing data extraction. 

\vspace{0.3cm}

\textbf{Research Objective}

\citet{Carrera-Rivera2022Review} suggest the PICOC structure to define the research objective and formulate search terms. Table \ref{table:SLRKeywordsfromPICOCCriteria} provides an overview of the keywords found using the PICOC criteria. Based on this breakdown, the following research question was formulated: How are the terms \textit{AI model} and \textit{AI system} defined?

\renewcommand{\arraystretch}{1.8}
\begin{longtable}{|p{2cm}|p{12cm}|}
\caption{SLR Keywords from PICOC Criteria}
\label{table:SLRKeywordsfromPICOCCriteria}\\
    \hline
    \textbf{Criterion} & \textbf{Keywords} \\
    \hline
    \textbf{Population} & 
    AI model and AI system 
    
    \vspace{0.3cm}
    
    \textit{Synonyms:}
    
    Machine learning system, autonomous system, automated system, algorithmic system
    
    Machine learning model, foundation model, algorithmic model, predictive model
    
    AI application, machine learning application, AI solution, machine learning solution, AI services \\
    \hline
    \textbf{Intervention} & 
    Definition 
    
    \vspace{0.3cm}
    
    \textit{Synonyms:}
    
    Ontology, conceptual framework, taxonomy, terminology, scope \\
    \hline
    \textbf{Comparison} & Not applicable \\
    \hline
    \textbf{Outcome} & Find different definitions, clarity on differences \\
    \hline
    \textbf{Context} & Computer science, engineering, social sciences, decision sciences, business, management and accounting, arts and humanities, multidisciplinary, psychology \\
    \hline
\end{longtable}

\textbf{Databases}

In order to sufficiently cover the area under investigation, different databases were selected. The following databases were used to obtain an overview of the literature: Scopus, Web of Science, IEEE Xplore. Google Scholar was used as a secondary source.

\vspace{0.3cm}

\textbf{Inclusion and Exclusion Criteria}

The following inclusion and exclusion criteria were chosen for the literature review. In particular, literature from 2012 onwards were included, as 2012 was the beginning of the modern AI era, with the influential AlexNet architecture \citep{Krizhevsky2012ImageNet} being released, marking the beginning of machine learning and deep neural networks as the dominant paradigm in AI. Moreover, only literature written in English was chosen, and their relevance to the research question was ensured by only including articles that provide a definition of the terms \textit{AI model}, \textit{AI system} or a concept synonymous to them in the final review. Lastly, only articles, reviews and conference papers were included. An overview of the criteria is provided in Table \ref{table:SLRInclusionandExclusionCriteria}.

\renewcommand{\arraystretch}{1.8}
\begin{longtable}{|p{4.5cm}|p{9.5cm}|}
\caption{SLR Inclusion and Exclusion Criteria}
\label{table:SLRInclusionandExclusionCriteria}\\
    \hline
    \textbf{Category} & \textbf{Selection Criteria} \\
    \hline
    \textbf{Period} & From 2012 to present \\
    \hline
    \textbf{Language} & English \\
    \hline
    \textbf{Relevance to research question} & The article provides a definition of \textit{AI model}, \textit{AI system}, or a concept synonymous with these terms \\
    \hline
    \textbf{Type of source} & Articles, reviews and conference papers \\
    \hline
\end{longtable}

\textbf{Quality Assessment and Data Extraction}

The quality of the articles is mainly assessed by manually evaluating for how clear and complete their definitions are. The relevance of each article’s definition to the research question is also evaluated.

Extracted data include the definitions of the relevant terms, as well as their sources. In some cases, the definitions in a source were attributed to another cited source. This was also noted during the manual review of the SLR results. 

\vspace{0.3cm}

\textbf{Search String}

Figure \ref{fig:SLRSearchString} shows the final search string. Synonyms of the terms \textit{AI system}, \textit{AI model} and ``definition'' were chosen as defined by the PICOC criteria. In order to ensure articles were discussing definitions, the proximity operator W/4, or NEAR/4, was used to only include articles where the definitional term is within a distance of four words of the AI term. To further narrow the search, articles with terms such as ``defined via'' and ``defined through'' in the abstract were excluded, as this would typically result in definitions of other terms in the context of AI models or systems, instead of the definition of AI models or systems themselves. Furthermore, only articles in the subject areas specified in the context criterion of the PICOC method were included. Lastly, as a first literature search showed high occurrence of article keywords that were not relevant for this search, the following keywords were excluded: diagnosis, female, adult, male, students, diseases, explainable ai, major clinical study, covid-19, fairness, energy.

\begin{figure}[h]
    \begin{tcolorbox}
        \footnotesize
        \begin{verbatim}
        (``AI system'' OR ``artificial intelligence system'' OR ``machine learning system'' 
        OR ``ML system'' OR ``Autonomous system'' OR ``Automated system'' 
        OR ``algorithmic system''
        
        OR ``AI model'' OR ``artificial intelligence model'' OR ``machine learning model'' 
        OR ``ML model'' OR ``foundation model'' OR ``algorithmic model'' 
        OR ``predictive model''
        
        OR ``AI application'' OR ``artificial intelligence application'' 
        OR ``machine learning application'' OR ``ML application''
        
        OR ``AI solution'' OR ``artificial intelligence solution'' 
        OR ``machine learning solution'' OR ``ML solution''
        
        OR ``AI services'' OR ``artificial intelligence services'')
        
        W/4 (``ontology'' OR ``defin*'' OR ``conceptual framework'' OR ``taxonomy'' 
        OR ``terminolog*'' OR ``conceptualis*'' OR ``conceptualiz*'' OR ``scope'')
        
        AND NOT (``defined via'' OR ``defined through'' OR ``well defined'' 
        OR ``user defined'')
        \end{verbatim}
    \end{tcolorbox}
    \caption{SLR Search String}
    \label{fig:SLRSearchString}
\end{figure}

\begin{figure}[h]
    \centering
    \includegraphics[width=12cm]{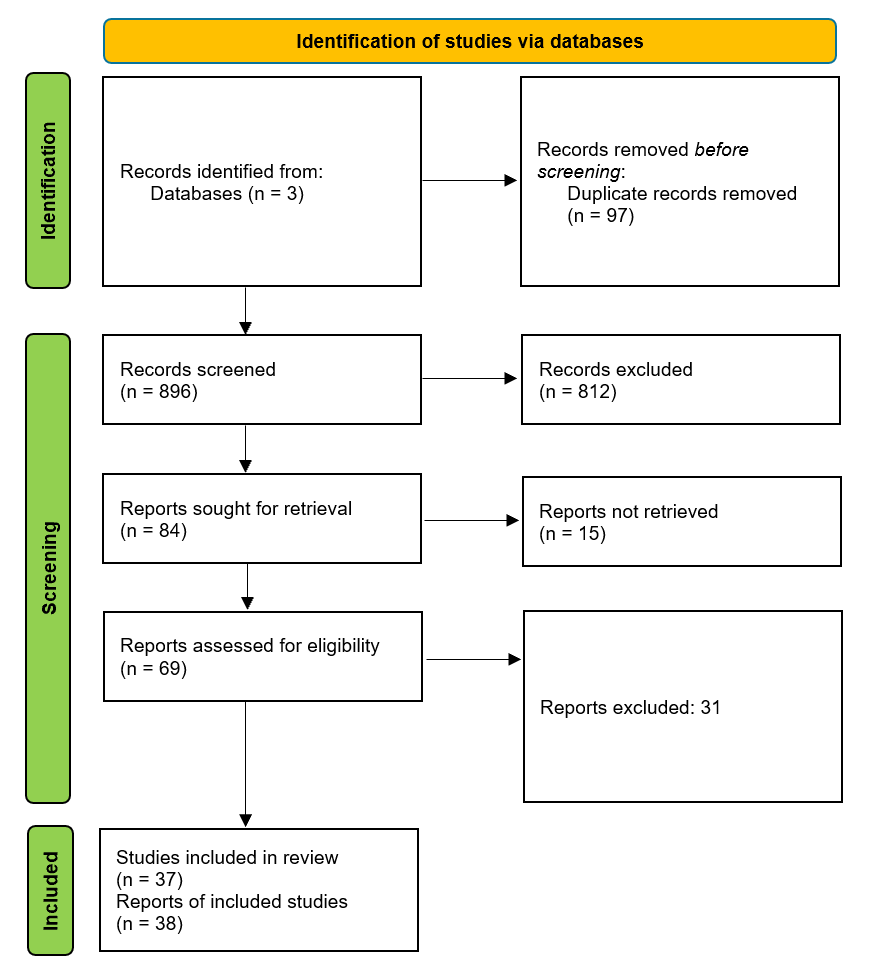}
    \caption{PRISMA Flow Diagram of the SLR}
    \label{fig:PRISMAFlow}
\end{figure}

This search was performed on October 15, 2025, and resulted in \textbf{864 papers} on Scopus, \textbf{113} on Web of Science, \textbf{16} on IEEE Xplore. After removing duplicates, a total of \textbf{896} papers were screened. Details can be found in Figure \ref{fig:PRISMAFlow}.

\section{Compilation of Definitions from the Manual Review}
\label{subsec:CompilationDefinitionManualReview}

As discussed in Section \ref{subsec:manual} (Manual Review of Global Regulatory and Standards Documents), we conducted a manual review of definitions from institutions that may not be found through the systematic literature review. Here, we group these institutions into four categories: (1) national/state-level government bodies, (2) intergovernmental organisations, (3) standards bodies, and (4) NGOs. Verbatim definitions were extracted and tabulated with source, date, and organisational category. This review illustrates how a small number of root formulations, most notably the OECD’s 2019 definition of an AI system, have been repeatedly adapted and embedded into legislation, standards and guidance (for example in the EU AI Act, the Council of Europe Framework Convention, ISO standards and NIST’s AI RMF). At the same time, the review highlights an asymmetry: while definitions of an AI system are largely based around a machine-based, goal-directed inference process conceptualisation of the term, definitions of an AI model remain sparse, fragmented, or purely derivative (e.g. ``essential components'' or ``general-purpose AI models'' defined without a prior base notion of ``model''). \textit{AI models} are alternately framed as mathematical objects, core components within systems, or bundles of capabilities, which in turn reinforces the conceptual ambiguity at the boundary between \textit{system} and \textit{model} that later sections of this paper analyse in more detail.

A compilation of these definitions, which are to be understood as samples of their respective categories, are in the following sections.

\subsection{Intergovernmental Organisations}

The OECD (Organisation for Economic Co-operation and Development) has standardised a definition of an \textit{AI system} since 2019, in line with that used in the popular textbook, \textit{Artificial Intelligence: A Modern Approach} \citep{Russell2009AI}. This root definition has, for the past few years, been a core influential definition of the term, factored into emerging AI legislation across the globe. Russel and Norvig’s definition has been interwoven into OECD documents since 2019. In 2023, the OECD expanded upon this choice in its \textit{Explanatory Memorandum On The Updated OECD Definition of an AI System}. Specifically, the objective can be ``implicit'', outputs can be ``inferred'', and the system may exhibit varying levels of ``adaptiveness'' after deployment. This same report also provides a definition of an AI model. In a separate report produced in 2022, the OECD situated its definition of an AI model as a process within the context of the AI system lifecycle.

\vspace{0.3cm}

\textbf{OECD definitions}

\begin{quote}
    \textbf{\citet[p. 2 \& 7]{OECD2017Foresight}}

    ``[...] because AI is an artifact, \textbf{AI systems} are constructed using architectures that limit AI to the knowledge and potential actions that make sense for a given application.''

    \vspace{0.2cm}
    ``[Dr. Joanna Bryson] characterised Artificial General Intelligence as a myth in that, like natural intelligence, AI is constrained by both the mathematics of combinatorics, and by the people who construct architectures limiting AI to actions that make sense. Stressing the difficulty of search if all possible options must be investigated (combinatorics), she attributed human intelligence to humans’ skill at communicating solutions once found, and --- more recently --- at mining these solutions from our culture as input to machines. Machine learning therefore exploits existing search to find the right thing to do at the right time.'' 
\end{quote}

\begin{quote}
    \textbf{\citet[p. 7]{OECD2019Scoping}}
    
    ``An \textbf{AI system} is a machine-based system that is capable of influencing the Environment by making recommendations, predictions or decisions for a given set of Objectives. It does so by utilising machine and/or human-based inputs/data to: i) perceive real and/or virtual environments; ii) abstract such perceptions into models manually or automatically; and iii) use Model Interpretations to formulate options for outcomes.'' 

    \vspace{0.2cm}
    ``A \textbf{Model} is an actionable representation of all or part of the external environment of an AI system that describes the environment’s structure and/or dynamics. The model represents the core of an AI system. A model can be based on data and/or expert knowledge, by humans and/or by automated tools like machine learning algorithms. Model Interpretation is the process of deriving an outcome from a model.'' 
\end{quote}

\begin{quote}
    \textbf{\citet[p. 23 \& 42]{OECD2022Classification}}
    
    ``An \textbf{AI system} is a machine based system that is capable of influencing the environment by producing recommendations, predictions or other outcomes for a given set of objectives. 

    \vspace{0.3cm}
    
    It uses machine and/or human-based inputs/data to: 
    \begin{enumerate}
        \item perceive environments;
        \item abstract these perceptions into models; and
        \item use the models to formulate options for outcomes.
    \end{enumerate}
    
    AI systems are designed to operate with varying levels of autonomy.''

    \vspace{0.2cm}
    ``\textbf{AI models} are actionable representations of all or part of the external context or environment of an AI system (encompassing, for example, processes, objects, ideas, people and/or interactions taking place in context). AI models use data and/or expert knowledge provided by humans and/or automated tools to represent, describe and interact with real or virtual environments.''
\end{quote}

\begin{quote}
    \textbf{\citet[p. 4, 6, 8]{OECD2024Explanatory}}
    
    ``An \textbf{AI system} is a machine-based system that, for explicit or implicit objectives, infers, from the input it receives, how to generate outputs such as predictions, content, recommendations, or decisions that can influence physical or virtual environments. Different AI systems vary in their levels of autonomy and adaptiveness after deployment.''

    \vspace{0.3cm}
    ``An \textbf{AI model} is a core component of an AI system used to make inferences from inputs to produce outputs […] while the parameters of an AI model change during the build phase, they usually remain fixed after deployment once the build phase has concluded.''    

    \vspace{0.3cm}
    ``\textbf{AI models} include, among others, statistical models and various kinds of input-output functions (such as decision trees and neural networks).''    
\end{quote}

\textbf{Intergovernmental organisations (Europe) definitions}

\begin{quote}
    \textbf{\citet[Article 3(1); Recital 97; Article 3(63); Article 51(1)]{EUParliament2024AIAct}}

    ``\textbf{AI system} means a machine-based system that is designed to operate with varying levels of autonomy and that may exhibit adaptiveness after deployment, and that, for explicit or implicit objectives, infers, from the input it receives, how to generate outputs such as predictions, content, recommendations, or decisions that can influence physical or virtual environments.''

    \vspace{0.3cm}
    ``Although \textbf{AI models} are essential components of AI systems, they do not constitute AI systems on their own. AI models require the addition of further components, such as for example a user interface, to become AI systems. AI models are typically integrated into and form part of AI systems.''

    \vspace{0.3cm}
    ``A `general-purpose AI model' is an \textbf{AI model}, including where such an AI model is trained with a large amount of data using self-supervision at scale, that displays significant generality and is capable of competently performing a wide range of distinct tasks regardless of the way the model is placed on the market and that can be integrated into a variety of downstream systems or applications, except AI models that are used for research, development or prototyping activities before they are placed on the market.''

    \vspace{0.3cm}
    ``1. A general-purpose textbf{AI model} shall be classified as a general-purpose AI model with systemic risk if it meets any of the following conditions:

    \vspace{0.1cm}
    (a) it has high impact capabilities evaluated on the basis of appropriate technical tools and methodologies, including indicators and benchmarks;

    \vspace{0.1cm}
    (b) based on a decision of the Commission, ex officio or following a qualified alert from the scientific panel, it has capabilities or an impact equivalent to those set out in point (a) having regard to the criteria set out in Annex XIII.

    \vspace{0.2cm}
    2. A general-purpose textbf{AI model} shall be presumed to have high impact capabilities pursuant to paragraph 1, point (a), when the cumulative amount of computation used for its training measured in floating point operations is greater than $10^{25}$.''
\end{quote}

\begin{quote}
    \textbf{\citet[2.1(17)]{EC2025Guidelines}}
    
    ``A general-purpose \textbf{AI model} is defined as any model trained using more than $10^{23}$ FLOPS (floating point operations per second) and capable of generating language (text/audio), text-to-image, or text-to-video outputs.''
\end{quote}

\begin{quote}
    \textbf{\citet[p. 1]{EC2018Communication}}
    
    ``\textbf{Artificial intelligence (AI) refers to systems} that display intelligent behaviour by analysing their environment and taking actions --- with some degree of autonomy --- to achieve specific goals.

    \vspace{0.2cm}
    \textbf{AI-based systems} can be purely software-based, acting in the virtual world (e.g. voice assistants, image analysis software, search engines, speech and face recognition systems) or AI can be embedded in hardware devices (e.g. advanced robots, autonomous cars, drones or Internet of Things applications).''
\end{quote}

\begin{quote}
    \textbf{\citet[p. 7]{HLEGAI2019Definition}}
    
    ``Artificial intelligence (AI) refers to systems designed by humans that, given a complex goal, act in the physical or digital world by perceiving their environment, interpreting the collected structured or unstructured data, reasoning on the knowledge derived from this data and deciding the best action(s) to take (according to pre-defined parameters) to achieve the given goal. \textbf{AI systems} can also be designed to learn to adapt their behaviour by analysing how the environment is affected by their previous actions. As a scientific discipline, AI includes several approaches and techniques, such as machine learning (of which deep learning and reinforcement learning are specific examples), machine reasoning (which includes planning, scheduling, knowledge representation and reasoning, search, and optimization), and robotics (which includes control, perception, sensors and actuators, as well as the integration of all other techniques into cyber-physical systems)''
\end{quote}

\begin{quote}
    \textbf{\citeauthor{HLEGAI2019Ethics} (\citeyear{HLEGAI2019Ethics}, p. 36; \citeyear{HLEGAI2020Assessment}, p. 24)}
    
    ``\textbf{Artificial intelligence (AI) systems} are software (and possibly also hardware) systems designed by humans that, given a complex goal, act in the physical or digital dimension by perceiving their environment through data acquisition, interpreting the collected structured or unstructured data, reasoning on the knowledge, or processing the information, derived from this data and deciding the best action(s) to take to achieve the given goal. AI systems can either use symbolic rules or learn a numeric model, and they can also adapt their behaviour by analysing how the environment is affected by their previous actions.''
\end{quote}

\begin{quote}
    \textbf{\citet[p. 13; p. 43]{EstevezAlmenzar2022Glossary}}
    
    ``\textbf{AI system}: A system based on artificial intelligence.''

    \vspace{0.2cm}
    ``\textbf{AI model}: In AI, this keyword mostly refers to a machine learning model or statistical model, which can make predictions/decisions over data. So only a subset of algorithms are models.'' 
\end{quote}

\begin{quote}
    \textbf{\citet[Article 2]{EUCouncil2024Convention}}
    
    ``23. The definition of an \textbf{artificial intelligence system} prescribed in this Article is drawn from the latest revised definition adopted by the OECD on 8 November 2023. The choice of the Drafters to use this particular text is significant not only because of the high quality of the work carried out by the OECD and its experts, but also in view of the need to enhance international co-operation on the topic of artificial intelligence and facilitate efforts aimed at harmonising governance of artificial intelligence at a global level, including by harmonising the relevant terminology, which also allows for the coherent implementation of different instruments relating to artificial intelligence within the domestic legal systems of the Parties.

    \vspace{0.2cm}
    24. The definition reflects a broad understanding of what artificial intelligence systems are, specifically as opposed to other types of simpler traditional software systems based on the rules defined solely by natural persons to automatically execute operations. It is meant to ensure legal precision and certainty, while also remaining sufficiently abstract and flexible to stay valid despite future technological developments. The definition was drafted for the purposes of the Framework Convention and is not meant to give universal meaning to the relevant term. The Drafters took note of the Explanatory Memorandum accompanying the updated definition of an artificial intelligence system in the OECD Recommendation on Artificial Intelligence (OECD/LEGAL/0449, 2019, amended 2023) for a more detailed explanation of the various elements in the definition. While this definition provides a common understanding between the Parties as to what artificial intelligence systems are, Parties can further specify it in their domestic legal systems for further legal certainty and precision, without limiting its scope.'' 
\end{quote}

\begin{quote}
    \textbf{\citet[p. 12]{Dunietz2024EUUSTaxonomy}}
    
    ``\textbf{Model}: A core component of an AI system used to make inferences from inputs in order to produce outputs. A model characterizes an input-to-output transformation intended to perform a core computational task of the AI system (e.g., classifying an image, predicting the next word for a sequence, or selecting a robot's next action given its state and goals)'' 
\end{quote}

\begin{quote}
    \textbf{\citet[p. 9]{EDPS2025Risk}}
    
    ``For the purposes of this Guidance, an \textbf{AI system} is understood within the meaning of Article 3(1) of Regulation 2024/1689 (AI Act) as `a machine-based system designed to operate with varying levels of autonomy, that may exhibit adaptiveness after deployment and that, for explicit or implicit objectives, infers, from the input it receives, how to generate outputs such as predictions, content, recommendations, or decisions that can influence physical or virtual environments'. The AI Act, however, does not contain a definition of an `AI model'. The terms AI model and AI system are often used as if they were synonyms, when they are not.''

    \vspace{0.2cm}
    \textbf{AI models} are mathematical representations that capture, in a set of parameters, the patterns underlying their training personal data. 16 Although AI models are essential components of AI systems, they do not constitute AI systems on their own, as they will always require other software components to be able to function and interact with users and the virtual or physical environment. In fact, an AI system can be composed of more than one AI model. For example, a voice translator AI system could be composed of a first model transcribing voice data into text, a second model translating the text from one language to another and a third model producing as output voice data from the translated text.'' 
\end{quote}

\textbf{Intergovernmental organisations (global) definitions}

\begin{quote}
    \textbf{\citet[p. 35]{Bengio2025International}}
    
    ``ChatGPT is a general-purpose \textbf{AI system} that combines the GPT-4o model with a chat interface, content processing, web access, and application integration to create a functional product. The additional components in an AI system also aim to enhance capability, usefulness, and safety. For example, a system might come with a filter that detects and blocks model inputs or outputs that contain harmful content. Developers are also increasingly designing so-called ‘scaffolding’ around general-purpose AI models that allows them to plan ahead, pursue goals, and interact with the world (see 1.2. Current capabilities).''
\end{quote}

\begin{quote}
    \textbf{\citet[p. 10]{UNESCO2022Ethics}}
    
    ``\textbf{AI systems} are information-processing technologies that integrate models and algorithms that produce a capacity to learn and to perform cognitive tasks leading to outcomes such as prediction and decision-making in material and virtual environments. AI  systems  are  designed  to  operate with varying degrees of autonomy by means of knowledge modelling  and representation and by exploiting data and calculating correlations. AI systems may include several methods, such as but not limited to:

    \vspace{0.1cm}
    (i)  machine learning, including deep learning and reinforcement learning;

    \vspace{0.1cm}
    (ii) machine reasoning, including planning, scheduling, knowledge   representation and reasoning, search, and optimization. 

    \vspace{0.2cm}
    AI  systems  can be used in cyber-physical  systems, including  the  Internet  of  things, robotic systems, social robotics, and human-computer  interfaces, which involve control, perception, the processing of data collected by sensors, and the operation of actuators  in  the  environment in which AI systems work''
\end{quote}

\subsection{Governmental Organisations}

\textbf{US state definitions}

\begin{quote}
    \textbf{\citet[22757.11(b)]{California2025SB53}}
    
    ``\textbf{`Artificial intelligence model'} means an engineered or machine-based \textbf{system} that varies in its level of autonomy and that can, for explicit or implicit objectives, infer from the input it receives how to generate outputs that can influence physical or virtual environments.''
\end{quote}

\begin{quote}
    \textbf{\citet[6-1-1701(2)]{Colorado2024SB24-205}}
    
    ``A \textit{machine-based} system that, for explicit or implicit goals, uses the inputs it receives to \textbf{infer} how to generate outputs (like content, decisions, predictions, or recommendations) that can influence \textbf{physical or virtual environments}.''
\end{quote}

\textbf{Chinese governmental definitions}

\begin{quote}
    \textbf{\citet[Article 22(1-2), translated from Mandarin Chinese]{CAC2023Measures}}
    
    ``\textbf{Generative Artificial Intelligence Technology}
    
    Refers to models and related technologies with the capability to generate content such as text, images, audio, or video.''

    \vspace{0.3cm}
    ``\textbf{Generative Artificial Intelligence Service Provider}
    
    Refers to organisations or individuals that provide generative AI services using generative AI technology (including through programmable interfaces).''
\end{quote}

\textbf{Korean governmental definitions}

\begin{quote}
    \textbf{\citet[Article 2(1-3)]{ROKAssembly2025Framework}}
    
    ``\textbf{`Artificial intelligence'} refers to the electronic implementation of human intellectual abilities such as learning, reasoning, perception, judgment, and language comprehension.''

    \vspace{0.3cm}
    ``\textbf{'Artificial intelligence technology'} refers to hardware, software technology, or its utilization technology required to implement artificial intelligence.''

    \vspace{0.3cm}
    ``\textbf{`Artificial intelligence system'} refers to an artificial intelligence-based system that infers results such as predictions, recommendations, and decisions that affect real and virtual environments for a given goal with various levels of autonomy and adaptability.''
\end{quote}

\subsection{Standards Agencies}

\textbf{ISO definitions}

\begin{quote}
    \textbf{\citet[ISO/IEC 22989:2022]{ISOIEC22989:2022}}

    ``3.1.1 \textbf{AI agent}
    
    automated (3.1.7) entity that senses and responds to its environment and takes actions to achieve its goals''

    \vspace{0.3cm}
    ``3.1.4 \textbf{AI system}
    
    engineered system that generates outputs such as content, forecasts, recommendations or decisions for a given set of human-defined objectives

    \vspace{0.2cm}
    Note 1 to entry: The engineered system can use various techniques and approaches related to artificial intelligence (3.1.3) to develop a model (3.1.23) to represent data, knowledge (3.1.21), processes, etc. which can be used to conduct tasks (3.1.35).

    \vspace{0.2cm}
    Note 2 to entry: AI systems are designed to operate with varying levels of automation (3.1.7).''

    \vspace{0.3cm}
    ``3.1.23 \textbf{model}

    physical, mathematical or otherwise logical representation of a system, entity, phenomenon, process or data

    \vspace{0.2cm}
    [SOURCE:ISO/IEC 18023-1:2006, 3.1.11, modified – Remove comma after ``mathematical'' in the definition. ``or data'' is added at the end.]''

    \vspace{0.3cm}
    ``3.3.7 \textbf{machine learning model}
    
    mathematical construct that generates an inference (3.1.17) or prediction (3.1.27) based on input data or information''

    \vspace{0.3cm}
    ``7.1 [...] Al systems contain a model which they use to produce predictions and these predictions are in turn used to successively make recommendations, decisions and actions, in whole or in part by the system itself or by human beings.''
\end{quote}

\begin{quote}
    \textbf{\citet[ISO/IEC/IEEE 29119-1:2022]{ISOIECIEEE29119-1:2022-3.6}}

    ``3.6 \textbf{AI-based system}
    
    system including one or more components implementing AI (3.7)''
\end{quote}

\begin{quote}
    \textbf{\citet[ISO/IEC TS 42119-2:2025]{ISOIECTS42119-2:2025-3.2}}

    ``3.2 \textbf{AI model}
    
    machine-readable representation of knowledge (3.15)

    \vspace{0.1cm}
    Note 1 to entry: An ML model (3.19) and the knowledge captured from experts as rules in an expert system are both forms of AI model.''

    \vspace{0.3cm}
    ``3.15 \textbf{knowledge}
    
    <artificial intelligence> abstracted information about objects, events, concepts or rules, their relationships and properties, organized for goal-oriented systematic use

    \vspace{0.2cm}
    Note 1 to entry: Knowledge in the AI (3.1) domain does not imply a cognitive capability, contrary to usage of the term in some other domains. In particular, knowledge does not imply the cognitive act of understanding.

    \vspace{0.2cm}
    Note 2 to entry: Information can exist in numeric or symbolic form.

    \vspace{0.2cm}
    Note 3 to entry: Information is data that has been contextualized, so that it is interpretable. Data is created through abstraction or measurement from the world.
    [SOURCE:ISO/IEC 22989:2022, 3.1.21]''
\end{quote}

\textbf{Chinese standards agencies definitions}

\begin{quote}
    \textbf{\citet[GB/T 41867-2022]{PRC2022GB/T41867-2022}}

    ``3.1.2 \textbf{artificial intelligence; AI}
    
    <subject> research and development of related mechanisms and applications of artificial intelligence system (3.1.8)''

    \vspace{0.3cm}
    ``3.1.8 \textbf{artificial intelligence system}
    
    a class of engineering systems that produce outputs such as content, prediction, recommendation, or decision for a given human-defined target
    
    \vspace{0.2cm}
    Note 1: This engineering system uses various technologies and methods related to AI (3.1.2) to develop models such as representing data, knowledge, processes, etc. for performing tasks.

    \vspace{0.2cm}
    Note 2: AI systems have different levels of automation.''
\end{quote}

\textbf{US standards agencies definitions}

\begin{quote}
    \textbf{\citet[NIST AI 100-1]{NIST2023RMF}}

    ``The AI RMF refers to an \textbf{AI system} as an engineered or machine-based system that can, for a given set of objectives, generate outputs such as predictions, recommendations, or decisions influencing real or virtual environments. AI systems are designed to operate with varying levels of autonomy (Adapted from: OECD Recommendation on AI:2019; ISO/IEC 22989:2022).''
\end{quote}

\subsection{Other Non-Governmental Organisations}

Non-governmental organisations (NGOs) were identified through a structured manual review of publicly available online sources. Starting from curated indexes and directories of AI-focused organisations, such as the AI Ethicist AI Organisations list, we iteratively followed outbound links to NGO websites and publications, and then applied inclusion criteria aimed at locating definitional or quasi-definitional language about AI models or AI systems/applications (e.g., glossaries, policy briefs, technical explainers, and governance reports). The resulting NGO list is therefore a sample, and does not represent all civil society actors.

\vspace{0.3cm}

\textbf{NGO definitions}

\begin{quote}
    \textbf{\citet[p. 17-19]{Poretschkin2023Guideline}}
    
    ``An \textbf{ML model} is a mathematical, abstract object that was created by a Machine Learning technique and serves to solve a task in the sense of creating an input-output relation. For example, regarding a neural network, the model consists of a list of hyperparameters, learned parameters and a description of how they interact when operating (architecture), among others. The ML model provides the functional basis of the AI application. For instance, a model can serve as the basis for performing a classification task, where the input is the object to be classified and the model output characterizes its class. In some cases, such as in generative models, the input (usually random numbers) may be of secondary importance to the actual task to be solved.''

    \vspace{0.2cm}
    ``An \textbf{AI application} is the input-output mapping in a given application context based on the implemented AI component. Importantly, the AI assessment catalog does not take an isolated view of the ML model or the AI component. In fact, the catalog examines whether the individual processing steps performed up to the results being generated by the AI component through interaction with other components of the embedding are meaningful and appropriate for the given application context. For example, it assesses whether they are sufficiently free of errors and discrimination or secure against attacks and manipulation. Thus, the assessment object of the catalog is not predominantly the mathematical concepts underlying the AI component, but the functionality i.e., the input-output mapping, that is performed based on the AI component (as a functional basis) in a given application context. An AI application can be a standalone system that, for example, communicates directly with users, or it can be integrated into a larger (IT) system or product.''

    \vspace{0.2cm}
    ``If an \textbf{AI application} is part of a larger system that is not entirely based on AI technologies, its boundaries within this surrounding system must be clearly defined. For example, ML-based object recognition (AI application) can be integrated into an autonomous vehicle, into a drone or into a site surveillance system (as larger system). Defining an AI application within a surrounding system is largely based on its functionality. For example, regarding an AI-based pedestrian detection system (AI application) in an autonomous car (larger system), the software modules that perform consistency checks on the outputs of the AI component should be seen as part of the AI application; however, other components that perform the overall planning of a driving route, for instance, would not be.''
\end{quote}

\begin{quote}
    \textbf{\citet{MindFoundry2026Model}}

    ``A \textbf{model} in AI and machine learning is a mathematical representation that captures patterns in data to make predictions or perform specific tasks. It is created through training, where the model learns from examples to recognise relationships between inputs and outputs. Models can vary in complexity, from simple linear regression to advanced neural networks, depending on the type of problem they are designed to solve.''
\end{quote}

\begin{quote}
    \textbf{\citet[p. 25]{CHT2024Society}}

    ``\textbf{AI systems} are more ``grown'' or ``trained'' than they are programmed. AI systems are produced by rewarding an AI for ``good behavior'' and punishing it for ``bad behavior'', but the final capabilities, intentions, and goals of an AI model are hard to characterize. (informal definition)''
\end{quote}

\begin{quote}
    \textbf{\citet[p. 2]{Moes2022Defining}}

    ``Benchmark-based definition is the most accurate and liked, but harder to roll out: `General purpose \textbf{AI systems} are AI systems that score above $x\%$ on the EU standardized testing suite for generality administered by the European Benchmarking Institute.'''

    \vspace{0.2cm}
    ``Self-reported task-based definition is easier to roll out, but has many loopholes: `General purpose \textbf{AI systems} are AI systems that can be reasonably foreseen to carry out a broad range of tasks (e.g. $\geq 10$) from the EU official list of tasks with minimal recalibration.'''
\end{quote}

\begin{quote}
    \textbf{\citet[p. 25]{Sharkey2024Causal}}

    ``The objects of our framework, \textbf{AI systems}, are a slight generalization of AI models. AI systems include not only the weights and architecture of the model, but also include a broader set of system parameters (Figure 1). These consist of retrieval databases and particular kinds of prompts.''

    \vspace{0.2cm}
    ``\textbf{AI models} are still \textbf{AI systems}. They are simply a special case of AI systems that have only model parameters and no other AI system parameters.''
\end{quote}

\section{Compilation of Definitions from the Systematic Literature Review}

In total, 25 definitions for the term ``AI model'' and related terms were retrieved through the systematic literature review (across 13 different papers). The total number of definitions for the term ``AI system'' and related terms is higher, with 57 (across 25 different papers). Terms related to ``AI model'' found in the literature were: foundation model (8 definitions), AI model (7 definitions), machine learning model (7 definitions), semantic machine learning model (2 definitions), general purpose AI model (1 definition). Terms related to ``AI system'' were: AI system (32 definitions), general purpose AI system (7 definitions), autonomous system (5 definitions), a system’s autonomous capabilities (3 definitions), automated decision system (2 definitions), machine learning system (1 definition), foundation model-based system (1 definition), Artificial General Intelligence system (1 definition), industrial autonomous system (1 definition), AI-based system (1 definition), intelligent system (1 definition), human-aware AI system (1 definition), cybernetic system (1 definition).

\subsection{AI Model}

Included in Table \ref{table:SLRFFindings-AIModel} are the authors and sources of definitions for the terms \textit{AI model} and ``machine learning model''. 10 authors provide a total of 13 unique definitions of these terms (the definition from ISO/IEC 22989:2022 \citep{ISOIEC22989:2022} is used by two separate authors). Despite the broad search from 2012 to 2025, the publication years of the papers that were found range between 2022 and 2025, with 6 of 9 papers published in 2024. When looking at the publication dates of the sources referenced, a similar pattern can be found, with all sources apart from one \citet{Ramanathan2016Symbolic} being published between 2021 and 2024.

\Needspace{12\baselineskip}
\begin{longtable}{|p{5cm}|p{4cm}|p{5cm}|}
\caption{SLR Findings for \textit{AI model} and \textit{machine learning model}}
\label{table:SLRFFindings-AIModel}\\
    \hline
        \textbf{Authors} & \textbf{Term defined} & \textbf{Source of definition} \\
    \hline
        \citet{Muhlhoff2025Updating} & 
        Machine learning model
    
        \vspace{0.3cm}
        
        Machine learning model &
        
        Authors themselves
    
        \vspace{0.3cm}
        
        \citet{Muhlhoff2024Regulating} \\
    \hline
        \citet{Kallab2024ML} & 
        Machine learning model
    
        \vspace{0.3cm}
        
        Machine learning model &
        
        Authors themselves
    
        \vspace{0.3cm}
        
        \citet{Janiesch2021Machine} \\
    \hline
        \citet{Fernandez-Llorca2025Account} & 
        AI model
    
        \vspace{0.3cm}
        
        Machine learning model &
        
        \citet{ISOIEC22989:2022}
    
        \vspace{0.3cm}
        
        \citet{ISOIEC22989:2022} \\
    \hline
        \citet{Lu2024Taxonomy} & AI model & Authors themselves \\
    \hline
        \citet{Hupont2024Cards} &

        AI model
        
        \vspace{0.3cm}
        AI model &

        \citet{ISOIEC22989:2022}

        \vspace{0.3cm}
        \citet{OECD2024Explanatory} \\
    \hline
        \citet{Park2024IMO} & AI model & Authors themselves \\
    \hline
        \citet{Trincado2024Legal} & AI model & \citet{EOPresident2023Safe} \\
    \hline
        \citet{Kallab2023SML} & Machine learning model & Authors themselves \\
    \hline
        \citet{Li2022SAIBench} & AI model & Authors themselves \\
    \hline
        \citet{Myllyaho2022Misbehaviour} & AI model & Authors themselves \\
    \hline
\end{longtable}

\subsection{AI System}

Table \ref{table:SLRFindings-AISystem} lists the authors and sources for the definitions of the terms \textit{AI system}, ``machine learning system'', and ``AI-based system''. 18 authors provided definitions for these terms, with many giving multiple definitions, leading to a total of 36 definitions. Some sources are referenced across different papers, with 9 references of the EU AI Act or other sources authored by EU institutions and 6 references of various OECD documents. This gives a total of 22 unique definitions, twice the number of definitions for \textit{AI model}. Publications are again skewed towards recent years, and 2024 is once again the publication year with the highest number of publications (5). However, some papers with earlier publication years than for the term AI model exist, with the earliest published in 2018. Some of the sources referenced by the authors show even earlier publication years, including: \citet{Rai2019Hybrids}, \citet{Korikov2011Mechatronics}, \citet{Russell2009AI}, \citet{Mitchell2006Discipline}, \citet{Makarov2001Intelligent}, and \citet{Backlund2000Definition}.

\begin{longtable}{|p{5cm}|p{4cm}|p{5cm}|}
\caption{SLR Findings for \textit{AI system} and ``machine learning system''}
\label{table:SLRFindings-AISystem}\\
    \hline
        \textbf{Authors} & \textbf{Term defined} & \textbf{Source of definition} \\
    \hline
        \citet{Ojanen2025Neutrality} & AI system & \citet{EUParliament2024AIAct} \\
    \hline
        \citet{Muller2025Transparency} & AI system & Authors themselves \\ 
    \hline
        \citet{Fernandez-Llorca2025Account} & 

        AI system
    
        \vspace{0.3cm}
        AI system

        \vspace{0.3cm}
        AI system
    
        \vspace{0.3cm}
        AI system &
        
        \citet{OECD2019Recommendation}
    
        \vspace{0.1cm}
        \citet{EC2021Proposal} 
            
        \vspace{0.1cm}
        \citet{EUParliament2023Amendments} 
            
        \vspace{0.1cm}
        \citet[Article 3(1)]{EUParliament2024AIAct} \\
    \hline
        \citet{Lu2024Taxonomy} & AI model & Authors themselves \\
    \hline
        \citet{Hupont2024Cards} &

        AI system
        
        \vspace{0.3cm}
        AI system &

        Authors themselves

        \vspace{0.3cm}
        \citet{OECD2019Recommendation} \\
    \hline
        \citet{Buiten2024Liability} & 
        
        AI system
        
        \vspace{0.3cm}
        AI system

        \vspace{0.3cm}
        AI system & 
        
        Authors themselves
        
        \vspace{0.3cm}
        \citet{Buiten2023Law}
        
        \vspace{0.3cm}
        \citet{EUParliament2024AIAct} \\
    \hline
        \citet{Trincado2024Legal} & 
        
        AI system  
        
        \vspace{0.65cm}
        AI system

        \vspace{0.3cm}
        AI system & 
        
        \citet{EOPresident2023Safe}
        
        \vspace{0.1cm}
        \citet{OECD2019Recommendation}
        
        \vspace{0.1cm}
        \citet{EUParliament2024AIAct} \\
    \hline
        \citet{Boone2023Challenge} & 
         
        AI system
        
        \vspace{0.8cm}
        AI system

        \vspace{0.6cm}
        AI system 

        \vspace{0.3cm}
        AI system
        
        \vspace{0.3cm}
        AI system

        \vspace{1cm}
        AI system & 
                
        \citet{EC2021Proposal}
        
        \vspace{0.1cm}
        \citet{EUCouncil2022AIAct}

        \vspace{0.1cm}
        \citet{EUParliament2023Amendments}

        \vspace{0.1cm}
        \citet{OECD2019Recommendation}
        
        \vspace{0.1cm}
        \citet[Adapted from \cite{OECD2019Recommendation}, \citet{ISOIEC22989:2022}]{NIST2023RMF}

        \vspace{0.1cm}
        \citet{OECD2024Explanatory} \\
    \hline
        \citet{Straub2023Government} & 
        
        AI system

        \vspace{0.3cm}
        AI system & 
        
        Authors themselves
        
        \vspace{0.3cm}
        \citet{Gil2019Roadmap}\\
    \hline
        \citet{Spivakovsky2023Policies} & AI system & \citet{CMUkraine2020Resolution1556} \\
    \hline
        \citet{Diaferia2022Standard} & 
        
        AI system

        \vspace{0.3cm}
        AI system & 
        
        \citet{Kaplan2019Siri}
        
        \vspace{0.1cm}
        \citet{Rai2019Hybrids}\\
    \hline
        \citet{Myllyaho2022Misbehaviour} & Machine learning system & \citet{Mitchell2006Discipline} \\
    \hline
        \citet{Krafft2021Action} & 
        
        AI/AI System/Automated Decision System

        \vspace{0.15cm}
        AI/AI System/Automated Decision System & 
        
        \citet{OECD2019Recommendation}
        
        \vspace{0.3cm}
        Authors themselves \\
    \hline
        \citet{Martinho2021Computer} & AI system & 
        
        Authors themselves

        \vspace{0.1cm}
        \citet{Backlund2000Definition} \\
    \hline
        \citet{Manser2021Servitization} & AI system & Authors themselves \\
    \hline
        \citet{Balasubramaniam2020Guidelines} & AI system & Authors themselves \\
    \hline
        \citet{Coleman2021Understanding} & AI system & \citet{Russell2009AI} \\
    \hline
        \citet{Korikov2018Robot} & 
        
        AI System

        \vspace{0.15cm}
        AI System & 
        
        Authors themselves
        
        \vspace{0.1cm}
        \citet{Makarov2001Intelligent} and \citet{Korikov2011Mechatronics} \\
    \hline
\end{longtable}

\section{References on Conceptual Perspectives of Definitions}
\label{subsec:ReferencesConceptualDefinitions}

As described in Section \ref{subsec:conceptual} (Conceptual Perspectives on Existing Definitions), we offer several conceptual perspectives on the existing definitions of AI models and AI systems, according to their parent categories and their features for both AI models and AI systems, and the relationship between the two. Below, we detail the references behind these conceptual perspectives.

\subsection{AI Model}

\paragraph{Parent Categories of AI Model}

\textbf{Model Artifact (Result of Building/Training)}

Several sources specify that an AI model is the result of building/training. The \citet{OECD2019Scoping} states that a model can be ``based on data and/or expert knowledge'', specifying that it is created either by humans or by automated tools like machine learning algorithms. Similarly, \citet{Hupont2024Cards}, citing \textit{Explanatory Memorandum on the Updated OECD Definition of an AI System} \citep{OECD2024Explanatory}, portrays an AI model as the outcome of training via machine learning algorithms or other AI methods. \citet{Trincado2024Legal}, referring to \textit{Executive Order 14110} \citep{EOPresident2023Safe}, further supports this notion, suggesting that a model uses ``computational, statistical, or machine-learning techniques.'' Moreover,  \citet{EstevezAlmenzar2022Glossary} specifies that the term \textit{AI model} often refers to either statistical models or machine learning models.

A few definitions for machine learning models also fall under the perspective of model artifact. \citet{Myllyaho2022Misbehaviour} cited \citet{Ramanathan2016Symbolic} for a description of machine learning models as non-deterministic statistical approximations. \citet{Kallab2024ML} define a machine learning model as consisting of an algorithm, a training dataset, an application domain, model performance metrics, and metadata. Lastly, \citet{Muhlhoff2024Regulating} (as cited by \citet{Muhlhoff2025Updating}) refer to weights and parameters as ``model data'', emphasising the informational essence of machine learning models.

\vspace{0.3cm}
\textbf{Model as Representation}

AI models have been described as different types of representations. Some definitions foreground the mathematical nature of models. The \citet{EDPS2025Risk} defines an AI model as a mathematical representation, capturing patterns found in their training data. This is further confirmed by \citet{MindFoundry2026Model} that defines a model as a mathematical representation. The \citet{EstevezAlmenzar2022Glossary}’s view of AI models as either statistical or machine learning models suggests a similar view, as statistical models are mathematical representations. While \citet{Hupont2024Cards} have a similar stance, defining an AI model as a mathematical construct, reviewing the source they are citing reveals an error in their citation. The researchers refer to \citet{EstevezAlmenzar2022Glossary} who provide a definition of a machine learning model as a mathematical construct, rather than an AI model, and refer to ISO/IEC 22989:2022 \citep{ISOIEC22989:2022}. We have reviewed ISO/IEC 22989:2022 \citep{ISOIEC22989:2022} in our manual review, and found that while ISO/IEC 22989:2022 \citep{ISOIEC22989:2022} does describe an AI model as possibly a mathematical representation, it can also be a physical or otherwise logical representation of a system, entity, phenomenon, process or data (also quoted like this by \citet{Fernandez-Llorca2025Account}). The definition of a machine learning model is provided as referenced by \citet{EstevezAlmenzar2022Glossary}.

Other definitions focus on the AI model as a representation of an external environment. The OECD, both in 2019 and 2022 \citep{OECD2019Scoping,OECD2022Classification} refers to an AI model as an ``actionable representation of all or part of the external environment of an AI system''. More broadly, ISO/IEC TS 42119-2:2025 \citep{ISOIECTS42119-2:2025-3.2} defines it as a machine-readable representation of knowledge, and  ISO/IEC 22989:2022 \citep{ISOIEC22989:2022} suggests it can be a number of different representations (as discussed above). This is also found by \citet{Sponheim2024Glossary}, who refer to models as approximate representations of natural phenomena or concepts.

Particularly for machine learning models, as already mentioned, ISO/IEC 22989:2022 \citet{ISOIEC22989:2022} defines them as mathematical constructs (also cited by \citet{Fernandez-Llorca2025Account}). This view is also held by \citet{Poretschkin2023Guideline}. \citet{Myllyaho2022Misbehaviour} citing \citet{Ramanathan2016Symbolic} to define machine learning models as non-deterministic statistical approximations suggests a similar viewpoint.

\paragraph{Features of AI Model}

\textbf{Production (How It Is Built/Trained)}

Some definitions specify how a model is built. In particular, several definitions by the OECD discuss the creation of an AI model. Its definition from both 2019 and 2022 specifies a model based on data and/or expert knowledge, provided by humans and/or automated tools \citep{OECD2019Scoping,OECD2022Classification}. This definition changes in 2023, describing more broadly that a model’s parameters change during its build phase, and specifying the inclusion of statistical models and other input-output functions \citep{OECD2024Explanatory}. \citet{Hupont2024Cards} also cited the \textit{Explanatory Memorandum on the Updated OECD Definition of an AI System} \citep{OECD2024Explanatory}, portraying an AI model as the outcome of training via machine learning algorithms or other AI methods. Other sources apart from the OECD provide similar definitions. \citet{Trincado2024Legal} cites \textit{Executive Order 14110} \citep{EOPresident2023Safe} to define an AI model as using ``computational, statistical, or machine-learning techniques.'' The \citet{EstevezAlmenzar2022Glossary} specifies that the term \textit{AI model} often refers to either statistical models or machine learning models, which provides information on how it is built. More explicitly, \citet{MindFoundry2026Model} suggests that an AI model is built through training, learning to recognise relationships between inputs and outputs through examples, which indicates a focus on machine learning methods as well. \citet{Lu2024Taxonomy} follow this focus on machine learning, defining a model as trained from scratch on a dataset for a specific task. 

\vspace{0.3cm}
\textbf{Function (What It Is Used for After Building)}

A recurring focus in several definitions is on the function or purpose of the model. ISO/IEC 22989:2022 \citep{ISOIEC22989:2022} (according to \citet{Hupont2024Cards} and \citet{Fernandez-Llorca2025Account}), \textit{Executive Order 14110} (according to \citet{Trincado2024Legal}), \textit{Glossary of human-centric artificial intelligence} \citep{EstevezAlmenzar2022Glossary}, and \textit{California Senate Bill 53} \citep{California2025SB53} describe a model as generating inferences or predictions from input data. Moreover, the \citet{OECD2022Classification} describes AI models as representing, describing and interacting with real or virtual environments, and \citet{MindFoundry2026Model} describes the function of an AI model as pattern recognition in order to make predictions or perform specific tasks. Some sources phrase the function of a model slightly differently, suggesting that a model is \textit{used} for inference \citep{Dunietz2024EUUSTaxonomy,OECD2024Explanatory,ISOIEC22989:2022}. One source specifies that a model has the capacity to adapt to other tasks, calling an AI model a ``universal approximator'', which might indicate a focus on machine learning \citep{Li2022SAIBench}.

Regarding machine learning models, three definitions address their functions. Similar to the broader definitions of AI model functions, ISO/IEC 22989:2022 \citep{ISOIEC22989:2022} defines a machine learning model as generating inferences or predictions based on input data or information. \citet{Kallab2023SML} and \citet{Muhlhoff2025Updating} identify pattern recognition as the primary function of a machine learning model, resembling the AI model definition by \citet{MindFoundry2026Model}.

\subsection{AI System}

\paragraph{Parent Categories of AI System}

\textbf{AI System as a System}

AI systems have been defined as several kinds of systems. Particularly prominent in definitions by the OECD and the EU is the definition of AI systems as a type of machine-based system \citep{OECD2019Scoping,OECD2022Classification,OECD2024Explanatory,EUParliament2024AIAct,EUCouncil2024Convention,EDPS2025Risk}. Several studies also reference various definitions by OECD and the EU when defining an AI system as a machine-based system \citep{Hupont2024Cards,Trincado2024Legal,Boone2023Challenge,Fernandez-Llorca2025Account,Krafft2021Action}. Apart from OECD and the EU, other sources also use machine-based systems as a parent category (\citealp{NIST2023RMF} also referenced by \citealp{Boone2023Challenge}); \citealp{Colorado2024SB24-205}; \citealp{Straub2023Government}).

Several other terms have been used to describe which type of system is the parent category of AI system. NIST defines AI systems as ``engineered systems'' (\citealp{NIST2023RMF} also referenced by \citealp{Boone2023Challenge}). This term is also used by ISO/IEC 22989:2022 \citep{ISOIEC22989:2022} and GB/T 41867-2022 \citep{PRC2022GB/T41867-2022}. Several EU documents define AI systems as software or hardware systems (\citealp{HLEGAI2019Ethics,HLEGAI2020Assessment}; \citealp{EC2021Proposal} also referenced by \citealp{Boone2023Challenge} and \citealp{Buiten2024Liability})). Broader terms found in the literature are ``artificial intelligence-based system'' (\citealp{EC2018Communication}; \citealp{ISOIECIEEE29119-1:2022-3.6}; \citealp{ROKAssembly2025Framework}; \citealp{Coleman2021Understanding} referring to \citealp{Russell2009AI}), ``computer system'' (\citealp{Straub2023Government} referring to \citealp{Gil2019Roadmap}), and most broadly ``system'' (\citealp{EstevezAlmenzar2022Glossary}; \citealp[Article 3(1)]{EUCouncil2022AIAct} as cited by \citealp{Fernandez-Llorca2025Account}; \citealp{EOPresident2023Safe} as cited by \citealp{Trincado2024Legal}; \citealp{EC2021Proposal} as cited by \citealp{Boone2023Challenge} and \citealp{Korikov2018Robot}).

\vspace{0.3cm}
\textbf{AI/AI system as a Technology}

While most definitions describe AI systems as systems, a few either broadly define them as technologies, or define AI as AI systems. \citet{UNESCO2022Ethics} uses the term ``information-processing technologies'' to describe AI systems. The European Commission refers to AI as systems in two of their communications \citep{EC2018Communication,HLEGAI2019Definition}. The SLR also found two national government documents defining AI as a system. \citet{Hjaltalin2024Public} refer to \textit{Leading the way into the age of artificial intelligence: Final report of Finland's Artificial Intelligence Programme 2019} \citep{Ministry2019Leading}, defining AI as ``devices, software and systems that are able to learn and make decisions in almost the same manner as people.'' Similarly, \citet{Spivakovsky2023Policies} cite the \textit{Resolution of the Cabinet of Ministers of Ukraine of December 2, 2020, No. 1556-r: On the Approval of the Concept for the Development of Artificial Intelligence in Ukraine} \citep[in Ukrainian]{CMUkraine2020Resolution1556}, calling AI an ``organized set of information technologies'', using a ``system of scientific research methods and algorithms for processing information.

\paragraph{Features of AI System}

\textbf{Production (How It Is Built)}

Different definitions specify how an AI system is built and what it is made of. The EU AI Act and related documents, as referenced by \citet{Buiten2024Liability} and \citet{Boone2023Challenge} identify three main categories of techniques and approaches underpinning AI systems: (i) machine-learning approaches, (ii) logic-and knowledge-based approaches, and (iii) statistical, Bayesian estimation, search, and optimisation methods. These categories encapsulate the methodological diversity of AI systems and reflect the broad regulatory scope intended to encompass current and emerging techniques. Similarly, the \citet{HLEGAI2019Ethics,HLEGAI2020Assessment} specify that an AI system can either use symbolic rules or learn a numeric model. \citet{UNESCO2022Ethics} provides some explanation on the methods of developing an AI system as well, grouping them into (i) machine learning, which entails deep learning and reinforcement learning, and (iii) machine reasoning, which is specified as planning, scheduling, knowledge representation and reasoning, search and optimisation. 

Other definitions are less specific about how a system is built. The \citet{CHT2024Society} describes AI systems as being ``more ‘grown’ or ‘trained’ than they are programmed, produced by rewarding an AI for ‘good behavior’ and punishing it for ‘bad behavior’'', suggesting a focus on machine learning. GB/T 41867 (2022) keeps the definition similarly broad, stating that AI systems use ``AI-related technologies and methods.'' The \citet{EstevezAlmenzar2022Glossary} communicates this even more broadly, describing AI systems as containing artificial intelligence.

Some studies provide further descriptions of what a system consists of. \citet{Hupont2024Cards} and \citet{Lu2024Taxonomy} both describe an AI system as consisting of one or more AI models, infrastructure, data pipelines, user interfaces, and other modules. Furthermore, an AI system has been described as an interacting ensemble of interdependent components operating within an integrated structure (\citealp{Martinho2021Computer} referring to \citealp{Backlund2000Definition}). Lastly, \citet{Buiten2023Law} describes AI systems as incorporating algorithms that require training data, model architecture, and decision-making rules.

\vspace{0.3cm}
\textbf{Function (What It Does)}

According to the literature, the defining function of an AI system, especially in regulatory and policy contexts, is to generate inferences or outputs from inputs and to influence its environment (physical or virtual). This is reflected in numerous sources (\citealp{OECD2019Recommendation} as cited by \citealp{Hupont2024Cards}); \citealp[Article 3(1)]{EUParliament2024AIAct}; \citealp[Article 3(1)]{EUCouncil2022AIAct} and \citealp[Article 3(1)]{EUParliament2023Amendments} as cited by \citealp{Fernandez-Llorca2025Account}); \citealp{EUCouncil2024Convention}; \citealp{EDPS2025Risk}; \citealp[NIST AI 100-1]{NIST2023RMF}; \citealp[ISO/IEC 22989:2022]{ISOIEC22989:2022}; \citealp[Article 2]{ROKAssembly2025Framework}; \citealp[Senate Bill 24-205]{Colorado2024SB24-205}; \citealp[GB/T 41867-2022]{PRC2022GB/T41867-2022}; \citealp{Poretschkin2023Guideline}; \citealp{Krafft2021Action}). The  \citet{OECD2019Scoping,OECD2022Classification,OECD2024Explanatory} has a very similar formulation, suggesting an AI system makes recommendations, predictions or decisions for a given set of objectives. Moreover, ISO/IEC 22989:2022 \citep{ISOIEC22989:2022} describes an AI system’s function as generating outputs such as content, forecasts, recommendations or decisions for a given set of human-defined objectives (similarly found in \textit{GB/T 41867-2022} \citep{PRC2022GB/T41867-2022}). 

Several other specifications are made regarding the function of an AI system. The EU AI Act (as referenced by \citet{Buiten2024Liability}) stipulates that systems operate toward human-defined objectives. Other sources specifying this requirement are ISO/IEC 22989:2022 \citep{ISOIEC22989:2022} and \textit{GB/T 41867-2022} \citep{PRC2022GB/T41867-2022}. \citet{UNESCO2022Ethics} describes a system’s function as prediction and decision-making. Other interpretations describe AI systems as mechanisms for interpreting external data (\citealp{Diaferia2022Standard} referring to \citealp{Kaplan2019Siri}; \citealp{Manser2021Servitization} referring to \citealp{Haenlein2019History}), or performing pattern recognition \citep{Krafft2021Action}. Further, systems are described as completing cognitive tasks (\citealp{Straub2023Government}; \citealp{Diaferia2022Standard} referring to \citealp{Kaplan2019Siri}; \citealp{UNESCO2022Ethics}), problem-solving (\citealp{Coleman2021Understanding} referring to \citealp{Russell2009AI}), perceiving, reasoning, learning, and interacting (both with their environment and internally) (\citealp{Diaferia2022Standard} referring to \citealp{Rai2019Hybrids}; \citealp{Martinho2021Computer} referring to \citealp{Backlund2000Definition}).

AI systems are often described as exhibiting varying degrees of autonomy (\citealp{OECD2019Recommendation} as referenced by \citealp{Hupont2024Cards}; \citealp[EU AI Act]{EUParliament2024AIAct} as referenced by \citealp{Trincado2024Legal} and \citealp{Ojanen2025Neutrality}; \citealp{NIST2023RMF} as cited by \citealp{Boone2023Challenge}; \citealp{Straub2023Government}; \citealp{OECD2022Classification,OECD2024Explanatory}; \citealp{UNESCO2022Ethics}; \citealp[Article 3(1)]{EUParliament2024AIAct}; \citealp[Article 2]{EUCouncil2024Convention}; \citealp{EDPS2025Risk}; \citealp[NIST AI 100-1]{NIST2023RMF}; \citealp[ISO/IEC 22989:2022]{ISOIEC22989:2022}; \citealp[Article 2]{ROKAssembly2025Framework}; \citealp[GB/T 41867-2022]{PRC2022GB/T41867-2022}). Definitions of autonomous systems expand on this by describing entities capable of acting on their own rules, with self-contained decision-making capacity (\citealp{Buiten2023Law} as referred to by \citealp{Buiten2024Liability}; \citealp{Song2021Concepts} referring to \citealp{Helle2016Testing}; \citealp{Flammini2021Twins}). Further attributes of autonomous systems found in the literature include self-awareness (\citealp{Song2021Concepts} referring to \citealp{Helle2016Testing}), adaptivity (\citealp{Song2021Concepts} referring to \citealp{Helle2016Testing} and \citealp{Sifakis2019Autonomous}; \citealp{Torkjazi2022Taxonomy} referring to \citealp{Antsaklis2018Control}), and the ability to learn, diagnose failures, reconfigure, and plan (\citealp{Torkjazi2022Taxonomy} referring to \citealp{Antsaklis2018Control}). \citet{Song2021Concepts} quote \citet{Sifakis2019Autonomous} to identify five core modules of autonomy: perception, reflection, planning, goal management, and self-adaptation. Similarly, \citet{Torkjazi2022Taxonomy} refer to \citet{Huang2007Unmanned} to outline ``Root Autonomous Capabilities (RACs)'': sensing, perceiving, analysing, communicating, planning, deciding, and acting toward goals. They also refer to Woudenberg et al. (2020) to highlight three dimensions of autonomy: intelligence, operational independence, and collaboration with other systems.

Multiple other attributes are ascribed to AI systems. Some definitions describe AI systems as showing various degrees of adaptivity (\citealp{Hupont2024Cards} referring to \citealp{OECD2019Recommendation}; \citealp{Trincado2024Legal} and \citealp{Ojanen2025Neutrality} referring to the \citealp[EU AI Act]{EUParliament2024AIAct}; \citealp{Manser2021Servitization} referring to \citealp{Haenlein2019History}; \citealp{OECD2024Explanatory}; \citealp{HLEGAI2019Definition,HLEGAI2019Ethics,HLEGAI2020Assessment}; \citealp[Article 2]{EUCouncil2024Convention}; \citealp{EDPS2025Risk}; \citealp[Article 2]{ROKAssembly2025Framework}). Another prominent feature is learning ability (\citealp{Straub2023Government}; \citealp{Diaferia2022Standard} referring to \citealp{Kaplan2019Siri}; \citealp{Manser2021Servitization} referring to \citealp{Haenlein2019History}; \citealp{Coleman2021Understanding} referring to \citealp{Russell2009AI}; \citealp{UNESCO2022Ethics}). Lastly, intelligence is mentioned by some sources as well (\citealp{Straub2023Government} referring to \citealp{Gil2019Roadmap}; \citealp{Korikov2018Robot} referring to \citealp{Makarov2001Intelligent} and \citealp{Korikov2011Mechatronics}; \citealp{EC2018Communication}). Further definition on the aspect of intelligence is provided by \citep{Coleman2021Understanding}, who defines an intelligent system as one that manipulates facts and concepts and displays behaviours across different levels of Bloom’s taxonomy \citep{Bloom1956Taxonomy}.

One definition describes the function of a machine learning system: producing correct outputs without explicit programming (\citealp{Myllyaho2022Misbehaviour} referring to \citealp{Mitchell2006Discipline}).

\vspace{0.3cm}
\textbf{Mechanism (How It Does It)}

Apart from analysing what a system does, it is important to consider how it does this. The OECD has a clear definition for this, explaining that AI systems use machine and/or human-based inputs for various activities, such as perceiving its environment, abstracting these perceptions into models, and using model interpretations to produce an outcome \citep{OECD2019Scoping,OECD2022Classification}. In a similar manner, the \citet{HLEGAI2019Definition,HLEGAI2019Ethics,HLEGAI2020Assessment} describes AI systems as acting in the physical or digital world and achieving their function by perceiving their environment, interpreting the collected structured or unstructured data, reasoning on the knowledge derived from this data and deciding the best action(s) to take (according to pre-defined parameters) to achieve the given goal. This is also found by \citet{EC2018Communication}, describing AI systems as analysing the environment and taking actions to achieve their goals. \citet{UNESCO2022Ethics} explains that knowledge modelling, representation, data and calculations are the means by which AI systems reach their goals. Broadly, AI systems have also been described as typically operating in whole or in part using AI (\citealp{Trincado2024Legal} referring to \citealp[Executive Order 14110]{EOPresident2023Safe}). \citet{MIRI2026Guide} focuses on the safety aspect of an AI system, explaining that some AI systems may possess the capability to resist intervention by programmers, while others may be ``corrigible''.

Lastly, machine learning systems have also been described in terms of its mechanism. They are defined as improving their behaviour through experience and having the ability to adapt and predict situations (\citealp{Myllyaho2022Misbehaviour} referring to \citealp{Mitchell2006Discipline}).

\subsection{Relationship between AI Model and AI System}

\paragraph{Model as a Core Component of a System}

Several sources describe the AI model as an integral part of a broader system. ISO/IEC 22989:2022 \citep{ISOIEC22989:2022} describe AI models as part of AI systems, used to make predictions. The \citep{OECD2019Scoping} and \citep{OECD2024Explanatory}, and \citep{Dunietz2024EUUSTaxonomy} share this perspective, describing an AI model as the core component of an AI system. Recital 97 of the EU AI Act \citep{EUParliament2024AIAct} and the \textit{Guidance for Risk Management of Artificial Intelligence Systems} \citep{EDPS2025Risk} confirm this view, by identifying AI models as essential components of AI systems, while also emphasizing that they require other software components to be able to function and interact with users and the virtual or physical environment. In addition, \citet{Trincado2024Legal} refers to \textit{Executive Order 14110} \citep{EOPresident2023Safe} when defining an AI model as a component of an information system, and both \citet{Hupont2024Cards} and \citet{Lu2024Taxonomy} describe AI systems as containing one or more models.

Regarding definitions of machine learning models, only \citet{Myllyaho2022Misbehaviour} consider the machine learning model as a core component, broadly referring to them as elements within a machine learning system.

\paragraph{Model as a Tool for a System}

Some definitions describe the relationship between AI model and AI system in terms of how an AI system uses an AI model as a tool. The definition of ISO/IEC 22989:2022 \citep{ISOIEC22989:2022} fits not just the perspective of the model as a core component of a system, but also this one, as it specifies AI systems use AI models to make predictions and conduct tasks. Two other sources apart from ISO/IEC 22989:2022 \citep{ISOIEC22989:2022} refer to AI models as tools for inference in an AI system, both also suggesting AI models are core components of systems \citep{Dunietz2024EUUSTaxonomy,OECD2024Explanatory}. The OECD’s definition provided in 2019 \citep{OECD2019Scoping} differs slightly, specifying that an AI system uses models to ``formulate options for outcomes.'' Another source that describes AI models not just as core components but also as tools is \citet{Trincado2024Legal}, citing  \textit{Executive Order 14110} \citep{EOPresident2023Safe}, and suggesting that AI models implement AI technology within an AI system, providing outputs from a set of inputs. Not defining AI models as core components of AI systems, \citet{UNESCO2022Ethics} defines AI systems as integrating models that give them the capacity to learn and perform cognitive tasks. Moreover, \citet{Park2024IMO} describe AI models as implementing AI functions within systems. Lastly, and referring to AI systems but only machine learning models, \citet{Poretschkin2023Guideline} defines machine learning models as the ``functional basis'' of an AI application.

\end{document}